\documentclass[twocolumn,twocolappendix]{aastex631}

	\usepackage{threeparttable}%
	\usepackage{mathtools}
	\usepackage[T1]{fontenc}
	\usepackage{tabularx}

	\shorttitle{
		Alignments in Powerful Radio Galaxies}
	\shortauthors{Breiding et al.}
	\graphicspath{{./}{figures/}}
\begin{document}
			
			%\title{The Alignment Effect at Cosmic Noon: Powerful Radio Jets and Host-Galaxy Assembly Share a Common Axis}
			
			%\title{Powerful Radio Jets Trace the Axis of Galaxy Assembly at Cosmic Noon}
			
			\title{Powerful Radio Jets and Galaxy Assembly Align at Cosmic Noon}

			\correspondingauthor{Peter Breiding}
			\email{pbreiding@gmail.com}\\
			
			\author[0000-0003-1317-8847]{Peter Breiding}
			\affiliation{Department of Physics, Applied Physics, and Astronomy, Binghamton University, Binghamton, NY 13902, USA}
			
			\author[0000-0003-1564-3802]{Marco Chiaberge}
			\affiliation{The William H.  Miller III Department of Physics \& Astronomy, Johns Hopkins University, Baltimore, MD 21218, USA}
			\affiliation{Space Telescope Science Institute for the European Space Agency (ESA), ESA Office, 3700 San Martin Drive, Baltimore, MD 21218, USA}

			\author[0000-0002-7676-9962]{Eileen T.  Meyer}
			\affiliation{Department of Physics, University of Maryland Baltimore County, Baltimore, MD 21250, USA}

			\author[0000-0002-6875-1543]{Bryan Hilbert}
			\affiliation{Space Telescope Science Institute, 3700 San Martin Drive, Baltimore, MD 21218, USA}
			
			\author[0000-0001-6004-9728]{Mojegan Azadi}
			\affiliation{Center for Astrophysics \textbar\ Harvard \& Smithsonian, 60 Garden Street, Cambridge, MA, 02138, USA}
			
			\author[0000-0002-4735-8224]{Stefi A.  Baum}
			\affiliation{Department of Physics and Astronomy, University of Manitoba, Winnipeg, MB R3T 2N2, Canada}
			\affiliation{Center for Space Plasma \& Aeronomic Research, University of Alabama in Huntsville, Huntsville, AL 35899, USA}
			
			\author[0000-0002-7284-0477]{Martin Haas}
			\affiliation{Astronomisches Institut, Ruhr-University, Bochum, Germany}
			
			\author[0000-0003-2884-7214]{George Miley}
			\affiliation{Leiden Observatory, Leiden University, P.O.Box 9513, NL-2300 RA, Leiden, The Netherlands}
			
			\author[0000-0001-6421-054X]{Christopher P.  O'Dea}
			\affiliation{Department of Physics and Astronomy, University of Manitoba, Winnipeg, MB R3T 2N2, Canada}
			\affiliation{Center for Space Plasma \& Aeronomic Research, University of Alabama in Huntsville, Huntsville, AL 35899, USA}
			
			\author[0000-0002-3099-1664]{Eric S.  Perlman}
			\affiliation{Florida Institute of Technology, 150 W.  University Blvd, Melbourne, FL 32901, USA}
			
			\author[0000-0003-1809-2364]{Belinda Wilkes}
			\affiliation{Center for Astrophysics \textbar\ Harvard \& Smithsonian, 60 Garden Street, Cambridge, MA, 02138, USA}
			\affiliation{H.  H.  Wills Physics Laboratory, University of Bristol, Bristol BS8 1TL, UK}

			\author[0000-0002-9895-5758]{S.\ P.\ Willner}
			\affiliation{Center for Astrophysics \textbar\ Harvard \& Smithsonian, 60 Garden Street, Cambridge, MA, 02138, USA}
			
			\author[0000-0002-5222-5717]{Colin Norman}
			\affiliation{The William H.  Miller III Department of Physics \& Astronomy, Johns Hopkins University, Baltimore, MD 21218, USA}
			\affiliation{Space Telescope Science Institute, 3700 San Martin Drive, Baltimore, MD 21218, USA}

			%% depreciated in this version as it is no longer necessary.  AASTeX 
			
			%% provide the collaboration status of a group of authors.  These commands 
			%% can be used either before or after the list of corresponding authors.  The
			%% argument for \collaboration is the collaboration identifier.  Authors are
			%% encouraged to surround collaboration identifiers with ()s.  The 
			

			\begin{abstract}
				
				Powerful radio galaxies at cosmic noon ($1<z<2$) offer a unique laboratory for investigating jet-driven supermassive black hole (SMBH) feedback on their host galaxies and surrounding environments.  Unlike broad-lined quasars, radio galaxies lack a dominant nuclear point source in the rest-frame optical--UV, enabling direct constraints on host-galaxy structure and its relation to the radio jet.  We analyze Hubble Space Telescope (HST) WFC3/IR (rest-frame optical) and WFC3/UVIS (rest-frame UV) imaging of the $z>1$ 3CR radio galaxies.  Beyond confirming the classical rest-frame UV alignment effect, we find that the radio jets preferentially align with the major axes of the smooth rest-frame optical hosts, extending the alignment to the older stellar component.  The projected axes of candidate mergers in eight systems
				preferentially follow the radio axes.  This multiscale coherence suggests a common projected axis linking the radio source to host assembly, perhaps established by repeated mergers or anisotropic infall.  We test scattered quasar light and jet-triggered star formation as origins of the UV emission along this axis.  The much shallower-than-expected radial decline in UV surface brightness disfavors simple scattering as the dominant population-wide origin, while ionized-gas masses further disfavor electron scattering.  The required dust masses are small relative to global reservoirs, so dust scattering remains feasible in individual regions.  Comparing stellar-population ages and jet-propagation times, we identify seven UV-emitting regions as strong candidates for recently initiated jet-triggered star formation.  Together, these alignments reveal an assembly-linked direction along which jets may shock or compress material and hidden-quasar radiation may illuminate it.
				
			\end{abstract}

		\keywords{Active galactic nuclei (16) --- AGN host galaxies (2017) ---
			Radio galaxies (1343) --- Radio jets (1347) ---
			High-redshift galaxies (734) --- Galaxy structure (622) ---
			Galaxy mergers (608) --- Star formation (1569) ---
			Stellar populations (1622)}
		
		%% tied to the reference list via symbolic KEYs.  The KEY corresponds
		
		\section{Introduction}
		\label{sec:intro}
		
		The robust correlations observed between supermassive black hole (SMBH) mass
		and the stellar properties of galactic bulges have long suggested that SMBH
		growth is closely linked to the assembly and evolution of the galaxies that
		host them
		\citep[e.g.,][]{magorrian+98,ferrarese+00,gebhardt+00}.  Feedback from an
		active galactic nucleus (AGN) is often invoked as a possible mechanism
		coupling SMBH growth to the evolution of its host galaxy
		\citep[see][for a review]{kormendy_and_ho_13}.  In radiatively efficient
		quasars, radiation pressure and associated accretion-disk winds can heat or
		expel cold gas, thereby reducing the fuel available for star formation and
		producing ``negative feedback''
		\citep[see reviews by][]{fabian12,heckman+14,heckman+23}.  Under some
		conditions, however, shocks driven by quasar outflows propagating through a
		dense, clumpy interstellar medium (ISM) may instead produce ``positive
		feedback'' by promoting star formation.  Although the shock initially heats
		and compresses the ambient gas, dense clouds with sufficiently short cooling
		times can radiate much of the deposited energy while remaining confined by
		the pressure of the outflow.  If the compressed gas cools and fragments before
		it is dispersed, it may become gravitationally unstable and collapse to form
		stars \citep[e.g.,][]{nayakshin+12, zubovas+13}.

		In radio-loud AGNs, jets provide another means of driving shocks into the
		host-galaxy ISM.  As a jet propagates through the ISM it
		drives a bow shock and inflates an overpressurized cocoon whose pressure can
		compress cool, dense clouds and trigger star formation
		\citep[e.g.,][]{rees89,silk+09,silk13}.  Numerical simulations show that this
		interaction can enhance star formation locally even as gas dispersal and
		turbulence suppress it elsewhere in the same galaxy.  The balance depends on
		the gas density, structure, and cooling efficiency, as well as the jet power,
		orientation, and interaction timescale
		\citep[e.g.,][]{gaibler+12,mukherjee+18,mandal+21}.  Cloud size is another
		important factor: smaller clouds are more efficiently ablated and dispersed,
		whereas larger clouds are more likely to remain intact and become susceptible
		to pressure-triggered collapse \citep{wagner+12}.
		
		Observations likewise suggest that both outcomes occur.  In 3C~326~N, a
		massive molecular disk forms stars at least a factor of 60 below the
		Kennicutt--Schmidt relation.  Its broad, shock-excited H$_2$ emission suggests
		that jet-driven turbulence keeps the gas gravitationally unbound, thereby
		suppressing star formation \citep{nesvadba+11}.  At the population level,
		far-infrared-derived star formation rates were found to be $\sim3$--$5$ times
		higher in radio-detected than X-ray-only AGNs.  Together with the higher
		inferred jet powers of the radio-detected samples, this association was
		interpreted as possible evidence for jet-enhanced star formation
		\citep{zinn+13}.
		
		Spatially resolved evidence consistent with localized triggering includes
		Centaurus~A, where young stars lie near the edge of a radio lobe within the
		host-galaxy halo \citep{mould+00}.  Minkowski's Object provides an external
		example: it is a young starburst galaxy located near the outer end of the jet
		from the neighboring radio galaxy NGC~541 \citep{croft+06}.  In another
		possible external case, the star-forming companion of HE~0450$-$2958 is
		aligned with a quasar radio lobe and exhibits shock-like emission-line ratios
		\citep{feain+07}.  Higher-resolution VLA imaging subsequently resolved a
		bipolar jet from the quasar and revealed spectral flattening at its projected
		intersection with the companion, providing further evidence of jet--ISM
		interaction \citep{molnar+17}.
		
		Studies of the relative orientations of radio galaxies and their
		optical host galaxies have a long history \citep[see][]{miley80}.  A longstanding phenomenon that may provide evidence for positive jet
		feedback is the alignment effect, in which the rest-frame optical/UV
		emission of radio galaxies at $z\gtrsim0.6$ tends to be elongated along
		the radio axis, as first shown by \citet{chambers+87} and
		\citet{mccarthy+87} \citep[see also the review by][]{miley+08}.  The effect appears to be more prominent at higher
		radio luminosities \citep{dunlop+93}.  The rest-frame optical/UV
		morphology also varies with radio extent: smaller radio sources exhibit
		more strongly aligned, knotty emission, whereas larger sources contain
		fewer and more diffuse components \citep{best+96}.  This dependence on
		radio size and luminosity, together with the more disturbed gas
		kinematics and shock-like line ratios observed in smaller sources, has
		been interpreted as evidence that young radio sources interact strongly
		with the dense ISM of their host galaxies \citep{best+00}.
		
		A recent HST study similarly found radio-aligned UV knots in
		five of seven compact steep-spectrum radio galaxies at $z<0.6$ attributable to recent star formation, while
		neither of two galaxies with larger radio sources showed UV emission
		attributed to star formation \citep{duggal+24}.  Although the small
		sample and uncertain relationship between the stellar and radio-source
		ages preclude a firm evolutionary interpretation, these results suggest
		that jet-induced star formation may be favored while the radio source
		is propagating through the dense host-galaxy ISM.
		
		Jet-driven shocks and the pressure of the expanding radio cocoon may
		compress the host-galaxy ISM and promote star formation along the radio
		axis, making jet-triggered star formation a leading explanation for the
		alignment effect.  A well-studied example is 4C~41.17 at $z=3.8$, where
		combined stellar, polarization, emission-line, and radio evidence points towards star formation triggered by jet activity
		\citep{dey+97,bicknell+00}.
		
		A second leading hypothesis for the alignment effect is scattered light from an obscured quasar.
		In the orientation-based scheme for type~2 radio galaxies, nuclear optical/UV light from an obscured quasar
		escapes preferentially along the polar directions and is scattered into our
		line of sight by dust or free electrons \citep[][]{antonucci_miller_85,barthel89}.  This can produce extended UV
		emission along the radio axis with high linear polarization and an
		electric-vector position angle approximately perpendicular to the radio and
		UV axes.  Detection of this signature in several high-redshift radio galaxies
		demonstrates that scattered AGN light can make an important contribution to the aligned rest-frame optical/UV continuum
		\citep{disergo+89,dey+96,vernet+01}.
		
		Warm ionized gas can also contribute to the aligned structures through strong
		rest-frame UV emission lines and free--free, free--bound, and two-photon
		nebular continuum emission.  In three powerful radio galaxies at $0.31\leq z\leq3.8$, nebular continuum contributed up to $\sim60\%$ of the rest-frame UV emission in individual extended regions \citep{dickson+95}.  The gas may be ionized by the obscured AGN, young stars, or jet-induced shocks.

		\subsection{The Radio Galaxy Sample}
		\label{sample}
		
		The primary goal of this study is to investigate how powerful radio jets interact with and potentially shape the structures, stellar populations, and environments of their host galaxies during the peak epoch of galaxy assembly.  To this end, we analyze HST WFC3/IR F140W imaging, which
		primarily traces the older stellar component, together with WFC3/UVIS F606W
		imaging, which is more sensitive to recently formed stars, for the 35 type~2
		radio galaxies in the $z>1$ 3CR sample.  We intentionally restrict our
		analysis to radio galaxies because the obscured nucleus minimizes
		contamination from a bright quasar point-spread function (PSF), enabling a
		more reliable characterization of the host-galaxy morphology and spatially
		resolved stellar populations.  We combine these HST observations
		with radio maps that define the jet axes and projected radio-source extents.
		Together, these data allow us to quantify the relationships among the radio
		jets, the rest-frame optical host galaxies, the UV emission, and
		nearby candidate companions, and to test whether jet-triggered star formation
		or scattered quasar light can account for the aligned UV structures.
		
		The parent $z>1$ 3CR sample comprises 64 radio-loud AGN drawn from the
		low-frequency, flux-limited Third Cambridge Revised catalog
		($F_{178\,\mathrm{MHz}}>9$~Jy; $\delta>-5^{\circ}$)
		\citep{spinrad+85}.  These powerful radio sources
		predominantly exhibit edge-brightened FR~II morphologies
		\citep{fanaroff_filey_74}.  All but two of the $z>1$ 3CR
		sources reside at $1\lesssim z\lesssim2$, within the epoch commonly referred
		to as cosmic noon, when both AGN activity and cosmic star formation were near
		their peak \citep{madau+14,aird+15,alberts+16}.  Consequently, the $z>1$ 3CR
		sample is well suited for studying the interplay between radio-jet feedback
		and host-galaxy evolution.
		
		Throughout this paper, unless otherwise specified, the term \emph{host
			galaxy} refers to the smooth rest-frame optical stellar component traced by
		the WFC3/IR F140W imaging and dominated by older stellar populations.  We refer separately to the rest-frame UV emission detected in the WFC3/UVIS F606W images, whose physical
		origin and relationship to the radio-source geometry are central to this
		work.
		
		The HST and radio observations are described in
		Section~\ref{imaging}, and the image analysis, stellar-population modeling,
		scattering analysis, and alignment measurements are presented in
		Section~\ref{analysis}.  We give
		the results in Section~\ref{results}, discuss their implications in
		Section~\ref{sec:discussion}, and summarize our conclusions in
		Section~\ref{conclusion}.  Throughout this paper we adopt a $\Lambda$CDM
		cosmology with $H_0=67.74$~km~s$^{-1}$~Mpc$^{-1}$,
		$\Omega_{\Lambda}=0.69$, and $\Omega_m=0.31$ \citep{planck+16}.  With this cosmology, an angular separation of $3\arcsec$ corresponds
		to a projected physical separation of approximately 25~kpc,
		accurate to within $\sim$5\% over the entire redshift range $1<z<2.5$.  All
		magnitudes are in the AB system \citep{oke_and_gunn83} and are corrected for
		Galactic extinction following the reddening maps of \citet{schlafly+11}.

		\section{HST and Radio Imaging}
		\label{imaging}

		\subsection{HST Imaging}
		\label{sec:hst_data}
		
		The radio galaxies analyzed in this work were observed as part of the HST-3CR programs SNAP-13023 and GO-16281 (PI: M.  Chiaberge), which together obtained WFC3/IR F140W and WFC3/UVIS F606W imaging of the entire $z>1$ 3CR sample.  The GO program completed the observations of the sources that were not imaged during the SNAP program.  The WFC3/IR observations used the F140W filter, which has a pivot wavelength of 1392.3 nm.  At the median redshift of the sample ($z\approx1.4$), this corresponds to a rest-frame wavelength of approximately 580 nm and primarily traces the optical stellar continuum of the host galaxies' older stellar populations.  The accompanying WFC3/UVIS observations used the F606W filter, which has a pivot wavelength of 588.9 nm and probes the rest-frame near-UV, making it sensitive to much younger stellar populations.
		
		The observational characteristics of the SNAP-13023 and GO-16281 imaging
		are summarized in Table~\ref{tab:hst_obs}, including the filters, pivot
		wavelengths, exposure times, and surface-brightness sensitivity limits.  The
		F140W exposure times were uniform within each program, while the individual
		F606W exposure times varied between visits.  We therefore report the range
		of F606W exposure times and calculate a conservative sensitivity limit for
		the shortest total integration in each program with the online HST WFC3 exposure time calculator (ETC).%: 980~s from two exposures for
		%SNAP-13023 and 1329~s from three exposures for GO-16281.

		\begin{table*}
			\centering
			\caption{HST-3CR Imaging Observations}
			\label{tab:hst_obs}
			
			\begin{threeparttable}
				
				\begin{tabular}{lcccc}
					\hline
					
					HST & Pivot & Project & Exposure$^{\dagger}$ &
					3$\sigma$ Surface \\
					Filter & Wavelength & ID & Time &
					Brightness Limit$^{\ddagger}$ \\
					& (nm) & & (s) & (mag arcsec$^{-2}$) \\
					
					\hline
					
					F140W & 1392.28 & SNAP-13023 &
					$2\times250$ & 24.5 \\
					
					F606W & 588.74 & SNAP-13023 &
					$2\times(490\text{--}530)$ & 23.7 \\
					
					F140W & 1392.28 & GO-16281 &
					$3\times300$ & 24.8 \\
					
					F606W & 588.74 & GO-16281 &
					$3\times(443\text{--}568)$ & 23.8 \\
					
					\hline
				\end{tabular}
				
				\begin{tablenotes}
					
					\item[$\dagger$] Exposure times are shown as the number of dithers multiplied by the exposure time per frame.  F606W exposure durations varied
					by visit; the listed ranges give the shortest and longest
					individual exposures in the sample. 
					
					\item[$\ddagger$] The limits are nominal predictions from the
					HST WFC3 ETC, assuming a
					$2\times2$ pixel extraction area, an elliptical-galaxy
					spectral energy distribution at the median sample redshift
					($z\approx1.4$), and no foreground Milky Way extinction.
					For F606W, we used the shortest total integration in each
					program.
					
				\end{tablenotes}
				
			\end{threeparttable}
		\end{table*}

		All WFC3/IR \texttt{flt} files and WFC3/UVIS \texttt{flc} files were downloaded from the Mikulski Archive for Space Telescopes (MAST) after processing through the standard \texttt{calwf3} reduction pipeline \citep{dressel22}.  The individual dithered exposures were combined using \texttt{AstroDrizzle} \citep{gonzaga+12}, which performs sky subtraction, cosmic-ray rejection, geometric-distortion correction, and image combination.  For the GO-16281 observations, the WFC3/UVIS F606W images were drizzled with \texttt{final\_pixfrac}=0.8 and
		\texttt{final\_scale}=0.03 arcsec pixel$^{-1}$, while the WFC3/IR F140W
		images used \texttt{final\_pixfrac}=0.9 and
		\texttt{final\_scale}=0.06 arcsec pixel$^{-1}$.  For the SNAP-13023
		observations, we used \texttt{final\_pixfrac}=1.0 and \texttt{final\_scale}=0.04 arcsec pixel$^{-1}$ for F606W, and the corresponding values of 0.9 and 0.06 arcsec pixel$^{-1}$ were used
		for F140W.
		
		Although both filters were obtained for all 35 radio galaxies, 3C~173
		and 3C~294 were excluded from the HST-based morphological and
		photometric analyses because reliable host-galaxy measurements could
		not be obtained.  In 3C~173, a prominent unresolved point-source
		component is present in both the F140W and F606W images, preventing a
		robust separation of the extended host-galaxy emission.  In 3C~294, the
		very faint host is strongly contaminated in both bands by the PSF wings
		of a nearby bright star.  Consequently, we did not perform
		\textsc{galfit} modeling or F606W/F140W photometry for either source,
		and neither galaxy was included in the HST-based alignment analyses.
		The usable HST analysis sample therefore comprises 33 radio galaxies.

		\subsection{Radio Imaging}
		\label{radio_imaging}

\begin{deluxetable*}{lcccccclcc}
	\tabletypesize{\scriptsize}
	\setlength{\tabcolsep}{3pt}
	\tablewidth{0pt}
	\tablecaption{Radio Imaging Data and Source Properties
		\label{tab:radio_data}}
	\tablehead{
		\colhead{Source} & \colhead{$z$} & \colhead{Freq.}  &
		\colhead{Project} & \colhead{Obs.  Date} & \colhead{RMS} &
		\colhead{Beam Size} & \colhead{Origin} &
		\colhead{$PA_{\rm radio}$} & \colhead{$D_{\rm radio}$} \\
		\colhead{} & \colhead{} & \colhead{(GHz)} &
		\colhead{Code} & \colhead{YYYY-MM-DD} &
		\colhead{(Jy beam$^{-1}$)} & \colhead{(arcsec)} &
		\colhead{} & \colhead{(deg)} & \colhead{(kpc)}
	}
	\startdata
	3C~13 & 1.35 & 4.86 & AC200 & 1987-10-06 & $5.69\times10^{-5}$ & $0.376\times0.357$ & NVAS & 145.1 & 245.8 \\
	3C~36 & 1.30 & 4.86 & AC114 & 1985-03-11 & $2.45\times10^{-4}$ & $0.391\times0.332$ & NVAS & 21.2 & 90.8 \\
	3C~65 & 1.18 & 1.46 & PERL & 1982-02-28 & $2.73\times10^{-4}$ & $1.284\times1.125$ & NVAS & 98.0 & 189.4 \\
	3C~68.2 & 1.58 & 4.86 & AV164 & 1990-05-10 & $6.46\times10^{-5}$ & $0.529\times0.392$ & NVAS & 154.8 & 221.0 \\
	3C~124 & 1.08 & 4.86 & AM213 & 1987-08-17 & $5.37\times10^{-5}$ & $0.445\times0.394$ & NVAS & 7.8 & 25.9 \\
	3C~194 & 1.18 & 4.86 & AV164 & 1990-05-10 & $2.56\times10^{-4}$ & $0.535\times0.376$ & NVAS & 159.3 & 136.9 \\
	3C~210 & 1.17 & 4.86 & AO230 & 2009-04-19 & $5.09\times10^{-5}$ & $1.214\times1.016$ & HST-3CR & 164.1 & 198.6 \\
	3C~222 & 1.34 & 8.46 & AB978 & 2000-12-09 & $3.42\times10^{-6}$ & $0.256\times0.236$ & This work & 62.2 & 18.1 \\
	3C~225A & 1.57 & 4.86 & AV164 & 1989-04-20 & $1.29\times10^{-4}$ & $1.233\times1.136$ & NVAS & 42.2 & 60.0 \\
	3C~230 & 1.49 & 8.44 & AK403 & 1995-07-27 & $6.95\times10^{-5}$ & $0.316\times0.229$ & HST-3CR & 160.8 & 77.4 \\
	3C~238 & 1.41 & 4.71 & AB978 & 2000-12-09 & $9.07\times10^{-5}$ & $0.459\times0.390$ & NVAS & 162.0 & 84.9 \\
	3C~239 & 1.79 & 4.86 & AM205 & 1987-07-18 & $1.36\times10^{-4}$ & $0.398\times0.331$ & NVAS & 76.6 & 107.4 \\
	3C~241 & 1.62 & 8.44 & AA149 & 1992-11-10 & $1.00\times10^{-4}$ & $0.204\times0.199$ & NVAS & 81.8 & 8.4 \\
	3C~249 & 1.55 & 4.86 & AM213 & 1987-08-17 & $5.44\times10^{-5}$ & $0.474\times0.418$ & This work & 138.7 & 425.8 \\
	3C~250 & 1.26 & 4.86 & AV133 & 1986-04-16 & $4.50\times10^{-5}$ & $0.362\times0.356$ & NVAS & 54.5 & 437.2 \\
	3C~255 & 1.36 & 8.44 & AV157 & 1988-12-22 & $2.88\times10^{-5}$ & $0.295\times0.230$ & HST-3CR & 151.7 & 5.9 \\
	3C~256 & 1.82 & 8.44 & AV164 & 1990-05-10 & $1.57\times10^{-4}$ & $0.316\times0.208$ & NVAS & 133.8 & 41.1 \\
	3C~257 & 2.47 & 8.44 & AV164 & 1990-05-11 & $2.13\times10^{-5}$ & $0.412\times0.226$ & HST-3CR & 133.8 & 101.5 \\
	3C~266 & 1.27 & 4.86 & AL454 & 1998-04-09 & $1.30\times10^{-4}$ & $0.397\times0.348$ & NVAS & 176.7 & 51.5 \\
	3C~267 & 1.14 & 8.44 & AL322 & 1994-02-27 & $6.90\times10^{-5}$ & $0.227\times0.221$ & NVAS & 78.8 & 322.9 \\
	3C~297 & 1.41 & 8.44 & AV164 & 1990-05-10 & $1.91\times10^{-5}$ & $0.357\times0.233$ & HST-3CR & 134.5 & 53.6 \\
	3C~300.1 & 1.16 & 8.44 & AK403 & 1995-07-27 & $3.32\times10^{-5}$ & $0.291\times0.240$ & HST-3CR & 13.6 & 103.1 \\
	3C~305.1 & 1.13 & 14.94 & AM141 & 1985-02-18 & $2.67\times10^{-4}$ & $0.168\times0.096$ & HST-3CR & 10.4 & 19.6 \\
	3C~322 & 1.68 & 4.86 & AV133 & 1986-04-16 & $5.92\times10^{-5}$ & $0.487\times0.297$ & HST-3CR & 2.6 & 309.8 \\
	3C~324 & 1.21 & 8.21 & AB755 & 1995-07-11 & $1.25\times10^{-5}$ & $0.200\times0.200$ & HST-3CR & 70.7 & 103.4 \\
	3C~326.1 & 1.83 & 14.94 & AV133 & 1986-04-16 & $1.35\times10^{-4}$ & $0.173\times0.117$ & HST-3CR & 93.6 & 63.4 \\
	3C~356 & 1.08 & 4.85 & AF186 & 1990-04-22 & $2.48\times10^{-5}$ & $0.350\times0.350$ & HST-3CR & 164.4 & 625.4 \\
	3C~368 & 1.13 & 4.71 & AC420 & 1995-07-31 & $1.13\times10^{-4}$ & $0.518\times0.434$ & NVAS & 16.5 & 149.9 \\
	3C~437 & 1.48 & 4.86 & AV164 & 1990-05-10 & $2.89\times10^{-4}$ & $0.575\times0.430$ & NVAS & 160.4 & 621.8 \\
	3C~454.1 & 1.84 & 14.94 & AM213 & 1987-08-17 & $1.33\times10^{-4}$ & $0.122\times0.106$ & HST-3CR & 167.0 & 17.0 \\
	3C~469.1 & 1.34 & 8.46 & AO170 & 2003-06-17 & $1.01\times10^{-4}$ & $0.321\times0.187$ & NVAS & 171.2 & 656.8 \\
	3C~470 & 1.65 & 8.44 & AL322 & 1994-02-27 & $8.18\times10^{-5}$ & $0.300\times0.201$ & NVAS & 38.0 & 225.7 \\
	4C~13.66 & 1.45 & 8.44 & AL0401 & 1996-12-31 & $1.17\times10^{-4}$ & $0.242\times0.219$ & NVAS & 7.7 & 15.8 \\
	\enddata
	\tablecomments{NVAS = NRAO VLA Archive Survey;
		HST-3CR = HST-3CR archive; This work = radio data reduced and
		imaged for this study.  Beam sizes give the restoring-beam FWHM
		major axis $\times$ minor axis.  $D_{\rm radio}$ is the full projected
		radio-source extent.  The table is also available
		in machine-readable form online.}
\end{deluxetable*}

		We used the radio images to measure the radio-axis position angles,
		nucleus-to-hotspot distances, and total source extents for the
		alignment and jet-timing analyses.  Radio images were assembled from three sources: the NRAO VLA Archive Survey (NVAS; \url{https://www.vla.nrao.edu/astro/nvas/}), archival radio images distributed by the HST-3CR collaboration (\url{https://hz3c.stsci.edu/Observations.html}; \citealt{hilbert+16}), and archival VLA observations of 3C~222 and 3C~249 that we reduced for this work.  The radio datasets we reduced for this work were calibrated, imaged using CLEAN deconvolution, and iteratively self-calibrated following standard radio continuum data reduction procedures.  For 3C 194, 3C 210, 3C 322, 3C 324, 3C 356, and 3C 368, we corrected the HST–radio registration by shifting the radio images to align the radio cores with the host-galaxy centers determined from \textsc{galfit} modeling of the F140W images.  Table~\ref{tab:radio_data} summarizes the radio imaging data and source
		properties for the 33 galaxies retained in the HST analyses, including
		their redshifts, radio-axis position angles, and full projected
		radio-source extents.
		
		%%%%%%%%%%%%%%%%%%%%%
		
		%%%%%%%%%%%%%%%%%%%%%
		
		The primary goal of the radio analysis is to determine the projected orientation of the radio jets.  Most of the $z>1$ 3CR radio galaxies in our sample exhibit classical FR~II morphology.  For these sources, the jet position angle was measured from the axis connecting the hot spot to counter hot spot.  In the majority of cases the radio structures are relatively straight on the scales considered here, resulting in a well-defined jet axis.  
		
Position angles are measured from north through east, with $0^\circ$ corresponding to north
and $90^\circ$ corresponding to east.  All images are displayed with
north up and east to the left.  Because the radio structures considered
here are orientation axes rather than directed vectors, the radio
position angles were folded into the range $0^\circ\leq PA<180^\circ$.

		\section{Data Analysis}
		\label{analysis}
		
		This section describes the primary analyses performed on the HST and radio imaging data described in Section~\ref{imaging}.  We first measure the structural properties of the host galaxies from their rest-frame optical emission and the photometric and morphological properties of the UV-emitting regions.  These measurements are then used to constrain the physical origin of the UV emission through stellar population synthesis modeling.  We then compare the host-galaxy, UV-emission, and merger geometries with the radio jets.

		\subsection{Rest-Frame Optical Host Galaxy Surface Brightness Modeling}
		
		\label{galfit}
		
		To constrain the structural parameters of the host galaxies’ older stellar populations, we performed two-dimensional surface brightness modeling of the rest-frame optical emission using the HST WFC3/IR F140W images and \textsc{galfit} \citep{peng+10}.  Following the same methodology employed in \cite{breiding+24}, we used single S\'ersic models to fit the host galaxy light profiles.  Prior to fitting, we created custom bad-pixel masks to exclude nearby
		sources and localized irregular features, such as tidal structures and
		star-forming regions, so that the fits were driven by the underlying
		undisturbed stellar components of the host galaxies.  In occasional
		cases where a massive companion lay in very close projection to the
		radio galaxy, we modeled the companion simultaneously with an
		additional S\'ersic component rather than masking its pixels.  The
		radio-galaxy host itself was still represented by a single S\'ersic
		component.  The S\'ersic model has the following radial dependence \citep{sersic63,sersic68}:
		
		\begin{equation}
			\mathrm{I(R)=I_{e}~exp\left[-\kappa \left(\left(\frac{R}{R_{e}}\right)^{1/n}-1\right)\right]}.
		\end{equation}
		
		In this expression, $I(R)$ is the surface brightness at radius $R$, $R_{\rm e}$ is the effective radius enclosing half of the total light, $I_{\rm e}$ is the surface brightness at $R_{\rm e}$, $n$ is the S\'ersic index, and $\kappa$ is a constant determined by the value of $n$.  The S\'ersic index governs the overall shape of the radial light profile, with larger values corresponding to more centrally concentrated light distributions. 
		
		For each galaxy in the usable HST analysis sample, we fit a single
		S\'ersic component to the HST/WFC3 F140W image, allowing the
		sky position, total magnitude, effective radius, S\'ersic index, axis
		ratio, position angle, and sky background to vary freely.  The
		best-fitting model was determined through $\chi^2$ minimization,
		and residual images were visually inspected to assess each fit.
		Table~\ref{tab:host_properties} lists potential F140W emission lines
		from the SDSS MaNGA data analysis pipeline reference list
		\citep{belfiore+19}, redshifted into 1.185--1.613~$\mu$m, the range
		defined at 1\% of peak throughput.  Although localized irregular
		features were masked, diffuse line emission overlapping the smooth
		host could remain and affect the fitted structural parameters.
		
		\subsection{UV Morphology from Image Moments}
		\label{image_moments}

		The rest-frame UV emission observed in the HST/WFC3 UVIS F606W images is generally clumpy and, in most sources, is considerably less spatially extended than the rest-frame optical emission detected in the F140W images.  We therefore used the image-segmentation tools in \texttt{photutils} \citep{larry_bradley_2022_6825092} to identify discrete regions of UV emission and assign pixels to each region.  The background level and RMS were estimated using a median background model computed over the full image, and UV-emitting regions were identified as contiguous groups of at least 60 pixels exceeding a detection threshold of $1.5\,\sigma$ relative to the background RMS.  Pixels affected by image artifacts or unrelated sources were masked prior to source detection.
		
		Because the UV emission is often irregular and poorly described by simple parametric profiles, we characterized the morphology of each detected region using the intensity-weighted second moments of its surface brightness distribution.  We used the \texttt{SourceCatalog} routine in \texttt{photutils} to calculate the intensity-weighted centroid and second moments for each region.  The second moments describe the spatial distribution of the emission about its centroid and define an equivalent ellipse whose semimajor axis, semiminor axis, and position angle characterize the extent, elongation, and orientation of the UV emission.  We constructed these ellipses with semimajor and semiminor axes equal to three times the corresponding second-moment dispersions.  These moment-defined ellipses are only used to characterize the extent and orientation of the UV emission and are not used for the photometric measurements described in Section~\ref{sec:photometry}.  %This non-parametric approach is well suited to the irregular, clumpy morphologies commonly observed in high-redshift radio galaxies. 
		We adopted the position angle of the semimajor axis,
		$PA_{\rm UV}$, as the orientation of each extended UV-emitting
		region for comparison with the radio jet axis
		(Section~\ref{sec:alignment}).  These UV-region PAs are listed in
		Table~\ref{tab:uv_region_properties}.
		
		\subsection{Aperture Photometry}
		\label{sec:photometry}

		In order to measure rest-frame optical and UV fluxes from the WFC3/IR F140W and WFC3/UVIS F606W images for the UV-defined emission regions identified in Section~\ref{image_moments}, we performed aperture photometry using the \texttt{photutils} package.  These measurements were used to test competing emission mechanisms for the UV emission, primarily distinguishing ongoing or recent star formation from electron or dust scattering of radiation from an obscured quasar.

		Photometric apertures were defined from the original, unconvolved F606W segmentation maps used for the image-moment analysis (Section~\ref{image_moments}).  Before measuring the fluxes, we convolved the F606W images to match the F140W PSF; the construction and validation of the PSF-matching kernels are described in Appendix~\ref{sec:psf_homogenization}.  The F140W images were reprojected onto the F606W pixel grid so that identical apertures could be applied to both filters.  For the PSF-matched F606W images, a local background was estimated independently for each region from a surrounding annulus.  The median count rate after 3$\sigma$ clipping within this annulus was subtracted on a per-pixel basis from the aperture flux.  For the F140W images, a global sky background was estimated from the full image after masking the host galaxy and other significant sources using a 3$\sigma$-clipped estimate of the remaining sky pixels.  The corresponding sky contribution within each aperture was subtracted from the observed aperture flux.  We then measured the flux of the best-fitting \textsc{galfit}
		host-galaxy model (Section~\ref{galfit}) within the same aperture and
		subtracted it from the sky-subtracted aperture flux.  This subtraction
		removes the contribution from the smooth, older stellar population of
		the host galaxy, allowing us to isolate any excess rest-frame optical
		emission above that expected from the underlying host galaxy.  We
		estimated the statistical uncertainty in each aperture-integrated
		F140W residual from the sigma-clipped background RMS and the equivalent
		number of native F140W pixels in the aperture.  Residuals with
		signal-to-noise ratios below 3 were treated as nondetections, for which
		the corresponding $3\sigma$ flux limit was converted to a lower limit
		on the F140W magnitude, $m_{\rm F140W}>m_{\rm F140W,lim}$. 
		
		To estimate the projected distances of the UV-emitting regions from the nucleus, we measured the separations between the UV-region centroids and the host-galaxy centroid determined from the best-fitting F140W \textsc{galfit} model.  We emphasize that these projected distances represent lower limits to the true three-dimensional separations.  Consequently, the inferred surface brightness of the quasar radiation field at the locations of the UV-emitting regions is an upper limit.  This choice is therefore conservative when evaluating scattering scenarios, since scattered quasar light is more easily produced at smaller distances where the incident nuclear radiation field is strongest.

		\subsection{Modeling Stellar Populations}
		\label{stellar_population_modeling}
		
		To investigate whether the colors of the UV-emitting regions are consistent with emission from young
		stellar populations, we compared the measured rest-frame UV--optical
		colors with stellar population synthesis (SPS) models generated using
		the Flexible Stellar Population Synthesis code
		(\texttt{FSPS}; \citealt{conroy+09,conroy+10}) through the
		\texttt{python-fsps} interface \citep{pythonfsps}.  The models assume a
		Chabrier initial mass function \citep{chabrier03}, solar metallicity, a
		constant star-formation history, and the \cite{calzetti+00} dust
		attenuation law.  Nebular emission and intergalactic absorption were not
		included, so the models represent stellar-continuum-only predictions.
		
		Synthetic photometry was calculated in the observer-frame HST/WFC3
		F606W and F140W filters over a redshift grid spanning
		$1.0\leq z\leq2.5$ in steps of $\Delta z=0.1$.  The duration since the
		onset of constant star formation was sampled at 80 logarithmically
		spaced values between 1 Myr and 3 Gyr.  At each redshift, durations
		exceeding the age of the Universe were excluded.  We considered dust
		attenuations of $A_V=0.0$, 0.3, 0.5, 1.0, 1.5, 2.0, and 3.0 mag.
		
		For each UV-emitting region, the observed color was defined as
		$C_{\rm obs}=m_{\rm F606W}-m_{\rm F140W}$ using the Galactic
		extinction-corrected photometry.  The F140W magnitudes correspond to
		the residual flux remaining after subtraction of the smooth
		host-galaxy model (Section~\ref{sec:photometry}).  These measurements
		therefore test whether the UV emission is accompanied by a spatially
		coincident rest-frame optical excess consistent with a young stellar
		population.
		
		For every combination of redshift, model duration, and dust
		attenuation, we calculated the corresponding synthetic color,
		$C_{\rm model}=m_{\rm F606W}^{\rm SPS}-m_{\rm F140W}^{\rm SPS}$.
		Each source was compared with the model grid evaluated at the nearest
		sampled redshift.  For regions detected in both filters, models were
		retained when their predicted colors agreed with the observed color to
		within 0.20 mag, such that
		$|C_{\rm model}-C_{\rm obs}|\leq0.20$ mag.
		
		We used this deliberately inclusive model-comparison tolerance
		because the host-subtracted F140W knot fluxes are sensitive to small
		uncertainties in the smooth host-galaxy model.  The residual knot flux
		is the difference between the total observed aperture flux and the
		host-model flux.  Consequently, a small fractional change in the host
		model can produce a much larger fractional change in the residual when
		the host dominates the aperture flux.  As an illustrative sensitivity
		test, we rescaled the aperture-integrated \textsc{galfit} host-model
		flux by $\pm1\%$ and recomputed the residual colors.  The larger absolute color change had a median of
		0.035 mag and a 90th percentile of
		0.125 mag.  Thus, the
		0.20-mag threshold encompasses the response to a small
		host-normalization perturbation.  This
		tolerance is an operational model-consistency criterion rather than a
		source-specific uncertainty or formal confidence interval.
		
		We also assessed whether finite sampling of the model grid materially affected the predicted colors.  The 80 model durations have a logarithmic
		spacing of $\Delta\log(t)=0.044$ dex.  Repeating the analysis with 800
		duration values and an interpolated attenuation spacing of
		$\Delta A_V=0.01$ mag produced a maximum coarse-grid color offset of
		0.071 mag.  Evaluating the models directly at each source redshift,
		rather than at the nearest $\Delta z=0.1$ grid point, changed the
		colors by at most 0.026 mag.  These effects, together with the small residual systematic uncertainties from
		PSF matching (Appendix~\ref{app:psf_validation}), are smaller than the
		0.20-mag tolerance.
		
		For the F140W nondetections defined in
		Section~\ref{sec:photometry}, the constraint
		$m_{\rm F140W}>m_{\rm F140W,lim}$ yields an upper limit on the color,
		$C_{\rm obs}<C_{\rm lim}$, where
		$C_{\rm lim}=m_{\rm F606W}-m_{\rm F140W,lim}$.  For these regions, we
		retained models satisfying
		$C_{\rm model}\leq C_{\rm lim}+0.20$ mag.
		
		For each region, we recorded the minimum and maximum model durations
		and dust attenuations among the accepted models.  Because only one
		UV--optical color is available, these ranges are neither unique
		solutions nor formal confidence intervals.  Instead, they describe the
		parameter space permitted under the stellar-continuum model
		assumptions.
		
		\subsection{Jet--Host, Jet--UV, and Jet--Merger Alignments}
		\label{sec:alignment}
		
		To quantify the projected relationships among the radio jets, the
		stellar structures of the host galaxies, the rest-frame UV emission,
		and the nearby companion environment, we compared the position angle
		of each component with the radio-axis position angle measured from the
		radio images described in Section~\ref{radio_imaging}.  For the
		predominantly FR~II radio galaxies in our sample, the radio axis was
		defined by the line connecting the terminal hot spot and counter-hot
		spot.  The only exception is the highly curved ``C-shaped'' jet in the radio galaxy 3C~297.  For this source, we define the radio jet axis to be the line connecting the host galaxy center to the hot spot on the galaxy side relevant to the UV emission.
		
		The rest-frame optical host-galaxy position angle and axis ratio were
		obtained from the best-fitting \textsc{galfit} S\'ersic model in the
		WFC3/IR F140W image and are listed in
		Table~\ref{tab:host_properties}.  The position angle and axis ratio of each UV-emitting region
		were derived from its light-weighted second moments
		(Section~\ref{image_moments}).
		Components with axis ratio $b/a\geq0.8$ were excluded from the
		alignment tests because their position angles are poorly
		constrained.
		
	The host-galaxy, UV-region, and merger-axis position angles were
	expressed using the same astronomical convention as the radio
	position angles (measured from north through east) and folded into
	the same range, $0^\circ\leq PA<180^\circ$.  For each component, we
	calculated the acute projected misalignment angle relative to the
	radio axis as
	
	\begin{equation}
		\Delta\theta =
		\min\left[
		\left|PA_{\rm comp}-PA_{\rm radio}\right|,
		180^\circ-\left|PA_{\rm comp}-PA_{\rm radio}\right|
		\right],
	\end{equation}
	
	where $PA_{\rm comp}$ denotes the host-galaxy, UV-region, or
	merger-axis position angle, and $PA_{\rm radio}$ denotes the
	radio-axis position angle.  The resulting misalignment angles lie
	in the interval $0^\circ\leq\Delta\theta\leq90^\circ$, with smaller
	values indicating closer alignment with the radio axis.
		
		We examined the UV alignment at both the region and galaxy levels.  In
		the region-level analysis, each retained UV-emitting region was treated
		as independent and thus assigned equal weight.  Galaxies
		containing more UV regions therefore contributed
		proportionally more measurements to this analysis.  To obtain one
		galaxy-level UV statistic per source, we also calculated the mean misalignment angles of all individual UV regions
		in that galaxy.  %For galaxies with a single retained UV region, the
		%galaxy-level value is equal to that region's misalignment angle.
		Host-galaxy alignments were represented by one misalignment angle per
		galaxy derived from the \textsc{galfit} major-axis position angle.
		
		To investigate whether the radio axes are preferentially aligned with
		the projected geometry of nearby candidate companions, we constructed
		source-segmentation maps from the F140W images using a $5\sigma$
		detection threshold and a minimum detection area of 20 connected
		pixels.  Overlapping detections were deblended using 16 levels and a
		contrast parameter of 0.01.  The host was identified as the segmented
		source closest to its coordinates.  Candidate companions were
		required to lie at host-centered projected distances
		$5<R_{\rm proj}<25$~kpc, have a segmented area of at least 100 pixels,
		and have an F140W AB magnitude within two magnitudes of the
		host magnitude.  The magnitude threshold corresponds to a
		minimum companion-to-host F140W flux ratio of approximately
		$1:6.3$.  We refer to these systems as candidate mergers and characterize
		their projected geometry using the merger axis defined below.
		This photometric selection does not establish stellar mass ratios
		or confirm physical association. 
		
		When exactly one candidate companion satisfied these criteria, the
		merger axis was defined by the line connecting the host and companion.
		When multiple companions were present, the merger axis was determined
		from a flux-weighted principal-component analysis of the host and
		companion positions.  The eigenvector corresponding to the largest
		weighted positional variance defined the dominant projected
		orientation.  %Its acute misalignment relative to the radio axis was
		%calculated using the same definition as above.
		
		Under the null hypothesis of random axial orientations, individual
		misalignment angles are uniformly distributed between $0^\circ$ and
		$90^\circ$.  We therefore used one-sided one-sample
		Kolmogorov--Smirnov tests to compare the individual UV-region,
		host-galaxy, and 25-kpc merger-axis misalignment distributions with
		this uniform distribution.  The one-sided alternative tested
		specifically for an excess of small misalignment angles.
		
		Because an average of multiple uniformly distributed angles is not
		itself uniformly distributed, the galaxy-level mean UV misalignments
		were evaluated using a Monte Carlo random-orientation test.  In each of
		$10^6$ trials, an independent misalignment angle was drawn from a
		uniform distribution over $0^\circ$--$90^\circ$ for every retained UV
		region.  The observed assignment and number of UV regions per galaxy
		were preserved, and the simulated angles were averaged within each
		galaxy in the same manner as the observations.  The test statistic was
		the mean of the resulting galaxy-level misalignments, with smaller
		values representing stronger alignment.  The one-sided Monte Carlo
		probability was calculated from the fraction of trials producing a
		test statistic no larger than the observed value.

		\section{Results}
		\label{results}

		\begin{deluxetable*}{lcccccccl}
			\tabletypesize{\scriptsize}
			\setlength{\tabcolsep}{3pt}
			\tablewidth{0pt}
			\tablecaption{Rest-Frame Optical Host-Galaxy Properties
				\label{tab:host_properties}}
			\tablehead{
				\colhead{Source} & \colhead{$z$} & \colhead{$m_{140}$} &
				\colhead{$b/a$} & \colhead{$n$} & \colhead{$R_{\rm e}$} &
				\colhead{$PA_{\rm host}$} & \colhead{$\Delta\theta_{\rm host}$} &
				\colhead{Potential F140W lines} \\
				\colhead{} & \colhead{} & \colhead{(AB mag)} &
				\colhead{} & \colhead{} & \colhead{(kpc)} &
				\colhead{(deg)} & \colhead{(deg)} & \colhead{}
			}
			\startdata
			3C~13 & 1.35 & 19.51 & 0.76 & 2.63 & 8.52 & 144.5 & 0.6 &
			\parbox[t]{5.8cm}{\strut H$\alpha$, [N~II], [S~II], [O~I], He~II, He~I,
			[Ar~III], [N~I], Na~I\strut} \\
			3C~36 & 1.30 & 19.62 & 0.77 & 3.03 & 4.32 & 65.8 & 44.6 &
			\parbox[t]{5.8cm}{\strut H$\alpha$, [N~II], [S~II], [O~I], He~II, He~I,
			[Ar~III], [N~I], Na~I\strut} \\
			3C~65 & 1.18 & 19.26 & 0.70 & 3.07 & 10.61 & 27.7 & 70.3 &
			\parbox[t]{5.8cm}{\strut H$\alpha$, [N~II], [S~II], [O~I], He~II, [O~II],
			He~I, [Ar~III], Na~I, [Ar~IV]\strut} \\
			\enddata
			\tablecomments{The complete table is available in machine-readable form.}
		\end{deluxetable*}

		\subsection{Rest-Frame Optical Host-Galaxy Properties}
		\label{sec:results_host_structure}
		
		We obtained successful single-S\'ersic profile fits to the
		HST/WFC3 F140W images of all 33 radio galaxies in this study.
		Table~\ref{tab:host_properties} reports the source redshift,
		F140W apparent magnitude, $m_{140}$, axis ratio
		(semiminor-to-semimajor axis), $b/a$, S\'ersic index, $n$,
		effective radius, $R_{\rm e}$, host-galaxy major-axis position
		angle, $PA_{\rm host}$, and acute host--radio misalignment angle,
		$\Delta\theta_{\rm host}$.  Host PAs use the north-through-east
		convention defined in Section~\ref{sec:alignment}; the radio PAs
		are listed in Table~\ref{tab:radio_data}.  The complete table
		retains all 33 hosts, including those excluded from the alignment
		tests by the $b/a<0.8$ requirement.
		
		To place these measurements in context, we compared the radio galaxies studied here with
		the 25 resolved $z>1$ 3CR quasar hosts analyzed in \citet{breiding+24}.  The two samples are well matched in
		redshift, with median values of 1.36 and 1.38 for the radio galaxies and
		quasars, respectively.  A two-sided two-sample Kolmogorov--Smirnov (KS) test
		does not indicate a difference between their redshift distributions
		($p=0.416$).
		
		We converted the F140W apparent
		magnitudes to rest-frame $V$-band absolute magnitudes, $M_V$.  For both radio galaxy and quasar samples, we used the solar-metallicity
		stellar-population grid from the 2016 update to the Bruzual--Charlot models
		distributed with \texttt{Bagpipes}, assuming a single-burst population formed
		at $z=5$ and evolved passively \citep{bruzual+03,carnall+18}.
		
\begin{figure*}[t]
			\centering
			\includegraphics[width=\textwidth]
			{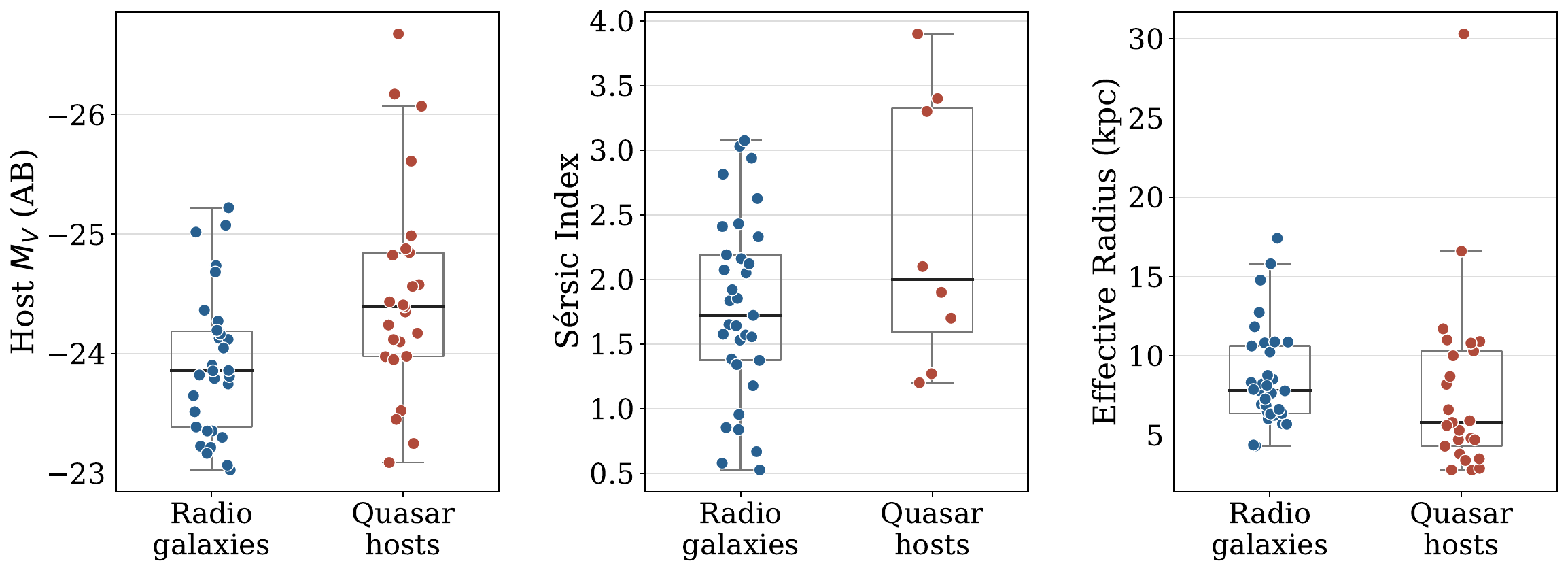}
			\caption{Comparison of the fitted rest-frame optical properties of the 33
				radio galaxies analyzed in this study and the 25 resolved $z>1$ 3CR
				quasar hosts from \citet{breiding+24}.  Individual objects are shown as
				horizontally jittered points.  Boxes span the 25th to 75th percentiles,
				horizontal lines mark the medians, and whiskers extend to the most
				extreme values within 1.5 times the interquartile range.  \textit{Left}:
				rest-frame $V$-band AB absolute magnitude inferred from the F140W host apparent magnitude.  \textit{Middle}: S\'ersic index; the quasar points and box include
				only the eight freely fitted indices.  \textit{Right}: semimajor-axis effective radius.}
			\label{fig:rg_qso_host_comparison}
		\end{figure*}

		Figure~\ref{fig:rg_qso_host_comparison} presents box-and-whisker comparisons
		of $M_V$, $n$, and $R_e$.  %The radio-galaxy and resolved quasar hosts have median absolute magnitudes of $M_V=-23.855$ and $-24.392$ AB, respectively.
		The fitted quasar hosts are brighter than the radio-galaxy hosts by
		0.537~mag at the median.  Their $M_V$ distributions differ according to a
		two-sided KS test ($p=4.58\times10^{-3}$).  A comparison using
		the apparent F140W magnitudes gives a similar
		median difference of 0.636~mag, with the radio-galaxy hosts again fainter than
		the resolved quasar hosts.
		
		Seventeen of the 25 resolved quasar hosts were fitted with $n=4$ fixed.
		Because these indices were assumed rather than measured, we exclude
		them from the middle panel of Figure~\ref{fig:rg_qso_host_comparison}
		and the statistical comparison of S\'ersic indices.  All 33 radio
		galaxies were fitted with $n$ free to vary.  Among the eight quasar hosts
		with freely fitted indices, the median is $n=2.0$, compared with
		$n=1.72$ for the radio galaxies.  A two-sided KS test does not indicate
		a difference between these distributions ($p=0.255$), although the
		quasar sample is small.
		
		The median effective radius of the resolved quasar hosts is 5.80~kpc,
		compared with 7.80~kpc for the radio galaxies.  The full distributions
		differ according to a two-sided KS test ($p=4.94\times10^{-3}$).
		Fixing $n$ could contribute to the smaller fitted quasar-host radii,
		but excluding these marginally resolved hosts may bias the comparison
		against compact quasar hosts \citep{breiding+24}.  Using only the
		freely fitted quasar hosts gives a median effective radius of
		10.45~kpc, and a two-sided KS test does not detect a significant
		difference from the radio-galaxy distribution ($p=0.162$).
		We therefore interpret the apparent size difference with caution.
		
		The greater fitted luminosities identified for the quasar host galaxies in
		comparison with the radio galaxies should not be interpreted as secure
		evidence that the quasar hosts are intrinsically more luminous.  The quasar
		images were dominated by a much brighter unresolved quasar point source
		superposed on a much fainter extended host, whereas the radio galaxies lack a
		comparable unresolved nuclear component.  Degeneracy between the quasar PSF
		and S\'ersic components may assign some nuclear light to the host model,
		biasing the recovered host magnitude brightward.  In addition, restricting
		the comparison to resolved quasar hosts may preferentially retain quasars
		with brighter or more readily detected hosts.  Establishing an
		intrinsic luminosity difference between quasar and radio-galaxy hosts would
		require a matched nucleus-injection-and-recovery analysis.
		
\begin{figure*}[t]
			\centering
			\includegraphics[width=0.88\textwidth]{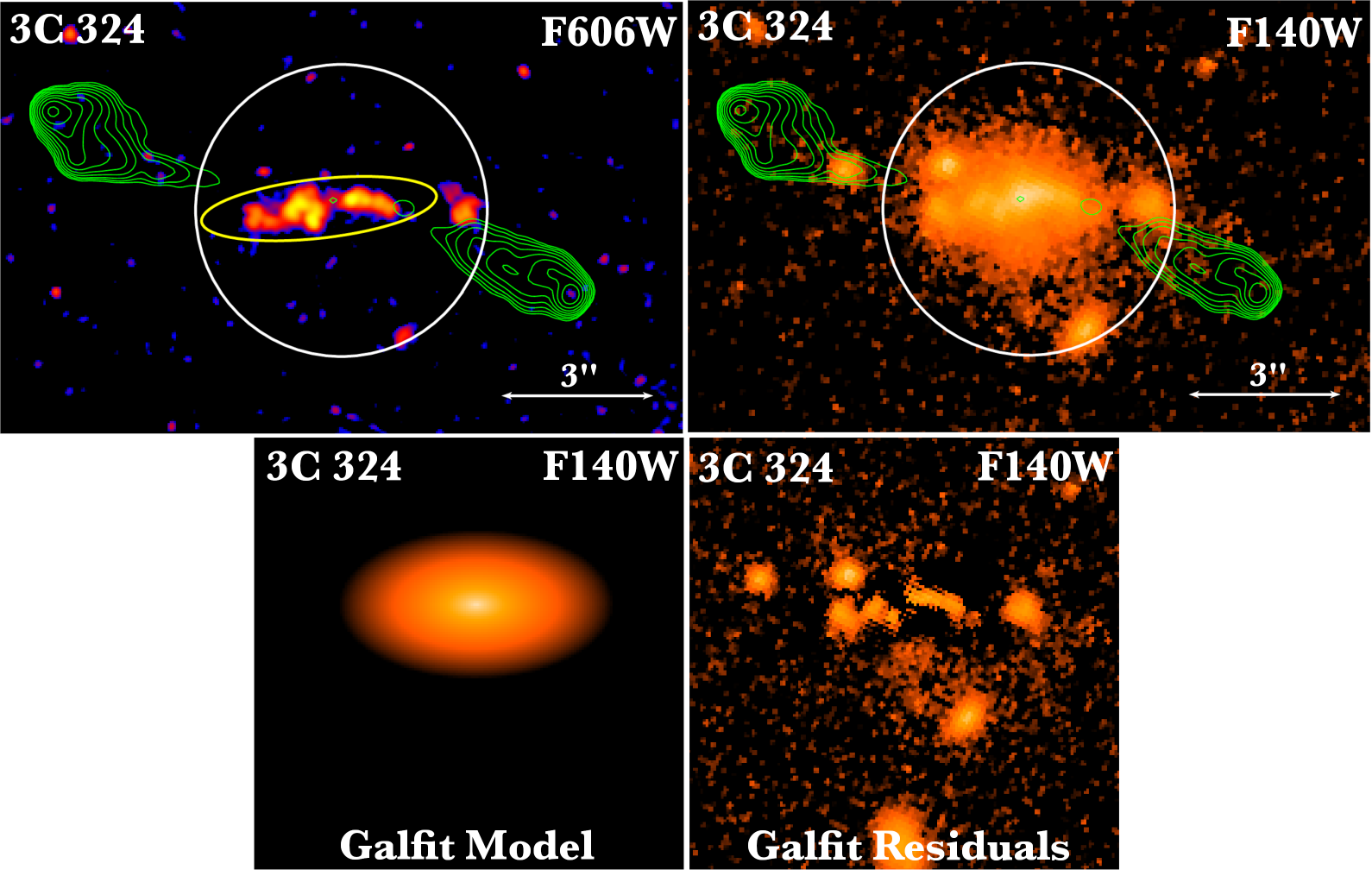}
			\caption{Example imaging and structural decomposition for 3C~324, which
				exhibits prominent knotty rest-frame UV emission aligned with its radio
				structure.  \textit{Top left}: HST/WFC3 F606W image with yellow equivalent
				ellipses derived from the intensity-weighted second moments of the segmented
				UV-emitting regions.  The semimajor and semiminor axes of each ellipse are
				three times the corresponding second-moment dispersions.  Regions are labeled
				by number when more than one region is detected.
				\textit{Top right}: HST/WFC3 F140W image tracing the rest-frame optical
				host galaxy.  \textit{Bottom left}: best-fitting S\'ersic
				\textsc{galfit} model of the F140W image.  \textit{Bottom right}: residual image
				after subtraction of the best-fitting model.  Radio contours are
				overlaid on the F606W and F140W images.  The circle is centered on the
				host galaxy and has a physical radius of 25~kpc, while the arrow
				indicates an angular scale of $3\arcsec$.  The model and residual panels each cover a $10.1\arcsec \times 10.1\arcsec$ field of view around the host galaxy.  The images are displayed with logarithmic intensity stretches spanning 24.7--31.5 AB~mag per pixel in F140W and 27.4--32.4 AB~mag per pixel in F606W.  All images are rotated north up and east to the left.  The F606W images are smoothed for display with a Gaussian kernel of $\sigma=1.5$ pixels and a radius of 3 pixels.  Equivalent panels for the other radio galaxies in our sample are available in the online figure set.}
			\label{fig:example_3c324}
		\end{figure*}

		Figure~\ref{fig:example_3c324} illustrates the F140W host-galaxy
		modeling for 3C~324, including the best-fitting single-S\'ersic model
		and its residual image.
		
		\subsection{Rest-Frame UV Morphologies and Segmented Regions}
		\label{sec:results_uv_morphology}
		
		\begin{deluxetable*}{lcccccccccc}
			\tabletypesize{\scriptsize}
			\setlength{\tabcolsep}{3pt}
			\tablewidth{0pt}
			\tablecaption{Photometric and Morphological Properties of the Rest-Frame UV Regions
				\label{tab:uv_region_properties}}
			\tablehead{
				\colhead{Source} & \colhead{Region} &
				\colhead{$m_{140}^{\rm obs}$} & \colhead{$m_{140}^{\rm res}$} &
				\colhead{$m_{606}$} & \colhead{$\mu_{606}$} &
				\colhead{$R_{\rm proj}$} & \colhead{$m_{606}-m_{140}^{\rm res}$} &
				\colhead{$b/a$} & \colhead{$PA_{\rm UV}$} &
				\colhead{$\Delta\theta_{\rm UV}$} \\
				\colhead{} & \colhead{} &
				\colhead{(AB mag)} & \colhead{(AB mag)} &
				\colhead{(AB mag)} & \colhead{(AB mag arcsec$^{-2}$)} &
				\colhead{(kpc)} & \colhead{(mag)} &
				\colhead{} & \colhead{(deg)} & \colhead{(deg)}
			}
			\startdata
			3C~13 & 1 & 22.65 & 23.44 & 24.68 & 23.67 & 9.9 & 1.24 & 0.64 & 131.6 & 13.5 \\
			3C~13 & 2 & 20.65 & 23.33 & 22.93 & 22.83 & 0.7 & -0.40 & 0.44 & 159.9 & 14.8 \\
			3C~36 & 1 & 20.45 & 22.79 & 22.77 & 21.57 & 0.5 & -0.02 & 0.95 & 53.7 & 32.5 \\
			\enddata
			\tablecomments{The complete table is available in machine-readable form.}
		\end{deluxetable*}
		
		The WFC3/UVIS F606W image-segmentation analysis identified 48 spatially
		resolved UV-emitting regions in 30 of the 33 radio galaxies analyzed.
		Eighteen of the 30 detected galaxies contain a single segmented UV
		region, while 12 contain multiple regions.  The rest-frame UV emission
		is generally clumpy and occupies only a subset of the smoother and
		more extended rest-frame optical emission traced by the F140W images.
		Table~\ref{tab:uv_region_properties} lists the region photometry,
		projected centroid distances from the host, intensity-moment axis
		ratios, major-axis position angles, $PA_{\rm UV}$, and acute
		radio-axis misalignment angles, $\Delta\theta_{\rm UV}$.  The PAs
		follow the north-through-east convention defined in
		Section~\ref{sec:alignment}. 
		
		The same figure illustrates the UV-morphology analysis: the F606W
		emission of 3C~324 contains several distinct components aligned with the
		radio source.  The yellow moment-defined ellipses characterize their
		extent, axis ratio, and orientation, while the segmentation boundaries,
		rather than the ellipses, define the photometric apertures used in
		Section~\ref{sec:photometry}.
		
		We also tested whether the strength or multiplicity of the rest-frame
		UV emission depends on projected radio size.  Among the 30 galaxies
		with the required measurements, Spearman rank tests show no significant
		correlation between radio size and either the summed luminosity of the
		segmented F606W regions ($\rho=-0.17$, $p=0.36$), their luminosity
		relative to the smooth F140W host ($\rho=-0.10$, $p=0.59$), or the
		number of segmented UV regions ($\rho=-0.19$, $p=0.30$).  We therefore
		do not recover evidence that the strongest or most multicomponent UV
		emission is preferentially associated with the most compact radio
		sources.
		
		Each segmented component was treated as an independent UV-emitting
		region for the photometric and morphological measurements.  We next quantify the orientations of the retained UV regions,
		the rest-frame optical host galaxies, and the candidate merger axes
		relative to the radio jets.

		\subsection{Alignment with the Radio Axis}
		\label{sec:results_alignment}

\begin{figure*}[t]
			\centering
			\includegraphics[width=\textwidth]
			{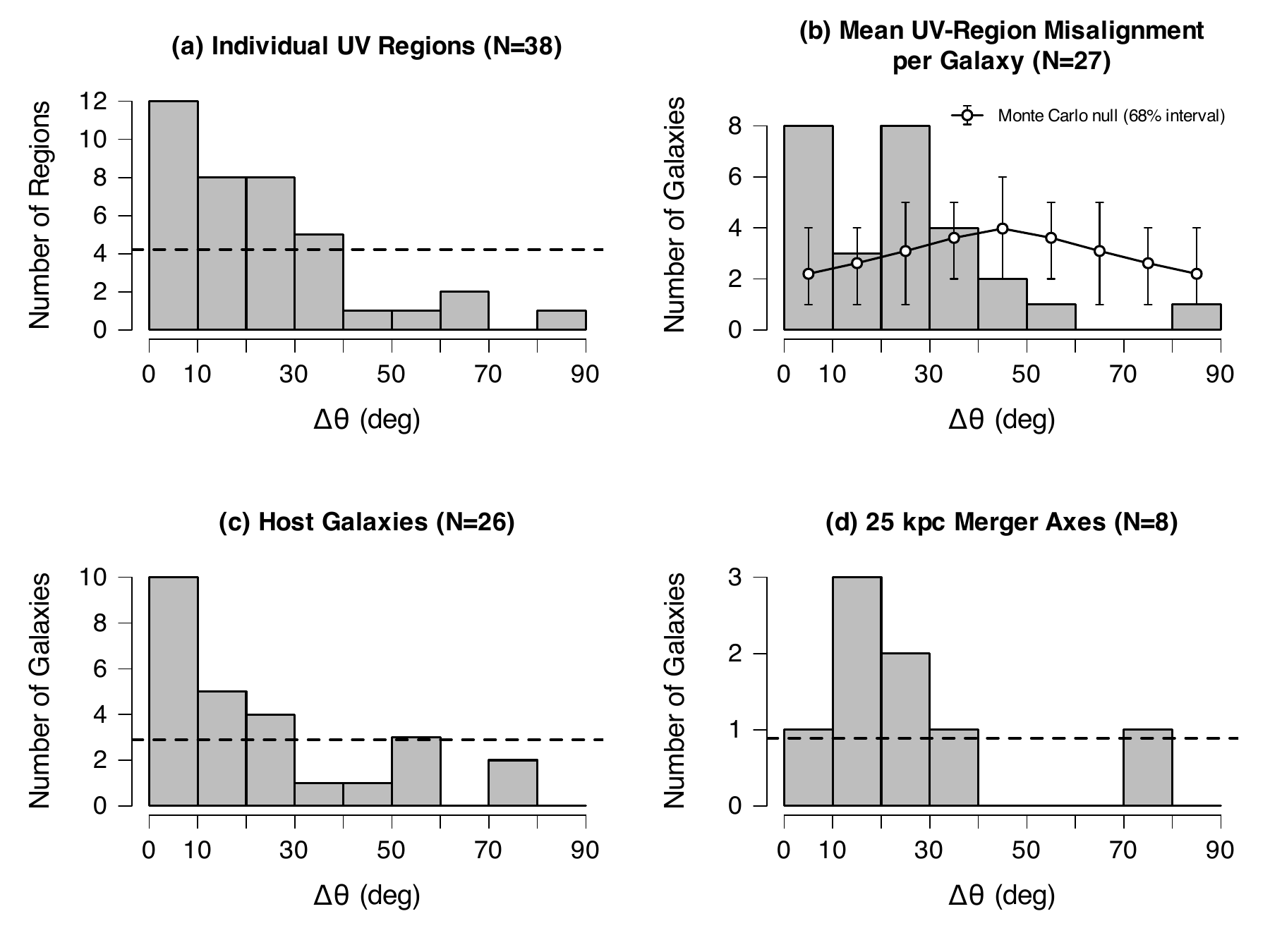}
			\caption{
				Radio-axis misalignment distributions for (a) individual
				UV-emitting regions, (b) the mean UV-region misalignment within
				each galaxy, (c) host-galaxy major axes, and (d) 25-kpc merger
				axes.  Gray histograms show the observed distributions.  The UV and
				host samples in panels (a)--(c) include only components satisfying
				$b/a<0.8$.  Dashed horizontal lines in panels (a), (c), and (d)
				show the expected number of objects per $10\arcdeg$ bin for a
				uniform distribution over $0\arcdeg$--$90\arcdeg$.  In panel (b),
				open circles show the mean bin counts expected from the Monte Carlo
				random-orientation simulations, and vertical bars show the central
				68\% simulation intervals (16th--84th percentiles).  The connecting
				line is included as a visual guide.  These simulation intervals
				represent variation under the null hypothesis and are not
				measurement uncertainties.
			}
			\label{fig:alignment_distributions}
		\end{figure*}

		Among the 33 galaxies with usable HST data, we apply the $b/a<0.8$
		criterion (Section~\ref{sec:alignment}) separately to the host galaxies
		and to the UV regions detected in 30 of those galaxies.  This leaves
		26 host orientations and 38 UV-region orientations in 27 galaxies.
		A galaxy can contribute to one alignment analysis even if it lacks
		a usable orientation for the other.  The misalignment-angle distributions
		for the individual UV regions and the rest-frame optical host
		galaxies are shown in panels (a) and (c) of
		Figure~\ref{fig:alignment_distributions}, respectively.  Both
		histograms are strongly weighted toward small misalignment angles,
		indicating that the UV emission and the stellar major axes of the
		host galaxies preferentially follow the radio axis.  The one-sided
		KS test comparing the UV-region misalignments with a uniform
		distribution gave $p=7.86\times10^{-8}$.  The same comparison for
		the rest-frame optical host galaxies gave $p=9.01\times10^{-5}$.
		The median misalignment angles are $16.7^\circ$ for the individual
		UV regions and $16.9^\circ$ for the host galaxies, compared with
		$45^\circ$ for random axial orientations. 
		
		Because galaxies containing multiple UV regions contribute more
		measurements to the region-level analysis, we also calculated one mean
		UV misalignment for each of the 27 galaxies
		(Figure~\ref{fig:alignment_distributions}, panel b).  In the
		galaxy-level Monte Carlo test described in
		Section~\ref{sec:alignment}, only one of the $10^{6}$
		random-orientation realizations produced an overall mean as small as or
		smaller than the observed value.  Including the finite-simulation
		correction gives $p=2.0\times10^{-6}$, demonstrating that the UV
		alignment remains significant when each galaxy is given equal weight.
		The observed mean of the galaxy-level mean UV misalignments is
		$23.3^\circ$, compared with the null expectation of $45^\circ$.  Holding the observed UV-region position angles fixed and drawing one uniformly
		random radio axis per galaxy in each of $10^7$ trials gave
		$p\simeq1.3\times10^{-6}$ for the same statistic.
		
		We also compared the host--radio misalignment with the mean
		host--UV misalignment within individual systems to test whether
		the host galaxies tend to align more strongly with the rest-frame
		UV regions than with the radio jets.  We would expect such a preference if emission lines from the same gas
		structures were the dominant influence on both the fitted F140W host
		orientations and the measured F606W region orientations.  We used the 22 systems in
		which the host and at least one UV region satisfy $b/a<0.8$,
		averaging only over UV regions that meet this cut.  Twelve of the
		22 hosts are more closely aligned with the UV-region axes than
		with the radio axis.  A one-sided sign test for an excess of hosts
		closer to the UV-region axes than to the radio axes gives
		$p=0.42$.  This finding is consistent with the host--radio alignment tracing the
		older stellar component, although diffuse line emission could still
		affect the fitted host orientations.
		
		Table~\ref{tab:merger_axes} lists the eight galaxies containing at
		least one candidate companion satisfying the selection criteria in
		Section~\ref{sec:alignment} at $5<R_{\rm proj}<25$~kpc, together with
		their merger- and radio-axis position angles and acute misalignments.
		Panel (d) of Figure~\ref{fig:alignment_distributions} shows the
		corresponding misalignment distribution.  As with the UV-region and
		host-galaxy distributions, the merger-axis histogram is weighted toward
		small misalignment angles.  Its departure from random orientations is
		significant ($p=0.0115$ in a one-sided KS test).  The eight merger axes
		have a median misalignment of $22.6^\circ$, with six of the eight lying
		within $30^\circ$ of the radio axis.
		
		\begin{deluxetable}{lcccc}
			\tabletypesize{\footnotesize}
			\setlength{\tabcolsep}{4pt}
			\tablewidth{0pt}
			\tablecaption{Candidate-Merger Axes and Radio Alignments
				\label{tab:merger_axes}}
			\tablehead{
				\colhead{Source} & \colhead{$N_{\rm comp}$} &
				\colhead{$PA_{\rm merger}$} & \colhead{$PA_{\rm radio}$} &
				\colhead{$\Delta\theta_{\rm merger}$} \\
				\colhead{} & \colhead{} & \colhead{(deg)} &
				\colhead{(deg)} & \colhead{(deg)}
			}
			\startdata
			3C~194   & 2 & 171.5 & 159.3 & 12.1 \\
			3C~210   & 1 &   2.5 & 164.1 & 18.5 \\
			3C~250   & 1 &  42.9 &  54.5 & 11.6 \\
			3C~267   & 2 &  52.0 &  78.8 & 26.7 \\
			3C~305.1 & 2 &  37.9 &  10.4 & 27.6 \\
			3C~469.1 & 1 & 137.3 & 171.2 & 33.9 \\
			3C~470   & 1 & 140.9 &  38.0 & 77.1 \\
			4C~13.66 & 1 &   2.0 &   7.7 &  5.7 \\
			\enddata
			\tablecomments{$N_{\rm comp}$ counts the companion candidates selected
				at $5<R_{\rm proj}<25$~kpc (Section~\ref{sec:alignment}).
				}
		\end{deluxetable}

		\subsubsection{Sensitivity to the Axis-Ratio Cut}
		\label{sec:alignment_axis_ratio_robustness}
		
		In order to examine how sensitively our results depend on the choice of
		axis-ratio cut, we repeated the host--radio and UV--radio alignment analyses
		using the stricter cut $b/a<0.7$ in place of $b/a<0.8$.  This leaves 15 host
		galaxies and 37 UV regions in 27 galaxies.  The median host--radio
		misalignment angle is $7.8^\circ$, with a one-sided KS-test p-value of
		$p=1.4\times10^{-5}$.  For the UV regions, the median misalignment angle is
		$16.5^\circ$, with a one-sided KS-test p-value of
		$p=2.0\times10^{-8}$.  Thus, both the host--radio and UV--radio alignment
		signals persist under the stricter axis-ratio cut.
		
		\subsection{Stellar-Population and Jet-Timing Constraints}
		\label{sec:results_sps}

		Of the 48 UV-emitting regions detected in the segmentation analysis
		described in Section~\ref{image_moments}, 47 had at least one
		acceptable SPS model solution.  The color limit for 3C~194 knot~2 lies blueward of every
		stellar-continuum model and therefore yields no accepted solution.
		
		For the following discussion, we define a recent onset of star
		formation as an elapsed duration since the onset of constant star
		formation of $\leq100$~Myr.  Recent onset is considered
		\emph{permitted} when at least one accepted model satisfies this
		criterion and \emph{required} when all accepted models satisfy it.
		Recent onset is permitted for all 47 regions with accepted
		stellar-continuum solutions, although this alone provides little
		constraint on their star-formation histories.  More importantly, it is
		required for 12 regions: none of their accepted models has a
		constant-SFR duration exceeding 100~Myr.  Both recent and
		longer-duration solutions remain allowed for the other 35 regions.
		
		The extinction constraints further distinguish these two groups.
		Among the 35 regions that permit longer-duration solutions, 27 allow
		both zero and nonzero extinction, while eight require nonzero
		extinction.  Of these eight, five have minimum permitted values of
		$A_V=0.3$ or 0.5~mag, while three require $A_V\geq1.5$~mag.  The 12
		recent-onset-required regions are more restrictive: all permit
		$A_V=0$, none permits $A_V>1$~mag, and two admit only models with
		$A_V=0$.  Thus, although the duration--extinction degeneracy remains
		substantial for most regions, the 12 UV-emitting regions for which a
		recent onset of star formation is required occupy a more restricted
		part of the model grid: none of their accepted models has a
		constant-SFR duration greater than 100~Myr or $A_V>1$~mag.  If their UV
		emission is stellar, they therefore constitute the strongest
		candidates for recently initiated, ongoing star formation.
		
		We next used the SPS constraints to determine whether the permitted
		constant-SFR durations are temporally consistent with jet-induced
		star formation.  In the SPS models, $t_{\rm SF}$ is the time
		elapsed since the onset of constant star formation.  We restricted
		this comparison to UV-emitting regions located within the projected
		host galaxy.
		
		For a localized UV-emitting region, we estimated the projected
		distance traveled by the advancing radio structure since passing the
		region as
		
		\begin{equation}
			D_{\rm timing}
			=
			D_{\rm hs}-R_{\rm UV},
			\label{eq:timing_distance}
		\end{equation}
		
		where $D_{\rm hs}$ is the projected distance from the nucleus to
		the hotspot on the same side of the radio source as the UV region,
		and $R_{\rm UV}$ is the projected nuclear distance of the
		UV-region centroid.  The corresponding available post-passage time
		was calculated as
		
		\begin{equation}
			t_{\rm avail}
			=
			\frac{D_{\rm timing}}{\beta_{\rm hs}c},
			\label{eq:available_jet_time}
		\end{equation}
		
		where $\beta_{\rm hs}=v_{\rm hs}/c$ denotes the assumed mean axial
		advance speed of the lobe head, for which the terminal hotspot
		provides an observational tracer.  This is distinct from both the
		relativistic bulk speed of the jet plasma and the backflow speed
		within the radio lobe.  This calculation treats the projected radial
		nucleus-to-region distance as a proxy for the distance already
		traversed along the relevant side of the radio source.  It is
		therefore a first-order geometrical approximation and does not
		establish that an off-axis region was physically crossed by the jet
		or lobe.  Regions with $R_{\rm UV}\geq D_{\rm hs}$ were excluded
		because the lobe head has not demonstrably propagated beyond the UV
		region in projection.
		
		For systems in which the UV emission is distributed broadly across
		the host galaxy rather than confined to a localized component, we
		instead used one-half of the total projected
		hotspot-to-counter-hotspot extent for $D_{\rm timing}$.  Assuming that
		the two lobe heads advance simultaneously at the same mean speed,
		this corresponds to the mean projected nucleus-to-hotspot distance and
		provides a host-wide estimate of the radio-source propagation time.
		These host-wide cases are identified by the geometry code H in
		Table~\ref{tab:sps_jet_timing_regions}.
		
		We evaluated Equation~\ref{eq:available_jet_time} for a range of
		$\beta_{\rm hs}=0.005$, 0.01, 0.02, 0.05, and 0.10.  Resolved spectral-aging studies of mature FR~II radio
		galaxies find characteristic lobe-head advance speeds of order
		$0.01c$, while analyses of very powerful FR~II sources generally
		indicate growth speeds of a few percent of $c$ and characteristic
		speeds below approximately $0.1c$
		\citep{scheuer95,odea+09,harwood+17}.  We use
		$\beta_{\rm hs}=0.01$ as the fiducial value and included
		$\beta_{\rm hs}=0.10$ as a deliberately fast limiting case.  No correction was applied for projection effects.  Deprojecting the
		distances would increase $D_{\rm timing}$ and therefore
		$t_{\rm avail}$.  Our values of $t_{\rm avail}$ are consequently
		lower limits and are conservative when evaluating whether the timing
		can accommodate jet-triggered star formation.
		
		We define $t_{\rm SF,min}$ and $t_{\rm SF,max}$ as the minimum and
		maximum constant-SFR durations among the accepted SPS models.  The
		timing permits jet-induced star formation for at least one accepted
		solution when
		
		\begin{equation}
			t_{\rm SF,min}\leq t_{\rm avail},
			\label{eq:timing_possible}
		\end{equation}
		
		and is consistent with every accepted solution when
		
		\begin{equation}
			t_{\rm SF,max}\leq t_{\rm avail}.
			\label{eq:timing_all}
		\end{equation}
		
		Table~\ref{tab:sps_jet_timing_regions} presents the permitted SPS
		duration and attenuation ranges for each UV-emitting region, together
		with the timing geometry, $D_{\rm timing}$, $t_{\rm avail}$,
		and the timing result for the fiducial $\beta_{\rm hs}=0.01$.
		The timing analysis includes 41 of the 48 UV-emitting regions in 28
		galaxies.  Of these, 33 use the localized post-passage distance in
		Equation~\ref{eq:timing_distance}, while eight use one-half of the
		total host-wide radio-source extent.  Four regions were excluded because they lie
		outside the projected host galaxy.  Three additional
		regions---3C~225A knot~1, 3C~255 knot~2, and 3C~266 knot~1---were
		excluded because their projected nuclear separations equal or exceed
		the corresponding same-side hotspot distances.
		
		For the fiducial hotspot speed $\beta_{\rm hs}=0.01$, at least one
		accepted SPS duration satisfies $t_{\rm SF}\leq t_{\rm avail}$ for 38
		of the 41 regions (92.7\%).  This high fraction is not strongly
		discriminating because short durations are permitted for nearly every
		region.  The more restrictive all-model condition, in which every accepted SPS
		duration satisfies $t_{\rm SF}\leq t_{\rm avail}$, is satisfied by
		seven regions (17.1\%).  All seven regions satisfying the all-model timing condition belong to
		the 12 regions whose colors require a recent onset of star formation.
		Ten of these 12 regions enter the timing analysis, and seven satisfy
		$t_{\rm SF,max}\leq t_{\rm avail}$ at the fiducial lobe-head advance
		speed.  These are 3C~13 region~2, 3C~210 region~1, 3C~238 region~1,
		3C~267 region~1, 3C~326.1 region~1, 3C~356 region~1, and 3C~469.1
		region~2.  If their UV emission is predominantly stellar, these seven
		regions constitute the strongest combined SPS and jet-timing
		candidates for recently initiated, jet-induced star formation.
		
		Six of these seven candidates were evaluated using the localized
		timing geometry; 3C~210 region~1 uses the host-wide comparison.  For
		localized regions, the timing condition is intrinsically easier to
		satisfy at smaller projected nuclear distances because
		$t_{\rm avail}$ increases as $R_{\rm UV}$ decreases.  The prevalence
		of near-nuclear regions among the candidates is therefore partly a
		consequence of the timing geometry and is not independent
		evidence for jet-induced star formation.  In this respect, 3C~356
		region~1 and 3C~326.1 region~1 are particularly notable because they
		satisfy the all-model condition at projected nuclear distances of
		2.0 and 5.9~kpc, respectively.

\begin{deluxetable*}{lccccccccc}
	\tabletypesize{\scriptsize}
	\tablewidth{0pt}
	\tablecaption{Stellar-Population and Fiducial Jet-Timing Constraints for the UV-Emitting Regions
		\label{tab:sps_jet_timing_regions}}
	\tablehead{
		\colhead{Source} &
		\colhead{Region} &
		\colhead{$t_{\rm SF,min}$} &
		\colhead{$t_{\rm SF,max}$} &
		\colhead{$A_{V,\min}$} &
		\colhead{$A_{V,\max}$} &
		\colhead{Geometry} &
		\colhead{$D_{\rm timing}$} &
		\colhead{$t_{\rm avail}$} &
		\colhead{Timing result} \\
		\colhead{} &
		\colhead{} &
		\colhead{(Myr)} &
		\colhead{(Myr)} &
		\colhead{(mag)} &
		\colhead{(mag)} &
		\colhead{} &
		\colhead{(kpc)} &
		\colhead{(Myr)} &
		\colhead{}
	}
	\startdata
	3C~13 & 1 & 4.57 & 3000.00 & 0.0 & 2.0 & L & 92.12 & 30.04 & Some \\
	3C~13 & 2 & 2.03 & 25.61 & 0.0 & 1.0 & L & 142.69 & 46.54 & All \\
	3C~36 & 1 & 1.00 & 143.45 & 0.0 & 1.0 & L & 37.77 & 12.32 & Some \\
	\enddata
	\tablecomments{The complete table is available in machine-readable form.
		Geometry codes: L = $D_{\rm hs}-R_{\rm UV}$;
		H = half the full projected radio extent;
		O = outside the projected host; R = $R_{\rm UV}\geq D_{\rm hs}$.
		None, Some, and All mean that none, some but not all, or all
		accepted SPS durations satisfy $t_{\rm SF}\leq t_{\rm avail}$.
		Not-tested denotes exclusion from the timing comparison;
		blank SPS entries indicate no accepted stellar-continuum model.}
\end{deluxetable*}

		Table~\ref{tab:jet_timing_fractions} shows how the fractions of
		regions satisfying $t_{\rm SF,min}\leq t_{\rm avail}$ or
		$t_{\rm SF,max}\leq t_{\rm avail}$ depend on the assumed lobe-head
		advance speed.  Because $t_{\rm avail}\propto\beta_{\rm hs}^{-1}$,
		faster advance speeds leave less time for star formation after the
		radio structure passes a region, and fewer regions satisfy either
		criterion.  At every tested speed, substantially fewer regions satisfy
		$t_{\rm SF,max}\leq t_{\rm avail}$, meaning that the available time
		is sufficient for every accepted SPS duration.  All eight host-wide regions have at least one accepted SPS model with
		$t_{\rm SF}\leq t_{\rm avail}$ for $\beta_{\rm hs}\leq0.02$, while
		one satisfies this inequality for every accepted SPS model at the
		fiducial speed.  At $\beta_{\rm hs}=0.05$ and 0.10, at least one
		accepted model remains consistent with the available time for four
		and three of the eight host-wide regions, respectively.
		
		\begin{table}
			\centering
			\caption{Sensitivity of the Jet--Stellar-Population Timing Test}
			\label{tab:jet_timing_fractions}
			\begin{tabular}{ccc}
				\hline
				$\beta_{\rm hs}$ & At least one SPS model & All SPS models \\
				\hline
				0.005 & 38/41 (92.7\%) & 8/41 (19.5\%) \\
				0.010 & 38/41 (92.7\%) & 7/41 (17.1\%) \\
				0.020 & 34/41 (82.9\%) & 5/41 (12.2\%) \\
				0.050 & 22/41 (53.7\%) & 4/41 (9.8\%) \\
				0.100 & 17/41 (41.5\%) & 3/41 (7.3\%) \\
				\hline
			\end{tabular}
		\end{table}
		
		These comparisons establish only whether the temporal ordering
		required for jet-induced star formation is permitted under the SPS, geometrical, and hotspot-speed assumptions.  They do not
		demonstrate that the jet caused the star formation.  Nevertheless, the
		combination of a required recent onset and consistency with the
		all-model timing condition identifies a restricted set of seven
		particularly strong candidates for further investigation.

		\subsection{Radial Trends in Optical-UV Color and Surface Brightness}
		\label{sec:results_radial}
		\label{sec:results_sb_radius}
		
\begin{figure}[t]
			\centering
			\includegraphics[width=\columnwidth]
			{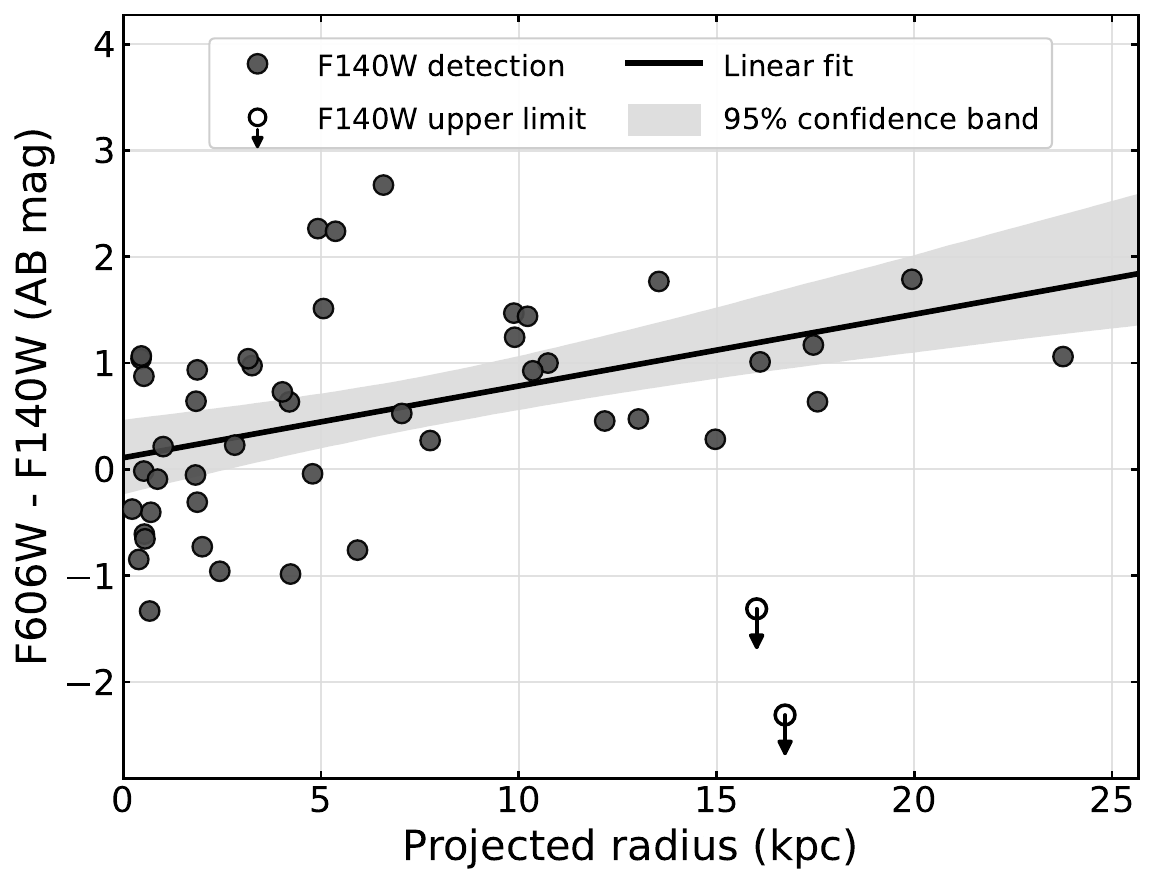}
			\caption{F606W--F140W color versus projected distance from the nucleus for
				the 48 UV-emitting regions.  Filled circles show the 46 regions
				with F140W detections.  Open circles with downward arrows show the two
				F140W nondetections and therefore represent color upper limits.  The
				solid line is the ordinary least-squares fit to the detections, and the
				shaded region is the corresponding 95\% confidence band.}
			\label{fig:color_radius}
		\end{figure}

		In Figure~\ref{fig:color_radius} we show the UV--optical color,
		F606W--F140W, as a function of projected distance from the galaxy nucleus.
		The two F140W nondetections are shown as color upper limits but are not
		included in the correlation or fit.  Among the 46 regions detected in
		F140W, the UV-emitting regions become redder with increasing projected
		distance, with Spearman rank correlation coefficient, $\rho=0.498$.  We
		evaluated the significance by randomly permuting the colors among the
		projected distances $10^6$ times and recalculating $\rho$.  Only 523
		permutations produced $|\rho|\geq0.498$, giving a two-sided $p$-value of
		$p=5.2\times10^{-4}$.
		
		We estimated the uncertainty using $10{,}000$ bootstrap realizations in
		which the 46 detected regions were resampled with replacement.  The
		resulting 95\% bootstrap percentile interval is
		$0.258<\rho<0.678$.  An ordinary
		least-squares fit gives a color gradient of
		$0.0675$~mag~kpc$^{-1}$, with a 95\% bootstrap interval of
		$0.0396$--$0.1040$~mag~kpc$^{-1}$.  The shaded region in
		Figure~\ref{fig:color_radius} is the pointwise 95\% confidence band
		obtained from these same bootstrap fits.  Several physical effects could produce this trend.  The bluer colors at
		smaller radii could reflect younger stellar populations or a stronger
		contribution from AGN-related UV emission, such as scattered quasar
		light or nebular continuum, whose importance is expected to decrease as
		the nuclear radiation field weakens with radius.  Alternatively, the
		redder colors at larger radii could reflect older stellar populations or
		greater dust extinction in the outer UV-emitting structures.
		
\begin{figure}[t]
			\centering
			\includegraphics[width=\linewidth]
			{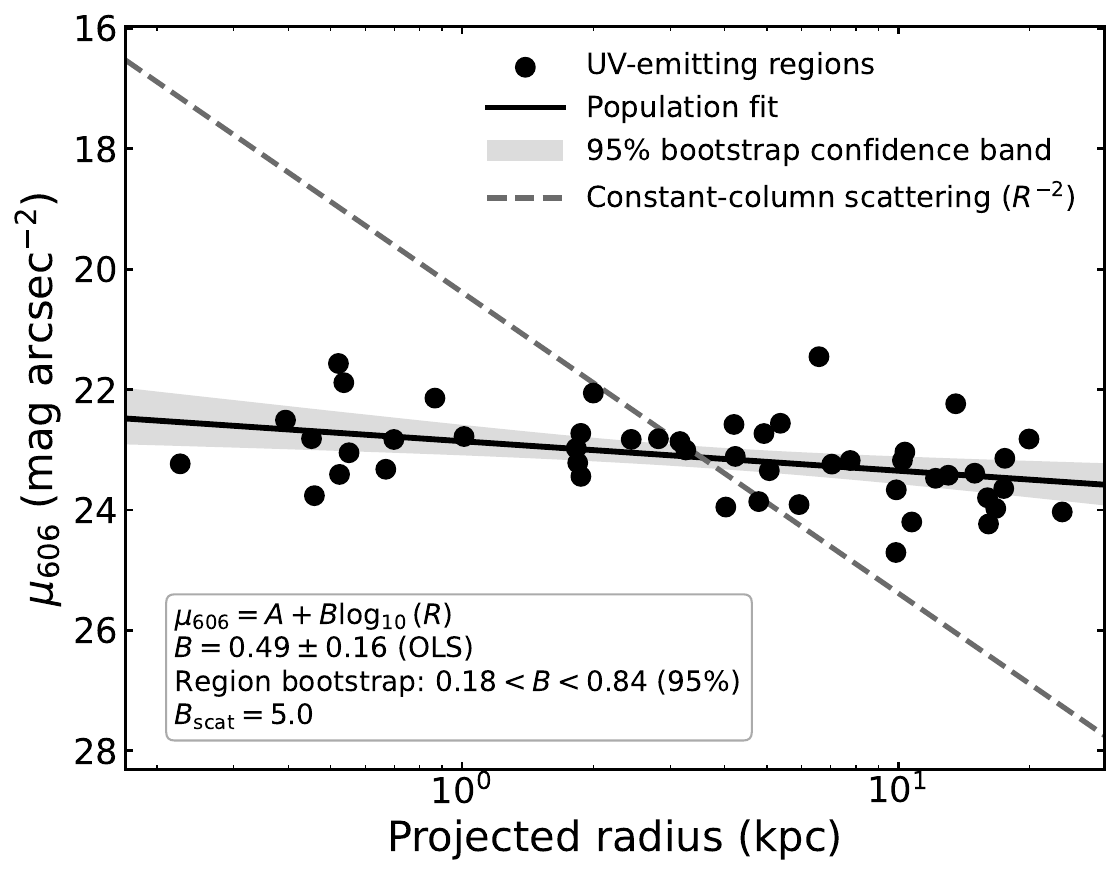}
			\caption{F606W surface brightness versus projected nuclear radius
				for all 48 UV regions.  Surface brightnesses were calculated from
				the native, pre-PSF-matching F606W aperture fluxes and native
				F606W segmentation areas.  The solid line is the population fit,
				and the shading is its pointwise 95\% mean-relation confidence
				band from $2\times10^{5}$ bootstrap realizations in which the 48
				regions were resampled with replacement.  The dashed line is the
				constant-column scattering reference
				$I_{606}\propto R_{\rm proj}^{-2}$, with its intercept fitted
				while fixing $B_{\rm scat}=5$.}
			\label{fig:sb_radius_population}
		\end{figure}

		We next examined the radial behavior of the F606W surface brightness.
		Given that the apertures were defined on the native F606W images, we
		calculated $\mu_{606}$ from the native, pre-PSF-matching F606W aperture
		fluxes and segmentation areas.  Figure~\ref{fig:sb_radius_population}
		shows the resulting rest-frame UV surface
		brightnesses as a function of projected distance from the host-galaxy
		center for all 48 UV-emitting regions identified in our segmentation
		analysis (\S~\ref{image_moments}).  The F606W surface brightness becomes
		fainter with increasing radius, with Spearman correlation coefficient $\rho=0.433$.  We evaluated
		the significance by randomly permuting the surface brightnesses among
		the projected radii $10^6$ times and recalculating $\rho$.  Of these,
		2,277 permutations produced $|\rho|\geq0.433$, giving a two-sided
		$p$-value of $p=2.28\times10^{-3}$.\footnote{Because some galaxies contribute more than one UV-emitting region, the
			permutation $p$-values for both radial relations treat all regions as
			independent.  As a robustness check, we resampled whole galaxies with
			replacement, retaining all UV-emitting regions belonging to each selected
			galaxy.  This yielded essentially unchanged confidence intervals for both
			relations.}
		
		The expected radial dependence for scattering follows from geometric
		dilution of the quasar radiation field.  At an observer-frame frequency
		$\nu_o$ in F606W, the corresponding emitted-frame frequency is
		$\nu_e=(1+z)\nu_o$.  For optically thin, single scattering, the observed
		specific intensity is then
		
		\begin{equation}
			I_{\nu_o}
			=
			\frac{L_{\nu_e,{\rm QSO}}}{4\pi(1+z)^3}
			N_s\sin^2\theta
			\frac{d\sigma_{\nu_e}}{d\Omega}
			T_{\nu_e}
			R_{\rm proj}^{-2},
			\label{eq:sb_radius_scattering_intensity}
		\end{equation}
		
		\noindent
		where $N_s$ is the scatterer column, $\theta$ is the scattering angle,
		$d\sigma_{\nu_e}/d\Omega$ is the differential scattering cross section
		evaluated at the emitted-frame frequency, and $T_{\nu_e}$ is the
		transmission (the full derivation is given in Appendix~\ref{app:scattering_formalism}).  We collect all terms other than the radial dependence into
		an effective scattering normalization, ${\cal K}_{606}$, such that
		
		\begin{equation}
			I_{606}
			=
			{\cal K}_{606}R_{\rm proj}^{-2},
			\label{eq:effective_scattering_normalization}
		\end{equation}
		
		where
		
		\begin{equation}
			{\cal K}_{606}
			=
			\frac{L_{\nu_e,{\rm QSO}}}{4\pi(1+z)^3}
			N_s\sin^2\theta
			\frac{d\sigma_{\nu_e}}{d\Omega}
			T_{\nu_e}.
		\end{equation}
		
		Thus, ${\cal K}_{606}$ contains the quasar luminosity, scatterer column,
		scattering geometry and phase function, cosmological dimming, and
		transmission.  We adopt this radius-independent scattering normalization as a
		simple reference model that isolates the radial dependence produced by
		geometric dilution.  This does not require every region to have the same
		scattering normalization; rather, it assumes that variations in the
		quantities entering ${\cal K}_{606}$ have no systematic dependence on
		projected radius and therefore add scatter without changing the mean
		radial slope.  The corresponding AB surface brightness is
		$\mu_{606}=-2.5\log_{10}I_{606}+C$, where $C$ is a
		radius-independent constant set by the AB zero point and
		intensity units.  Combining this relation
		with Equation~\ref{eq:effective_scattering_normalization} gives
		
		\begin{equation}
			\mu_{606}(R_{\rm proj})=
			\mu_{606}(R_0)+5\log_{10}
			\left(\frac{R_{\rm proj}}{R_0}\right).
			\label{eq:sb_radius_scattering_magnitude}
		\end{equation}

		We take $R_0=1$~kpc, so that $\mu_{606}(R_0)$ corresponds to the
		intercept $A$ in the fitted relation below.  Thus, the constant-column
		reference model, with the other scattering terms held fixed, predicts
		a coefficient $B_{\rm scat}=5$ in a fit of surface brightness against
		logarithmic radius.  An ordinary least-squares fit to the observed
		regions gives
		
		\begin{equation}
			\mu_{606}=A+B\log_{10}
			\left(\frac{R_{\rm proj}}{\mathrm{kpc}}\right),
			\qquad B=0.49\pm0.16.
			\label{eq:sb_radius_observed_fit}
		\end{equation}
		
		The difference $B-B_{\rm scat}=-4.51\pm0.16$ is 27.8 times the nominal
		ordinary-least-squares standard error.  Resampling the 48 regions with
		replacement in $2\times10^5$ bootstrap realizations gives a 95\%
		bootstrap percentile interval of $0.18<B<0.84$, which remains far below
		$B_{\rm scat}=5$.  The shaded region in
		Figure~\ref{fig:sb_radius_population} is the 95\% confidence
		band obtained from these same bootstrap fits.  The observed radial
		relation is therefore much shallower than predicted by simple scattering models.
		
		Allowing the effective scattering normalization to vary with radius,
		the observed relation
		$\mu_{606}=A+B\log_{10}(R_{\rm proj}/{\rm kpc})$ corresponds to
		$I_{606}\propto R_{\rm proj}^{-0.4B}$.  Because
		$I_{606}=\mathcal{K}_{606}R_{\rm proj}^{-2}$, the required
		normalization scales as
		$\mathcal{K}_{606}\propto
		R_{\rm proj}^{2-0.4B}$.  The measured slope therefore requires
		$\mathcal{K}_{606}\propto R_{\rm proj}^{1.80}$.  If all other scattering
		quantities are fixed, $\mathcal{K}_{606}$ is proportional to $N_s$, and
		the required scatterer column follows the same radial scaling.

		The inferred scaling corresponds to an increase in
		$\mathcal{K}_{606}$ by a factor of approximately 33 between 2.3 and
		16~kpc.  Such a strong outward increase is not expected for a smooth,
		centrally concentrated scattering medium.  For example, mass conservation
		in a steady radial outflow with fixed solid angle gives
		$\dot{M}=\Omega_{\rm flow}r^2\rho v=\mathrm{constant}$
		\citep[see Section~4.4]{obied+16}, where $\dot{M}$ is the mass-outflow
		rate, $\Omega_{\rm flow}$ is the solid angle subtended by the flow, $r$
		is the physical distance from the nucleus, $\rho$ is the gas mass
		density, and $v$ is the radial flow speed.  If the number of scattering particles per unit gas mass is constant
		with radius, the scatterer number density traces the gas density:
		$n_s\propto\rho\propto(r^2v)^{-1}$.  Thus, for a constant flow speed,
		$n_s\propto r^{-2}$.  A cone with a fixed opening angle widens in proportion to its distance
		from the nucleus, so the characteristic path length through the
		scattering material scales as $\ell\propto r$.  This offsets one power
		of the density decline: if $n_s\propto r^{-2}$, then
		$N_s\sim n_s\ell\propto r^{-1}$.  The scatterer column therefore still
		decreases outward.  A smooth
		radial flow would therefore tend to reduce, rather than increase, the
		effective scattering normalization at larger radii.  Clumpy gas accretion through the surrounding halo, gas associated with
		or stripped from satellite galaxies, and tidal debris or shells produced
		by galaxy encounters could create localized enhancements in the scatterer
		column or cloud filling factor within individual galaxies at large radii.
		However, these localized structures provide no clear physical basis for
		the large, systematic increase in the effective scattering normalization
		with radius inferred from the full galaxy sample.

		One possible selection bias is incompleteness near our
		surface-brightness detection limit: we may be missing faint outer
		UV-emitting regions, which would bias the measured relation toward a
		shallower slope.  However, a robustness test for incompleteness at large radii shows that
		adding 20 hypothetical undetected regions distributed evenly between 16
		and 24~kpc with $\mu_{606}=26$~mag~arcsec$^{-2}$ increases the fitted
		slope only from $B=0.49$ to $B=1.64$.  Even adding 40 hypothetical regions over
		the same radial range with $\mu_{606}=28$~mag~arcsec$^{-2}$ gives only
		$B=3.17$.  With 40 such regions, obtaining $B\simeq5$ would require
		$\mu_{606}\simeq31$~mag~arcsec$^{-2}$, more than 6~mag fainter than the
		faintest detected region.  Moreover, the absolute residuals from the
		fitted relation show no significant radial trend (Spearman
		$\rho=-0.039$, $p=0.79$), providing no indication that the scatter among
		the detected regions changes systematically with radius.  These tests
		indicate that modest faint-end incompleteness can not account for the
		large difference from the constant-normalization scattering prediction.
		
		A separate possible bias arises from differences in intrinsic
		hidden-quasar luminosity among the galaxies.  A more luminous quasar
		could make scattered emission detectable at larger radii.  Quasar
		luminosity, however, affects all regions within a given galaxy and
		therefore acts primarily as a galaxy-wide shift in surface brightness,
		rather than producing a radial trend within that galaxy.  Indeed, 30 of
		the 48 regions belong to galaxies with multiple regions at different
		projected radii.  Luminosity differences would substantially flatten the
		combined-sample relation only if the most luminous quasars
		systematically occurred in the galaxies contributing the largest-radius
		regions.
		
	One final consideration is that quasar variability, coupled with
	differences in light-travel time, could in principle flatten the profile
	if more distant clouds sample an earlier, more luminous phase of AGN
	activity.  As an illustration, consider clouds at nuclear distances of
	1 and 11~kpc, both lying in the plane of the sky.  Their light-echo delays
	differ by $10~{\rm kpc}/c\simeq3.3\times10^4$~yr.  The difference in AB surface brightness between two radii is
	$\mu_{606}(R)-\mu_{606}(R_0)
	=-2.5\log_{10}[I_{\nu_o}(R)/I_{\nu_o}(R_0)]$.
	Combining this with the fitted relation
	$\mu_{606}=A+B\log_{10}(R/R_0)$ yields
	$I_{\nu_o}(R)/I_{\nu_o}(R_0)=(R/R_0)^{-0.4B}$.
	Holding the scatterer columns, scattering geometry, and transmission
	fixed, the scattering intensity scales as
	$I_{\nu_o}\propto L_{\nu_e,{\rm QSO}}R^{-2}$, where
	$\nu_e=(1+z)\nu_o$ and $L_{\nu_e,{\rm QSO}}$ is the nuclear specific
	luminosity at the epoch sampled by each cloud.  Consequently,
	$L_{\nu_e,{\rm QSO}}\propto I_{\nu_o}R^2\propto R^{2-0.4B}$.
	For $B=0.49$, the nuclear specific luminosity would therefore have to
	decrease by a factor of $(11/1)^{2-0.4B}\simeq76$ between the epochs
	illuminating the outer and inner clouds.  Explaining the population
	trend in this way would invoke ubiquitous dimming of the nuclear UV
	continuum by similar factors over comparable light-travel timescales.
	We have no independent evidence for such a common fading trend;
	stochastic variability without it would primarily add scatter without
	changing the mean radial slope.

		Taken together, the much shallower observed profile provides strong
		evidence against scattered quasar light as the dominant population-wide
		origin of the UV emission, under the physically motivated expectation
		that the effective scattering normalization does not increase strongly
		with radius.  This result does not exclude a subdominant scattered
		component, scattering in individual regions, or more complex scattering
		models involving systematic variations in scatterer column, extinction,
		or geometry.
		
		\subsection{Scattering Energetics of the UV Regions}
		\label{sec:scattering_energetics}
		
		The radial surface-brightness analysis above tests whether the population of
		UV regions follows the radial behavior expected if scattered quasar light is
		the dominant emission mechanism.  We next consider the complementary
		region-by-region question of whether a representative hidden quasar could
		produce the observed F606W surface brightnesses without requiring an
		implausibly large mass of scattering material.  Specifically, if all of the
		F606W emission in each UV region were scattered quasar continuum, what
		ionized-gas or dust mass would be required within the corresponding UV
		aperture? This calculation tests the feasibility of scattering in individual
		regions and does not assume that all regions share the same emission
		mechanism.
		
		We calculate the required line-of-sight scatterer column in each UV
		region from the optically thin scattering relation derived in
		Appendix~\ref{app:scattering_energetics}.  Expressing the observed
		specific intensity in terms of the F606W surface brightness and
		normalizing the intrinsic quasar spectrum at 2500~\AA\ gives
		
		\begin{equation}
			\begin{aligned}
				N_s
				={}&
				\frac{4\pi(1+z)^3 k_{2500}\nu_{2500}R_{\rm proj}^{2}}
				{\Omega_{\rm as}L_{\rm bol}\sin^2\theta
					\left[d\sigma(\lambda_{\rm rest,piv},\theta)/d\Omega\right]
					T(\lambda_{\rm rest,piv})}
				\\
				&\times
				f_{\rm UV,scat}\,
				10^{-0.4(\mu_{606}+48.60)}
				\left(
				\frac{\lambda_{\rm rest,piv}}
				{2500~\text{\AA}}
				\right)^\alpha ,
			\end{aligned}
			\label{eq:required_scatterer_column_general}
		\end{equation}
		
		where $\mu_{606}$ is the observed F606W surface brightness,
		$f_{\rm UV,scat}$ is the fraction of the measured F606W emission
		attributed to scattered quasar light,
		$\Omega_{\rm as}=2.35044\times10^{-11}\ {\rm sr}$ is the solid angle
		of one square arcsecond, $L_{\rm bol}$ is the intrinsic quasar
		bolometric luminosity, and
		$k_{2500}=L_{\rm bol}/(\nu_{2500}L_{\nu,2500})$ is the 2500~\AA\
		bolometric correction.  The emitted-frame wavelength corresponding to
		the F606W pivot wavelength is
		$\lambda_{\rm rest,piv}=5887~\text{\AA}/(1+z)$, $\alpha$ is the
		intrinsic continuum slope defined by $L_\nu\propto\nu^\alpha$, $\theta$
		is the scattering angle, and $T$ is the combined transmission along
		the quasar--scatterer and scatterer--observer paths.
		
		For the fiducial calculation, we assign all of the measured F606W
		emission to scattered quasar light, so that $f_{\rm UV,scat}=1$, and
		neglect attenuation along both light paths, so that $T=1$.  We adopt a
		representative hidden-quasar luminosity of
		$\log(L_{\rm bol}/{\rm erg\,s^{-1}})=47.28$, a bolometric correction
		of $k_{2500}=2.75$, and an intrinsic continuum slope of
		$\alpha=-0.44$.  Equation~\ref{eq:required_scatterer_column_general}
		then becomes
		
		\begin{equation}
			\begin{aligned}
				\frac{N_{s,\rm fid}}{{\rm cm}^{-2}}
				={}&
				8.81\times10^{22}
				(1+z)^3
				\left(\frac{R_{\rm proj}}{{\rm kpc}}\right)^2
				10^{-0.4(\mu_{606}+48.60)}
				\\
				&\times
				\left(
				\frac{\lambda_{\rm rest,piv}}
				{2500~\text{\AA}}
				\right)^{-0.44}
				\left\{
				\sin^2\theta
				\left[
				\frac{d\sigma(\lambda_{\rm rest,piv},\theta)/d\Omega}
				{1\ {\rm cm^2\,sr^{-1}}}
				\right]
				\right\}^{-1}.
			\end{aligned}
			\label{eq:required_scatterer_column}
		\end{equation}
		
		We evaluate Equation~\ref{eq:required_scatterer_column} separately for
		electron and dust scattering and convert the inferred columns into
		ionized-gas and dust masses within the UV apertures.  For each model, we
		adopt the favorable geometry described in Sections~\ref{sec:electron_scattering} and \ref{sec:dust_scattering}, thereby minimizing the
		column required to account for all of the observed F606W emission.  The
		full derivation and justification of the fiducial assumptions are given
		in Appendix~\ref{app:scattering_energetics}.  %The scaling with scattered
		%fraction, attenuation, bolometric correction, and hidden-quasar
		%luminosity is given in
		%Equation~\ref{eq:scattering_scaling_simple}.
		
		\subsubsection{Electron Scattering}
		\label{sec:electron_scattering}
		
		For electron scattering, the scatterer column is the free-electron
		column, $N_s=N_e$.  The Thomson differential cross section is
		
		\begin{equation}
			\frac{d\sigma_{\rm T}}{d\Omega}
			=
			\frac{3\sigma_{\rm T}}{16\pi}
			(1+\cos^2\theta),
			\label{eq:thomson_differential_simple}
		\end{equation}
		
		where $\sigma_{\rm T}$ is the Thomson cross section.  At fixed $R_{\rm proj}$, the electron-scattering intensity depends on
		the scattering angle through the factor
		
		\begin{equation}
			{\cal E}_{e}(\theta)
			=
			\sin^2\theta
			\frac{d\sigma_{\rm T}}{d\Omega}
			=
			\frac{3\sigma_{\rm T}}{16\pi}
			(1-\cos^4\theta).
			\label{eq:electron_angular_efficiency}
		\end{equation}
		
		Unlike dust scattering, Thomson scattering is independent of
		wavelength.  Equation~\ref{eq:electron_angular_efficiency} is maximized
		at $\theta=90\arcdeg$, so we adopt this favorable geometry to obtain
		the smallest required electron column.
		
		If the scattering material lies along the radio axis, then
		$\theta=i$ on the near side and
		$\theta=180\arcdeg-i$ on the far side, where $i$ is the jet inclination
		to our line of sight.  Because
		Equation~\ref{eq:electron_angular_efficiency} is symmetric about
		$\theta=90\arcdeg$, the corresponding near- and far-side locations
		have the same Thomson-scattering efficiency.  Type~2 3CR radio galaxies
		are expected to have $i\gtrsim45\arcdeg$ \citep{barthel89}, so
		geometries reasonably close to the plane of the sky are plausible.
		For example, $i=60\arcdeg$, corresponding to
		$\theta=60\arcdeg$ or $120\arcdeg$, retains
		
		\begin{equation}
			\frac{{\cal E}_{e}(60\arcdeg)}
			{{\cal E}_{e}(90\arcdeg)}
			=
			1-\cos^4(60\arcdeg)
			=
			0.9375
		\end{equation}
		
		of the maximum efficiency.  Thus, although $\theta=90\arcdeg$ provides
		the most favorable limiting case, moderate departures from this
		geometry change the required electron column only modestly.
		
		To convert this column into a gas mass, let $\rho$ be the total gas
		mass density, $n_e$ the free-electron number density, and $m_p$ the
		proton mass.  We define
		
		\begin{equation}
			\mu_e\equiv\frac{\rho}{n_em_p},
		\end{equation}
		
		so that $\mu_e$ is the dimensionless mean gas mass per free electron
		in units of $m_p$.  The ionized-gas mass within an aperture of projected
		physical area $A_{\rm sky}$ is then
		
		\begin{equation}
			M_{\rm ion}
			=
			\mu_e m_pN_eA_{\rm sky}.
			\label{eq:electron_mass_simple}
		\end{equation}
		
		To obtain a lower limit on $M_{\rm ion}$ at fixed $N_e$, we assume
		full ionization\footnote{Partial ionization would provide fewer free electrons per unit mass and would therefore require more gas.}.  Fully ionized hydrogen supplies one electron per nucleon, while helium
		supplies two electrons per four nucleons.  For the dominant metals, the
		numbers of protons and neutrons are approximately equal, giving about
		one electron per two nucleons when fully ionized.  Approximating the
		nucleon mass by $m_p$ and adopting the protosolar mass fractions of
		hydrogen ($X=0.7154$), helium ($Y=0.2703$), and all elements heavier
		than helium ($Z=0.0142$) from \citet{asplund+09} gives
		
		\begin{equation}
			\mu_e
			\simeq
			\left[X+\frac{Y+Z}{2}\right]^{-1}
			=1.17.
		\end{equation}
		
\begin{figure*}[t]
			\centering
			\includegraphics[width=\textwidth]
			{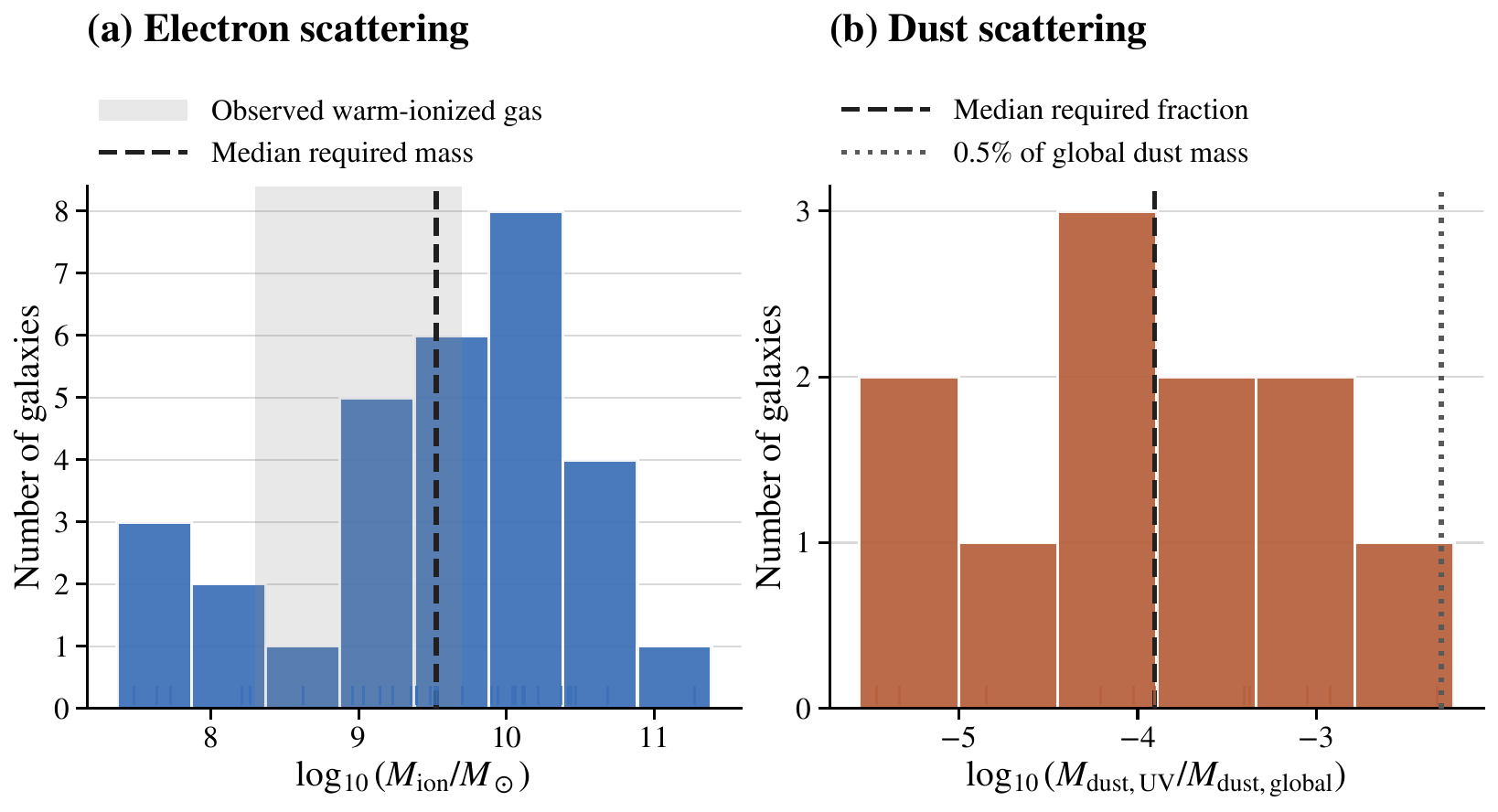}
			\caption{Scattering requirements for the UV-emitting regions.  Left:
				distribution of the ionized-gas mass required for electron scattering,
				summed over the disjoint UV apertures in each of the 30 galaxies.  The
				shaded region shows the warm-ionized gas masses measured for powerful
				high-redshift radio galaxies by \citet{nesvadba+17}, assuming
				$n_e=500\ {\rm cm}^{-3}$, and the dashed line marks the median required
				mass in our sample.  Right: distribution of the dust mass required
				within the UV apertures relative to the global SED-based dust mass for
				the 11 galaxies with available measurements.  The dashed line marks the
				median required fraction, and the dotted line marks 0.5\% of the global
				dust mass.  Short vertical marks along the bottom of each panel show the
				individual galaxies.}
			\label{fig:scattering_requirements_histograms}
		\end{figure*}

		The left panel of Figure~\ref{fig:scattering_requirements_histograms}
		shows the distribution of the required ionized-gas mass after summing
		the disjoint UV apertures within each galaxy.  For comparison, \citet{nesvadba+17} studied 33 powerful radio
		galaxies spanning $1.4<z<4.8$, and
		estimated jet kinetic powers ranging from a few $10^{46}$ to nearly
		$10^{48}\ {\rm erg\,s^{-1}}$.  Although their sample extends to higher
		redshift than our predominantly $1<z<2$ 3CR sample, both comprise
		exceptionally powerful radio galaxies with energetic jets.  Their
		measurements therefore provide a relevant empirical benchmark for the
		ionized-gas reservoirs associated with this broad class of systems,
		rather than a redshift- or radio-power-matched control sample.
		
		From the spatially integrated Balmer-line emission of the detected
		nebulae, \citet{nesvadba+17} inferred warm-ionized gas masses of
		approximately $2\times10^8$--$5\times10^9\ M_\odot$, adopting
		$n_e=500\ {\rm cm}^{-3}$.  This warm gas supplies free electrons capable
		of Thomson scattering, although Balmer emission preferentially traces
		dense, warm material, whereas scattering depends on the total electron
		column and may also include diffuse or hotter plasma.  Moreover, the
		SINFONI masses encompass the detected line-emitting nebulae, whereas
		our estimates include only the much more restricted UV apertures.  The
		comparison is therefore conservative: 14 of our 30 galaxies require
		more than $5\times10^9\ M_\odot$ within the UV apertures alone.  If the
		scattering medium extends beyond these apertures, its total ionized-gas
		mass would be larger still.  Dominant electron scattering would
		therefore require ionized-gas reservoirs at or beyond the upper end of
		those measured in other powerful high-redshift radio galaxies, despite
		being concentrated within only a spatial subset of each galaxy.
		
		\subsubsection{Dust Scattering}
		\label{sec:dust_scattering}
		
		Dust is a much more efficient UV scatterer than free electrons.  We
		adopt the optical/UV differential scattering cross sections calculated
		by \citet{draine03} for the carbonaceous--silicate grain-size
		distribution of \citet{weingartner_draine01} that reproduces the
		average Milky Way extinction curve with
		$R_V\equiv A_V/E(B-V)=3.1$.  For each source, we evaluate
		$d\sigma_{\rm dust}/d\Omega$ at its rest-frame F606W pivot wavelength
		and scattering angle by interpolating between neighboring
		entries in the public numerical tables accompanying \citet{draine03}.
		
		The tabulated differential cross sections are normalized per Hydrogen
		nucleus and already incorporate the assumed abundance and size
		distribution of dust grains relative to the gas.  Accordingly, the
		scatterer column in Equation~\ref{eq:required_scatterer_column} is
		$N_s=N_{\rm H}$.
		
		Because dust scattering is anisotropic and forward-peaked, its most
		favorable geometry at fixed $R_{\rm proj}$ is not
		$\theta=90\arcdeg$.  The relevant angular efficiency is
		
		\begin{equation}
			{\cal E}_{\rm dust}(\lambda,\theta)
			=
			\sin^2\theta
			\frac{d\sigma_{\rm dust}(\lambda,\theta)}{d\Omega}.
			\label{eq:dust_angular_efficiency}
		\end{equation}
		
		The wavelength dependence of the dust phase function causes the
		optimum angle to vary slightly across the sample.  Maximizing
		Equation~\ref{eq:dust_angular_efficiency} at each source's rest-frame
		pivot wavelength gives
		$\theta_{\rm max}=35\arcdeg$--$41\arcdeg$.  For a near-side cloud lying
		along the radio axis, the scattering angle equals the jet inclination
		to the line of sight, $\theta=i$.  In the orientation-based unification
		model for 3CR sources, \citet{barthel89} inferred a critical
		inclination of approximately $45\arcdeg$, with quasars viewed at
		smaller inclinations and radio galaxies at larger inclinations.  We
		therefore adopt $\theta=45\arcdeg$ as a favorable boundary case for an
		on-axis scattering cloud.  At this angle, ${\cal E}_{\rm dust}$ retains
		96.6--99.3\% of its source-specific maximum.  The adopted angle is not
		intended to represent the inclination of every source, and scattering
		material that does not lie along the radio axis need not have
		$\theta=i$.  Adopting $\theta=60\arcdeg$ or $90\arcdeg$ increases the required $N_{\rm H}$ by median per-region factors of 1.15 or 1.94,
		respectively.
		
		Equation~\ref{eq:required_scatterer_column} then gives the
		Hydrogen-nucleus column $N_{\rm H}$ required by the measured UV surface
		brightness of each region.  To keep the mass conversion consistent with
		our choice of grain model, we use its tabulated dust mass per Hydrogen
		nucleus,
		$m_{\rm dust/H}=1.398\times10^{-26}\ {\rm g\ H^{-1}}$.\footnote{See
			the \href{https://www.astro.princeton.edu/~draine/dust/extcurvs/kext_albedo_WD_MW_3.1_60_D03.all}
			{public Draine model table}.}
		The dust mass within an aperture of projected physical area
		$A_{\rm sky}$ is therefore
		
		\begin{equation}
			M_{\rm dust}
			=
			m_{\rm dust/H}N_{\rm H}A_{\rm sky}
			=
			\left(1.398\times10^{-26}\ {\rm g\ H^{-1}}\right)
			N_{\rm H}A_{\rm sky}.
			\label{eq:dust_mass_simple}
		\end{equation}
		
		Across the rest-frame wavelengths sampled by F606W, the dust
		differential cross section per Hydrogen nucleus at
		$\theta=45\arcdeg$ is of order $10^3$ times the Thomson differential
		cross section per electron at $\theta=90\arcdeg$.
		
		As a consistency check on the optically thin solution, we use the
		inferred Hydrogen-nucleus columns to calculate the corresponding
		optical depths:
		
		\begin{equation}
			\tau_{\rm sca}
			=
			N_{\rm H}\sigma_{\rm sca},
			\qquad
			\tau_{\rm ext}
			=
			N_{\rm H}\sigma_{\rm ext}.
		\end{equation}
		
		Here $\sigma_{\rm sca}$ and $\sigma_{\rm ext}$ are the total scattering
		and extinction cross sections per Hydrogen nucleus, respectively,
		supplied by the grain model at the source's rest-frame
		pivot wavelength.  The extinction cross section is
		
		\begin{equation}
			\sigma_{\rm ext}
			=
			\sigma_{\rm sca}+\sigma_{\rm abs},
		\end{equation}
		
		where $\sigma_{\rm abs}$ is the absorption cross section per Hydrogen
		nucleus.  The resulting
		optical depths have median values of $\tau_{\rm sca}=0.008$ and
		$\tau_{\rm ext}=0.015$, with maxima of 0.135 and 0.220, respectively.
		
		These small values support the assumptions used in the calculation.
		The maximum $\tau_{\rm sca}$ indicates that multiple scattering should
		remain a minor correction, while the maximum $\tau_{\rm ext}$
		corresponds to a minimum transmission of
		$T=e^{-\tau_{ext}}=e^{-0.22}\simeq0.80$.  Thus, neglecting attenuation is an excellent
		approximation for a typical region and remains a favorable limiting
		assumption for the most opaque case.  These optical depths are
		model-dependent consistency checks rather than independent
		measurements: the Draine cross sections do not impose $\tau<1$, and a
		larger inferred column could instead have invalidated the optically
		thin solution.  They also do not constrain additional attenuation along
		the quasar--cloud or cloud--observer paths.
		
		The right panel of
		Figure~\ref{fig:scattering_requirements_histograms} compares the
		inferred dust requirements with the galaxy-wide dust reservoirs.  Of
		the 12 $z>1$ 3CR radio galaxies with global SED-based dust masses from
		\citet{podigachoski+16}, 11 have detected UV
		regions and enter this comparison; 3C~68.2 has no such detection.  For
		each of these 11 galaxies, we sum the required dust mass over the
		disjoint UV apertures and normalize it by the galaxy-wide dust mass.
		The median required fraction is only $1.2\times10^{-4}$ and remains
		below 0.5\% in every case.  Even allowing one magnitude of UV
		attenuation keeps the maximum below 1.2\%.  Changing to an SMC-bar
		grain model increases the required Hydrogen-nucleus column by a median
		factor of 4.9 but still does not strain the available dust reservoir.
		
		Table~\ref{tab:scattering_constraints} presents the region-by-region
		quantities underlying these results, including the projected distance,
		aperture area, and F606W surface brightness.  The electron column and
		ionized-gas mass are calculated for the favorable
		$\theta=90\arcdeg$ geometry.  The hydrogen-nucleus column and dust mass
		are calculated for the fiducial Milky Way $R_V=3.1$ dust model at
		$\theta=45\arcdeg$.

\begin{deluxetable*}{lcccccccc}
	\tabletypesize{\scriptsize}
	\tablewidth{0pt}
	\tablecaption{Electron- and Dust-Scattering Requirements for the
		UV-Emitting Regions
		\label{tab:scattering_constraints}}
	\tablehead{
		\colhead{Source} &
		\colhead{Region} &
		\colhead{$R_{\rm proj}$} &
		\colhead{$A_{\rm sky}$} &
		\colhead{$\mu_{606}$} &
		\colhead{$\log N_e$} &
		\colhead{$\log M_{\rm ion}$} &
		\colhead{$\log N_{\rm H}$} &
		\colhead{$\log M_{\rm dust}$} \\
		\colhead{} &
		\colhead{} &
		\colhead{(kpc)} &
		\colhead{(kpc$^2$)} &
		\colhead{(AB mag arcsec$^{-2}$)} &
		\colhead{(cm$^{-2}$)} &
		\colhead{($M_\odot$)} &
		\colhead{(cm$^{-2}$)} &
		\colhead{($M_\odot$)}
	}
	\startdata
	3C~13 & 1 & 9.90 & 38.19 & 23.67 & 22.54 & 10.10 & 19.55 & 4.96 \\
	3C~13 & 2 & 0.70 & 78.46 & 22.83 & 20.58 & 8.44 & 17.59 & 3.31 \\
	3C~36 & 1 & 0.52 & 27.84 & 21.57 & 20.79 & 8.21 & 17.81 & 3.08 \\
	\enddata
	\tablecomments{All logarithms are base 10.  The complete table, including the dust
		scattering and extinction optical depths, is available in
		machine-readable form.}
\end{deluxetable*}

\subsection{UV Strength and Host--Radio Alignment}
\label{sec:results_alignment_uv_strength}

We next tested whether the luminosity of the UV-emitting regions
depends on the orientation of the host galaxy relative to the
jet axis.  The host-alignment sample contains 26 radio galaxies
satisfying $b/a<0.8$, of which 23 contain at least one resolved
UV-emitting region and enter this analysis.  We derive a power-law
spectral slope for each region from its PSF-matched F606W flux
density and host-subtracted F140W residual.  For regions lacking
an F140W detection, we used the median spectral slope of the
detected regions in the same galaxy.  We normalized each power
law to the native, pre-PSF-matching F606W aperture flux density
and estimated the rest-frame specific luminosity at
$2500$~\AA.  We then summed the luminosities of all detected UV regions within each
galaxy, including regions too round to enter the UV--radio alignment tests.  For both relations below, we calculated Spearman
$\rho$ across the 23 galaxies and obtained the two-sided
$p$-values from the standard Student-$t$ approximation to its
null distribution, with $n-2=21$ degrees of freedom.

\begin{figure*}[t]
			\centering
			\includegraphics[width=\textwidth]
			{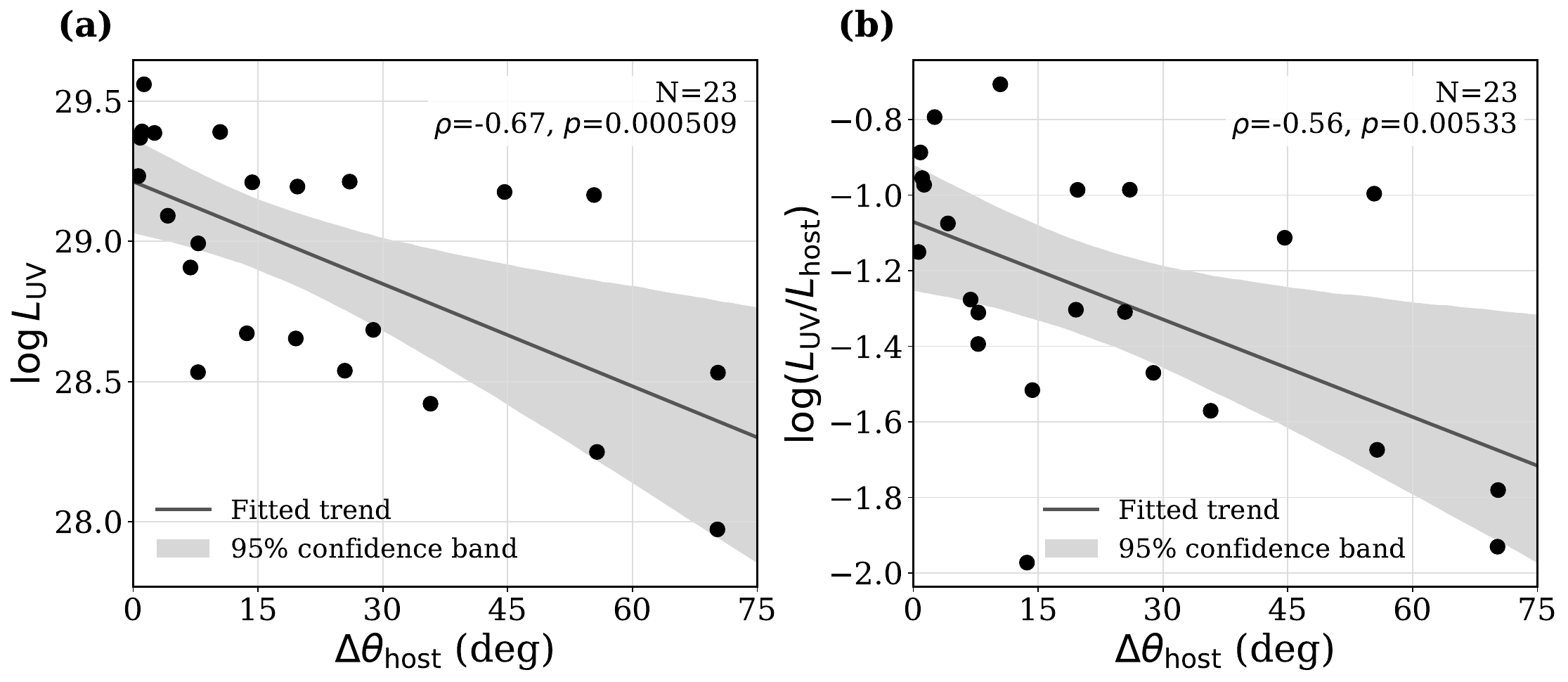}
			\caption{Rest-frame 2500~\AA{} specific luminosity of the UV-emitting regions as a function of the misalignment
				angle between the host-galaxy semimajor axis and the radio axis.
				Panel (a) shows the summed rest-frame
				$L_\nu(2500\,\text{\AA})$ of the resolved UV
				regions, and panel (b) shows this quantity normalized by the
				passive-template K-corrected rest-frame $V$-band host luminosity.  Black
				points represent individual galaxies.  Solid lines are ordinary
				least-squares fits, and the shaded regions are 95\% confidence
				bands for the fitted mean relations from 10,000
				source-bootstrap realizations.  The displayed $\rho$ values and two-sided $p$-values are from Spearman
				rank tests, with the $p$-values calculated using the Student-$t$ approximation.}
			\label{fig:host_alignment_uv_strength}
		\end{figure*}

The total UV luminosity decreases with increasing host--radio
misalignment (Figure~\ref{fig:host_alignment_uv_strength}a), with
Spearman $\rho=-0.667$ and $p=5.09\times10^{-4}$.  Because the
total rest-frame UV luminosity may scale with the galaxy's
rest-frame optical luminosity, we also normalized $L_{\rm UV}$
by the rest-frame $V$-band host luminosity derived from the
F140W photometry using the passive stellar-population K correction
described in Section~\ref{sec:results_host_structure}.  The
anti-correlation remains for $L_{\rm UV}/L_{\rm host}$
(Figure~\ref{fig:host_alignment_uv_strength}b), with Spearman
$\rho=-0.561$ and $p=5.33\times10^{-3}$.  The aligned systems
therefore exhibit stronger total UV emission not only in
absolute terms, but also relative to their rest-frame optical
stellar continuum.
		
		\section{Discussion}
		\label{sec:discussion}
		
		The principal result motivating the broader analysis presented in this
		paper is that the large-scale radio jets of the $z>1$ type~2 3CR radio
		galaxies are preferentially aligned with three distinct components of
		their host galaxies and surrounding environments: the major axes of
		their older stellar populations, their clumpy rest-frame UV emission,
		and the projected axes of close candidate mergers.  The alignment between the radio and
		rest-frame UV emission confirms the well-established UV manifestation
		of the alignment effect in a statistically complete high-redshift
		radio-galaxy sample at high significance
		\citep{mccarthy+87,chambers+87}.  We further find a strong alignment
		between the radio axes and the major axes of the smooth rest-frame
		optical hosts dominated by older stellar populations.
		
		While alignment at rest-frame optical wavelengths has been
		reported previously
		\citep{dunlop+93,best+98,pentericci+01,zirm+03}, most of these studies
		measured the orientation of the total light distribution which
		contains both the underlying stellar host \textit{plus any additional aligned
			structures}.  In particular, \citet{best+98} found that the measured (observer-frame) near-IR K-band alignment was strongest in the galaxies for which their
		SED modeling attributed the largest fraction of the K-band flux to a
		relatively flat-spectrum component associated with the rest-frame UV
		alignment effect.  After the four galaxies with the strongest
		contributions from this component were excluded, the remaining K-band
		alignment was only weakly significant.  The clearest previous evidence
		specific to the older stellar population came from \citet{zirm+03}, who
		analyzed nine 3CR radio galaxies with near-IR HST imaging.  They masked
		neighbors and bright aligned features before measuring host position
		angles, broadly paralleling our approach (Section~\ref{galfit}), and
		found non-random host--radio alignment at 99.9\% confidence.
		
		Our uniform two-dimensional S\'ersic analysis extends that result to
		the statistically complete $z>1$ 3CR type~2 parent sample.  Of 33
		usable hosts, 26 satisfy the $b/a<0.8$ requirement; their host--radio
		alignment significance, $p=9.01\times10^{-5}$, establishes a
		population-wide correspondence between the radio axis and the
		large-scale distribution of older stars.
		
		The strength of the UV emission is also related to the
		host--radio geometry.  Galaxies with closer host--radio alignment contain
		stronger UV emission, both in absolute terms and relative to the
		rest-frame optical host
		(Figure~\ref{fig:host_alignment_uv_strength}).  The common projected axis
		is therefore associated not only with the orientation but also with the
		prominence of the UV component.  One possible interpretation is that gas
		and dust are already distributed along the host or assembly axis.  When
		the radio source follows the same axis, more of this material may lie
		within the jet-driven cocoon or quasar ionization cone, enhancing
		jet-induced star formation, nebular emission, or scattering from a hidden quasar.  Beyond the trend with UV luminosity, our exploratory comparisons found
		no robust association between host--radio misalignment and any of the
		following: the presence of a selected companion, projected radio size,
		rest-frame 178~MHz radio luminosity, the UV--optical color measured from
		the combined light of the detected UV regions, or the number of those
		regions.

		A preference for neighboring galaxies to lie near the radio axis has
		also been reported previously \citep{pentericci+01}, although their
		analysis differed substantially from the galaxy-merger-focused
		measurement presented here.  In a heterogeneous sample of 19 radio
		galaxies at $1.7<z<3.2$, they defined companions as compact sources with
		$H<23.5$ at projected radii of $30$--$50$~kpc.  They found a non-random
		distribution of candidate companion orientations at 99\% confidence,
		with the companions preferentially located along the radio axis, or
		along the optical axis when the radio source was unresolved.  Their
		result established an anisotropy in the relatively faint, larger-scale
		neighboring population, but did not specifically test the geometry of
		close, relatively luminous candidate mergers.
		
		Our analysis extends this earlier environmental result into the inner,
		close-companion regime explicitly excluded by \citet{pentericci+01}.
		We require candidate companions to lie at projected separations of
		$5$--$25$~kpc and within two F140W magnitudes of the host, selecting
		relatively luminous candidate mergers.  Although our merger-axis sample contains only eight systems and
		spectroscopy is required to confirm the physical association of the
		candidate companions, the simultaneous alignment of the close-companion
		axes, smooth stellar hosts, and UV-emitting regions with the radio axes
		suggests that these orientations may not be independent.
		
		There is also a morphological degeneracy between a single elongated
		host galaxy and two strongly overlapping galaxies in a late-stage merger.
		Our requirement that candidate companions have $5<R_{\rm proj}<25$~kpc
		excludes systems with centers separated by less than 5~kpc, where
		overlapping stellar envelopes can cause two galaxies to be modeled as
		one elongated host.  Such systems would not enter our companion-selected
		merger-axis sample, although their projected separation axes could
		already contribute to the host--radio alignment.  The measured
		merger--radio alignment may therefore be conservative, since identifying
		additional radio-aligned systems of this type would tend to strengthen
		the signal.

		\subsection{A Common Axis for the Radio Source, Host Galaxy, and Galaxy Mergers}
		\label{sec:discussion_common_axis}
		
		While the clumpy rest-frame UV emission could, in principle, reflect a
		local and rapid response to the jet, the smooth rest-frame optical light
		traces the galaxy-wide distribution of older stars.  This stellar
		structure can not represent the same rapid response to the current radio
		episode.  Star formation associated with earlier episodes of jet activity
		may have contributed to the older stellar population, but the host--radio
		alignment could instead indicate that both orientations trace a
		longer-lived structure established during galaxy assembly.  
		
		\subsubsection{Stellar Rotation and the Jet Axis}
		\label{subsub:stellar_rotation}
		
		The physical origin of the host--radio alignment is not immediately
		obvious.  The launch direction of a radio jet is expected to be linked to
		the angular-momentum axis of the innermost accretion flow and/or the spin
		axis of the black hole.  These axes determine the preferred outflow
		direction in accretion-disk-launched and black-hole-spin-powered models,
		respectively \citep{blandford_payne_82,blandford_znajek_77}.  In a
		rotationally supported oblate galaxy, the stellar angular-momentum axis
		is perpendicular to the disk plane.  If the nuclear gas disk is coplanar with
		the stellar disk, this simple geometry predicts that the jet should be
		preferentially aligned with the host minor axis rather than its major
		axis.  Such a minor-axis preference has been observed among some
		lower-redshift early-type galaxies \citep{battye+09}.  Among
		the $z<0.5$ radio AGN studied by \citet{zheng+24}, this alignment along the host minor axis is
		stronger at lower radio luminosities and in more elongated hosts.
		They suggest that coherent and secular accretion between galaxy mergers could align
		black hole spins with the angular momentum of their host galaxies
		and thereby favor minor-axis jets in oblate hosts.  If the larger-scale ISM is also
		distributed in a disk, the minor axis provides a lower-column-density
		path, while jets propagating through the disk plane are expected to
		interact more strongly with the surrounding gas \citep{mukherjee+18}.
		
		The major-axis alignment found here is therefore opposite to this simple
		disk-based expectation.  This distinction is important because the hosts
		in our 3CR sample span a broad range of S\'ersic indices, including many
		low values consistent with disk-like surface-brightness profiles.
		However, a low S\'ersic index alone does not establish that the host
		galaxy's stellar component is a rotationally supported oblate disk
		\citep{schulze+18}.  If some of these hosts are genuine stellar disks,
		the observed alignment would instead suggest that the inner accretion
		flow and/or black-hole spin is preferentially misaligned with the stellar
		angular-momentum vector, with its projected direction lying closer to
		the stellar disk plane.  A strongly misaligned or polar nuclear gas disk
		could produce such a geometry.  Merger-driven torques, externally
		accreted gas, and instabilities within the inflow can generate this
		decoupling between galactic and nuclear scales
		\citep{hopkins+12,talbot+24}. 
		
		Dust absorption can trace the distribution of cold gas.  In nearby
		radio galaxies, the radio axis has been found to be perpendicular in
		projection to compact dust features \citep{dekoff+00} or to dust
		lanes misaligned with the host major axis
		\citep{verdoes+05,weller+26}.  If the dust
		traces a rotating disk, this projected radio--dust geometry is
		consistent with a jet directed along the gas angular-momentum axis. 
		
		A different possibility is that some hosts are triaxial or prolate spheroids rather than oblate stellar disks.  In a general triaxial system, the projected photometric axes do not uniquely determine the intrinsic stellar angular-momentum direction, and substantial photometric–kinematic misalignments are possible \citep{krajnovic+11}.  One particular configuration that could produce the observed geometry is prolate rotation.  In a prolate rotator, the stars rotate about the intrinsic long axis, so their angular-momentum vector lies approximately along that axis.  Except for unfavorable viewing angles, the intrinsic long axis projects near the observed photometric major axis; the projected stellar angular-momentum direction can therefore lie closer to the host major axis than to its minor axis.  This is seen in the nearby
		radio galaxy NGC~4261, where the stellar rotation axis is
		substantially misaligned with the radio axis
		\citep{davies+88}.  In a sample of nearby radio galaxies, \citet{heckman+85} found
		little evidence for a general alignment between the radio and
		stellar rotation axes.  If the gas feeding the black hole rotates about the same long axis
		as the stars and retains this orientation as it flows inward, a
		disk-launched jet could follow the host major axis in projection.
		Accretion that consistently supplies angular momentum along this
		axis could also bring the black-hole spin into alignment with it
		\citep{dotti+13}, giving a spin-powered jet the same orientation.
		
        Deep JWST/NIRSpec integral-field spectroscopy could test for short-
        or long-axis rotation by using Doppler shifts in rest-frame optical
        absorption features to map the motion of the older stars that contribute
        to the radio--host alignment observed in this study
        \citep{heckman+85,smith+90,forrest+26}.  NIRSpec has resolved stellar rotation in
		a post-starburst AGN host at $z=3.06$ \citep{deugenio+24} and in the
		central $\sim 2$~kpc of the nearby radio galaxy 3C~293
		\citep{demellos+26}.  For ordered stellar rotation, the  line-of-sight velocity gradient is nearly perpendicular to the projected rotation axis.  If the jet follows the host major
		axis, a gradient along that axis would favor short-axis rotation,
		with the projected rotation axis nearly perpendicular to the jet.
		A gradient along the host minor axis would be consistent with
		long-axis rotation, with the rotation axis close to the jet in
		projection. 
		
		\subsubsection{Merger History and the Large-Scale Environment}
		\label{subsub:merger_history}
		Major galaxy mergers provide one possible route to long-axis stellar rotation, although the outcome depends strongly on the merger configuration.
		Idealized simulations show that a major polar merger can produce a
		prolate remnant in which long-axis rotation is inherited primarily from
		the progenitor disk that was initially perpendicular to the merger
		orbital plane \citep{tsatsi+17}.  Cosmological simulations likewise find
		that the emergence of prolate shape and rotation is frequently associated
		with a significant merger, and that the major axis of the remnant often
		follows the direction from which the progenitors approached during the
		final stages of coalescence \citep{ebrova+lokas17}.
		
		The high incidence of major mergers among the $z>1$ 3CR radio galaxies
		makes prolate rotation an attractive possible connection among the host,
		merger, and radio alignments.  A preferred merger or infall direction
		could establish the long axis of a prolate stellar system, while angular
		momentum retained from the progenitors could produce rotation around that
		axis.  If the inner accretion flow and black-hole spin subsequently
		became aligned with the same long axis, the jet could emerge along the
		host major axis and remain aligned with the preferred merger direction.
		
		However, these simulations constrain the stellar remnant and do not
		demonstrate that either the nuclear gas or black-hole spin shares its
		long-axis orientation.  Nor does the projected companion axis reveal
		either the polar progenitor configuration modeled by
		\citet{tsatsi+17} or the orbital angular momentum, which also depends
		on the relative velocity.
		
\begin{figure*}[t]
			\centering
			\includegraphics[width=\textwidth]{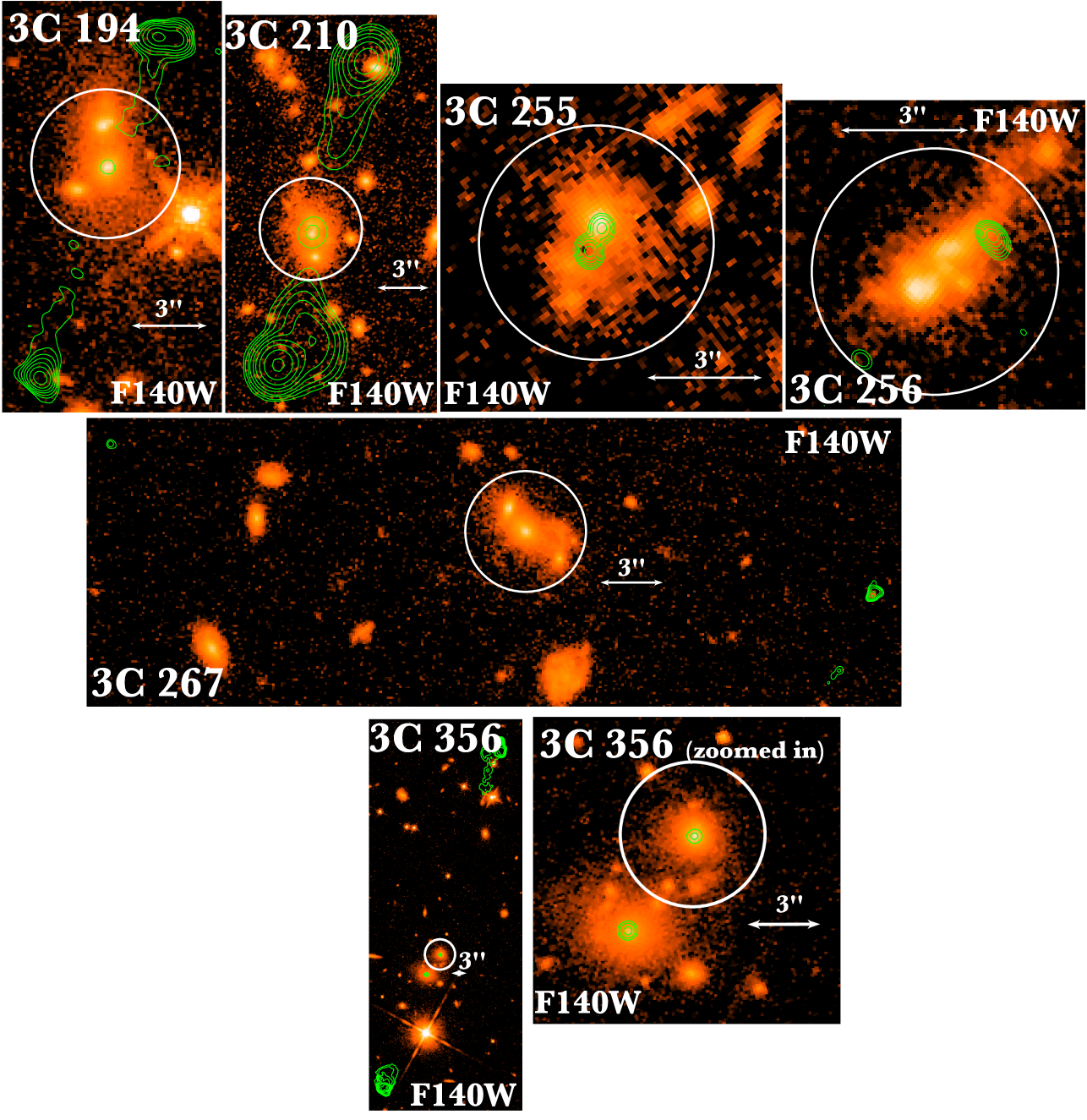}
			\caption{Selected multicomponent systems with rest-frame optical
				morphologies relevant to the merger--radio alignment.  The panels
				show HST/WFC3 F140W imaging with radio contours overlaid;
				white circles mark projected radii of 25~kpc from the host galaxies.
				3C~194, 3C~210, and 3C~267 enter the quantitative merger-axis
				analysis.  3C~255 and 3C~256 are included as illustrative
				morphological examples but do not enter that subsample.  For
				3C~356, the close and wide views show the two principal galaxies
				and place them within the large-scale radio structure.  The images are displayed with logarithmic intensity stretches spanning 24.7--31.5 AB~mag per pixel.  All images are rotated north up and east to the left.  }
			\label{fig:aligned_merger_examples}
		\end{figure*}

				Figure~\ref{fig:aligned_merger_examples} presents six multicomponent
		systems whose rest-frame optical morphologies illustrate geometries
		relevant to the merger--radio alignment.  Among the systems entering the quantitative subsample
		(Table~\ref{tab:merger_axes}), 3C~194 and 3C~267 each contain two selected
		companion candidates, producing suggestive three-component configurations.  Although 3C~255 and 3C~256 do not satisfy all of the
		companion-selection criteria, 3C~255 shows a possible three-component
		configuration broadly along the radio axis, while the central components
		and lower-surface-brightness extensions of 3C~256 share approximately the
		same orientation.

		3C~356 provides a wider-separation comparison: its two massive galaxies
		are separated approximately along the radio axis and are both associated
		with compact radio emission, although the identity of the galaxy
		powering the large-scale source has historically been ambiguous.
		\citet{simpson+02} interpreted 3C~356 as a late-stage merger between two
		galaxy groups in which an interpenetrating collision shock-heated the
		intragroup gas and funneled material into the northern source, which
		they identified as powering the large-scale jet.
		
		The close-companion systems considered here have generally not yet
		coalesced and therefore can not be the fully merged prolate remnants
		described by these simulations, although previous close passages may
		already have disturbed their structures.  Nor can the current
		early-stage interactions have produced the binary-SMBH coalescence
		proposed by \citet{breiding+24} to enable the present radio jet.  An
		earlier major merger could instead have supplied the binary SMBH, with
		the observed companions tracing a subsequent episode of assembly.
		Connecting their geometry to the radio axis would then require
		successive galaxy mergers to follow a longer-lived preferred direction.  
		
		Anisotropic assembly along the surrounding cosmic web provides one
		possible origin for this longer-lived preferred direction.  Using SDSS
		DR7, \citet{zhang+13} found that the major axes of red central galaxies
		preferentially follow cosmic-web filaments, with the alignment becoming
		stronger toward higher luminosities and stellar masses.  More recently,
		\citet{jung+25} found a similar alignment among massive radio galaxies
		at $z<0.6$, particularly for the most massive systems located closest
		to the filament axes.  These results are consistent with massive
		galaxies and their halos assembling through repeated accretion and
		mergers directed preferentially along filaments.  A long-lived filament
		could therefore establish the host major axis and provide a preferred
		direction for subsequent companions, without requiring the currently
		observed companion to have produced the existing stellar structure.
		Extending this coherence to the radio axis would still require the
		connection between galaxy assembly and the central engine discussed
		above.  Wide-field imaging and spectroscopy around the 3CR galaxies
		could determine whether their host, companion, and radio axes trace a
		common filamentary environment.
		
		If a cosmic web filament supplies gas along this common assembly axis, radio lobes expanding in this direction could
		be more tightly confined, increasing the magnetic field energy  density and synchrotron luminosity \citep{hardcastle+18}.  Since the 3CR survey is flux-limited, this effect could favor the
		inclusion of sources whose jets point along the assembly axis
		\citep{eales92}.
		
		One way to examine this possibility is to repeat the alignment
		measurements presented here at lower radio luminosities for a suitable
		subset of radio galaxies at $1\lesssim z\lesssim2$ in the LOFAR
		Two-metre Sky Survey Deep Fields \citep{best+23}.  The comparison would
		require rest-frame optical imaging from HST or JWST comparable to our
		HST data and high-angular-resolution radio maps that resolve both
		terminal hot spots.  Low-frequency radio maps sensitive to diffuse
		emission would also allow one to measure the fraction of the
		integrated flux density contributed by the lobe plasma.  This would allow
		one to test how the host--radio and companion--radio angle
		distributions change with the flux limit for samples matched in
		redshift, host galaxy properties, larger-scale environmental
		properties where possible, and radio morphology.  The lobe flux
		measurements would also test whether the alignment is stronger
		among sources for which the lobes contribute a larger fraction of
		the flux used for selection.

		The merger geometry is also important when interpreting the aligned UV
		emission.  Galaxy interactions can trigger star formation through tidal
		compression, gas inflows, and shocks.  Because the merger axes in this
		sample are themselves preferentially aligned with the radio axes,
		star-forming knots produced in tidal bridges or debris could appear
		radio-aligned even if the jet did not trigger them.  The radio--UV
		alignment alone is therefore insufficient to establish positive jet
		feedback.  Merger- and jet-triggered star formation need not, however,
		be mutually exclusive: the merger may supply and organize the cold gas,
		while the jet or its overpressured cocoon compresses particular clouds
		and determines where within that gas the most recent stars form.

		\subsection{Jet-Triggered Star Formation}
		\label{sec:discussion_jet_sf}
		
		The combined SPS and jet-timing analysis identifies seven regions as
		the strongest candidates for jet-triggered star formation
		(Section~\ref{sec:results_sps}).  For each, all accepted
		stellar-continuum models require a recent onset of star formation and
		satisfy $t_{\rm SF,max}\leq t_{\rm avail}$ at the fiducial lobe-head
		advance speed of $0.01c$.
		
		This temporal consistency is necessary but not sufficient evidence for
		jet triggering.  The comparison depends on the assumed lobe-head speed
		and projected geometry and establishes that the inferred sequence of
		events is possible, but not that the radio structure physically
		interacted with the star-forming gas.
		
\begin{figure*}[t]
			\centering
			\includegraphics[width=0.88\textwidth]{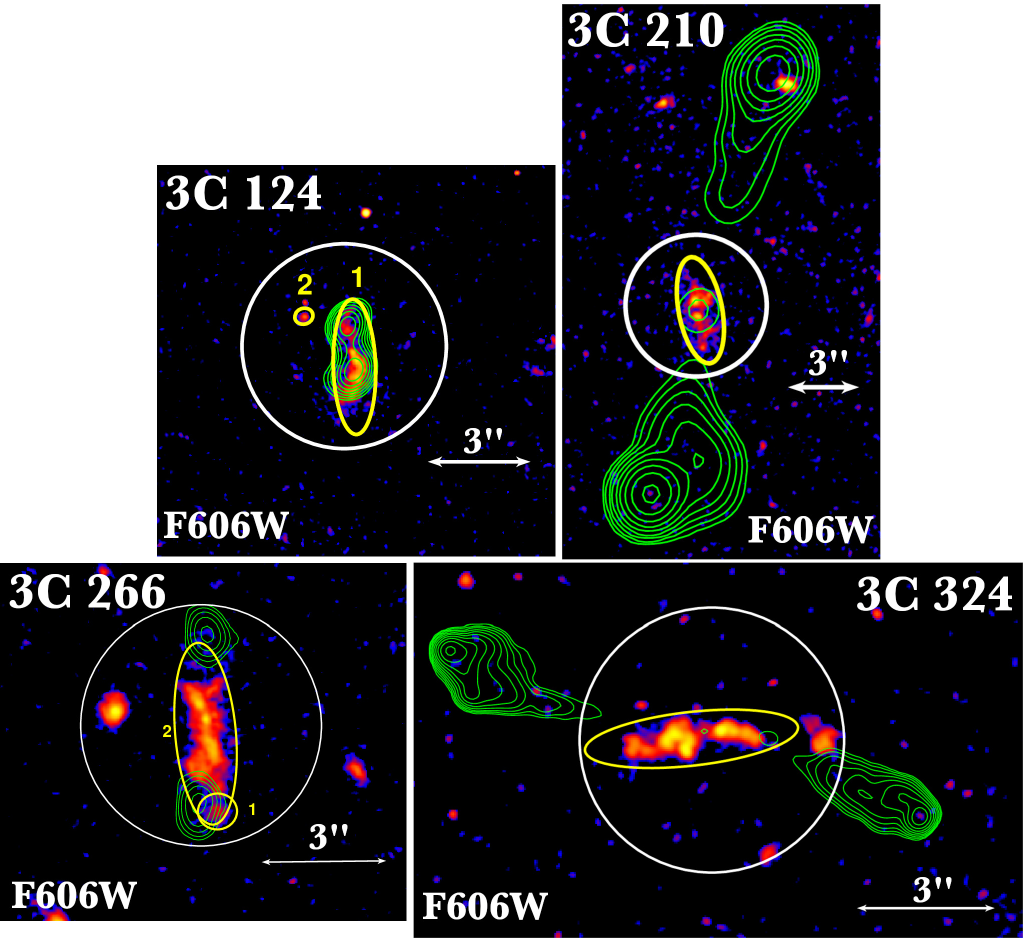}
			\caption{
				Examples of particularly clumpy, radio-aligned rest-frame UV
				morphologies.  Shown are the HST/WFC3 UVIS F606W images
				of 3C~124, 3C~210, 3C~266, and 3C~324, with the radio emission
				overlaid as green contours.  Yellow ellipses show the
				intensity-weighted second-moment regions used to characterize the
				UV morphology, and the white circles mark projected radii of
				25~kpc from the host galaxies.  All panels are oriented with
				north up and east to the left; the white arrows show
				$3\arcsec$ angular scales.  The images are displayed with logarithmic intensity stretches spanning  27.4--32.4 AB~mag per pixel. 
			}
			\label{fig:clumpy_uv_examples}
		\end{figure*}

		A direct spatial relationship with specific radio features would
		strengthen the causal interpretation.  If the UV emission is stellar,
		regions located at jet bends, hot spots, lobe edges, or independently
		identified shock fronts would be more difficult to explain through
		merger-triggered star formation alone.  A UV structure that follows a
		bend in the radio jet would favor a local jet--gas interaction over
		illumination, particularly if the structure extends beyond the ionizing radiation cone from the hidden quasar.  Exact coincidence with the narrow
		radio-jet spine is not required, because the jet-driven bow shock
		and overpressured cocoon can affect gas over a substantially wider area
		\citep{gaibler+12,mukherjee+18}.  A spatial sequence in which star
		formation began earlier in inner regions than in regions farther
		outward, as expected in the simplest picture of an advancing radio
		source, would provide stronger evidence still.
		Figure~\ref{fig:clumpy_uv_examples} presents the F606W images of
		3C~124, 3C~210, 3C~266, and 3C~324 with the radio contours overlaid.
		These galaxies were selected for their suggestive rest-frame UV
		morphologies rather than by the combined SPS and jet-timing criteria:
		their emission is resolved into multiple discrete knots or narrow,
		irregular structures containing several surface-brightness peaks, with
		the ensemble of components distributed approximately along the radio
		axis.  Among the four galaxies, only 3C~210 satisfies the strict
		all-model timing condition.
		
		The resolved, clumpy F606W morphologies are qualitatively suggestive of
		recent star formation, most clearly in 3C~324.  Its chain of UV-emitting knots is reminiscent of the ``beads-on-a-string''
		complex in SDSS~J1531+3414, where the individual beads are young stellar
		clusters associated with a reservoir of multiphase gas \citep{omoruyi+24}.
		Similar structures also clearly extend across the host galaxy and its companion in
		3C~356 (Figure~\ref{fig:aligned_merger_examples}).  Although this morphological resemblance does not
		by itself establish a stellar origin, it provides additional support
		for identifying at least some of the clumpy UV components as candidate
		regions of recent star formation.  However, even if the UV knots trace star formation, their linear
		arrangement alone does not establish jet triggering.
		\citet{elmegreen+04} identified galaxies with chains of large clumps
		in HST images of the Tadpole field and interpreted the clumps as
		star-forming regions in nearly edge-on disks.  In simulations without a jet, gravitational fragmentation of a gas-rich
		disk produces star-forming clumps that appear as a chain when the disk
		is viewed edge-on \citep{bournaud+07}.

		Even if confirmed, jet-triggered star formation in these regions would
		demonstrate a localized positive response rather than a net increase in
		the galaxy-wide star-formation rate.  Simulations show that jets can
		compress some clouds while turbulence and gas dispersal suppress star
		formation elsewhere \citep{mukherjee+18,mandal+21}.  The present UV
		measurements therefore can not determine the net effect of the jet on
		star formation throughout the host galaxy.
		
		The effect of the radio source on star formation may change as the jets
		propagate through and eventually beyond the host ISM.  While the radio
		source remains confined, its bow shock and overpressured cocoon can
		compress dense gas throughout the host; after breakout, the interaction
		may instead become increasingly dominated by turbulence, heating, and
		gas removal.  \citet{falkendal+19} found no significant relation between
		global star-formation rate and radio size, but noted that their robust
		detections all occurred in sources smaller than approximately
		$3\arcsec$, suggesting enhanced star formation before the radio source
		breaks out of the host ISM.  We likewise found no significant relation
		between projected radio size and either the luminosity or number of the
		segmented UV regions.  This provides no evidence for such evolution in
		our sample, although projected radio size is an imperfect age indicator
		and the UV emission is not necessarily stellar.

		\subsection{Scattered Quasar Light}
		\label{sec:electron_scat_discussion}
		
		Our finding that the surface brightness of the UV-emitting regions
		depends only weakly on projected radius argues against a simple
		scattering model as their dominant population-wide origin.  Geometric
		dilution of the incident quasar radiation predicts a substantially
		steeper decline unless the properties of the scattering material change
		systematically with radius.  Explaining the observed profile entirely
		through scattering would therefore require the effective scattering
		efficiency to increase strongly outward, which is not a natural
		expectation for a smooth medium or steady radial outflow.  
		
		The large ionized-gas masses required by electron scattering provide a
		second constraint.  Even assuming a favorable scattering geometry and no attenuation, many of our radio galaxies require ionized-gas reservoirs exceeding those typically inferred for other high-redshift radio-galaxy samples.
		Previous studies of extended UV emission in high-redshift radio galaxies
		have found similarly large mass requirements and have generally favored
		dust over electron scattering \citep[e.g.,][]{cimatti+94,cimatti+96,dey+96}.  Our results do not rule
		out a contribution from electron scattering in individual regions, but
		make it an unlikely explanation for the UV emission across the sample.

		Dust scattering is energetically much more feasible.  Spectropolarimetric
		studies have demonstrated that dust-scattered quasar light can dominate
		the rest-frame UV continuum of individual high-redshift radio galaxies
		and that the scattering medium can be highly clumped
		\citep[e.g.,][]{vernet+01,solorzano+04}.  In our study, the required dust mass within the UV apertures is a small fraction of the global dust reservoir for every galaxy with an SED-based dust-mass estimate.  The principal requirement is therefore not an unusually
		large dust mass, but the presence of illuminated dust-bearing structures
		at projected distances reaching tens of kpc from their host galaxy centers.  The high incidence of mergers in the sample provides a natural means of
		distributing dusty gas over these scales.  Observations of nearby
		interacting systems directly show that tidal stripping can displace dust into extended tidal structures
		\citep{walter+11}.
		
		\subsection{Nebular Line and Continuum Emission}
		\label{sec:discussion_nebular}
		
		Nebular line and continuum emission may contribute to the measured
		F606W and F140W fluxes.  In the HST snapshot sample examined by
		\citet{hilbert+16}, published line-flux measurements were available for
		only two of 12 radio galaxies:
		Mg~II in 3C~356 and [O~III] in 3C~297, which contributed 1.6\% and
		0.03\%, respectively, of the integrated broadband flux.  These results show that those particular lines do not dominate the
		corresponding images, but they do not constrain other lines or the
		nebular continuum across the sample.  Diffuse F140W line emission could
		also contribute to the fitted host component
		(Section~\ref{galfit}).
		
		The nebular continuum can be constrained from spatially matched
		spectroscopy because H$\beta$ and the nebular continuum are produced by
		the same recombining gas.  After correction for reddening, the measured
		H$\beta$ flux can normalize the expected free-bound, free-free,
		two-photon, and high-order Balmer emission across the HST bandpass.
		Using this method, \citet{tadhunter+02} inferred contributions of
		approximately 3--40\% immediately below the Balmer break in nuclear
		spectra of 2-Jy radio galaxies.  In 3C~368, which is included in our
		sample, spectroscopic analyses estimated that nebular continuum
		contributes approximately 60--80\% of the UV continuum in its brightest
		aligned components \citep{dickson+95,stockton+96}.
		
		The ionization required to produce the nebular continuum and
		recombination lines can be supplied by young stars, radiation from the
		hidden quasar, or jet-driven shocks.  Young stars and the quasar
		photoionize the gas, whereas shocks driven by the jet, bow shock, or
		expanding cocoon can heat and collisionally ionize it
		\citep{bicknell+00}.  As the gas recombines, it produces nebular
		continuum and emission lines regardless of the original ionization
		source, with different implications for the alignment effect.  Emission
		powered by young stars would still trace recent star formation, but
		would bias ages inferred by treating the broadband flux as pure stellar
		continuum.  Quasar radiation escaping through an ionization cone could
		instead produce aligned emission without a mechanical interaction with
		the jet.  Shock-powered emission would remain causally associated with
		the radio source, but the F606W flux would not directly measure recent
		star formation.  These mechanisms may also coexist within the same
		galaxy.
		
		Because the relevant lines vary with redshift and spatially matched
		recombination-line measurements are not presently available, we can not
		apply a reliable nebular correction.  This uncertainty does not affect
		the principal geometric result that the F606W emission is aligned with
		the radio source, but it does affect its physical interpretation and the
		stellar ages inferred from its broadband color.  We therefore treat the
		SPS and jet-timing results as conditional on the F606W emission being
		predominantly stellar.  The observations required to quantify the
		nebular contribution and distinguish its ionization source are discussed
		in the following section.
		
		\subsection{Distinguishing the UV Emission Mechanisms}
		\label{sec:future_scattering_tests}
		
		The stellar-population analysis presented in this work is limited by a
		single rest-frame UV--optical color.  Although F606W--F140W spans the
		wavelength interval containing the Balmer and 4000~\AA{} breaks, neither
		filter measures the continuum immediately around these features.
		Consequently, an older stellar population and a younger, dust-reddened
		population can produce the same color.  Nor can this color establish
		that the F606W emission is stellar.  Additional HST or JWST bands
		redward of F140W would constrain the less dust-sensitive rest-frame
		optical-to-near-infrared spectral shape, restricting the allowed
		combinations of stellar age and extinction and permitting more flexible
		star-formation histories than the constant-SFR models used here.  Rest-frame UV spectroscopy could reveal photospheric absorption and
		stellar-wind features from massive young stars.  Spectroscopy covering
		the Balmer and 4000~\AA{} breaks and the Balmer absorption lines
		would help constrain the ages of the stellar populations.
		
		Rest-frame UV imaging polarimetry provides the most direct spatial test
		for scattered quasar light.  Scattering of light from a hidden quasar predicts polarization vectors oriented perpendicular to the
		projected direction from the nucleus, producing a centrosymmetric
		pattern.  For an extended scattering region aligned with the radio
		source, the integrated electric-vector position angle should therefore
		lie broadly perpendicular to the radio axis \citep{disergo+89}.
		Spectropolarimetry provides a stronger spectral test: broad permitted
		lines detected in polarized flux directly reveal the hidden quasar
		spectrum.  
		
		Rest-frame UV
		spectropolarimetry has shown that dust-scattered nuclear light dominates
		the UV continuum in some powerful radio galaxies
		\citep{tran+98,vernet+01,solorzano+04}.

		HST/ACS narrow-band imaging of
		H$\alpha$+[N~II] and [O~III] in 19 3CR radio galaxies at $z<0.3$
		revealed kiloparsec-scale line-emitting structures preferentially
		aligned with the radio axes \citep{tremblay+09,baldi+19}.  These maps
		locate ionized gas and variations in its excitation, but do not uniquely
		identify star-forming regions because AGN photoionization and jet-driven
		shocks can also produce strong line emission.
		
		JWST/NIRSpec integral-field maps of six radio galaxies at $z>3.5$
		show that regions of enhanced [O~III] line width tend to coincide
		with radio structures \citep{roy+26}.  Similar observations of our
		strongest candidates could test whether the UV emission
		coincides with disturbed ionized gas.  H$\beta$, [O~III]~$\lambda5007$, H$\alpha$, and [N~II] are accessible
		over the sample redshift range, permitting spatially and spectrally resolved excitation
		and kinematic maps.  The Balmer decrement would provide a
		model-dependent constraint on reddening, while the corrected H$\beta$
		luminosity would constrain the nebular contribution to F606W.  Line
		ratios, widths, velocity fields, and radio morphology would help
		distinguish H~II-region-like emission from AGN photoionization and
		jet-driven shocks.  H~II-region-like excitation co-spatial with
		disturbed gas at a radio interaction site would substantially strengthen
		the case for jet-induced star formation, particularly if rest-frame UV
		spectroscopy detected young-star features and polarimetry limited
		scattered quasar light.

		Establishing jet-triggered star formation requires both evidence for
		young stars and a physical connection between those stars and the radio
		source.  The observations of 4C~41.17 provide a well-studied example of
		this combination.  Keck/LRIS spectropolarimetry found low polarization,
		photospheric absorption and stellar-wind features from massive young
		stars, together with strongly disturbed emission-line profiles
		\citep{dey+97}.  HST imaging registered against VLA maps placed bright
		UV knots and Ly$\alpha$ emission near a bright radio knot
		\citep{vanbreugel+98}.  Building on these observations,
		\citet{bicknell+00} modeled the interaction as a jet-driven bow shock
		and found that the compressed gas could become gravitationally unstable
		within 2--6~Myr, well within the estimated age of the radio source.  These stellar, spatial, kinematic, and timescale constraints jointly
		support a jet-triggered origin.
		
		In the nearby radio galaxy Coma~A (3C~277.3), the southern jet bends
		beside a bright emission-line cloud.  Chandra observations support a
		jet--cloud interaction at this site \citep{worrall+16}.  VLT/MUSE line
		ratios identify star-forming knots along filaments at the edges of
		the radio lobes, where the surrounding line emission is consistent
		with a contribution from shocks \citep{capetti+22}.  At candidate
		jet--gas interaction sites in our sample, spectroscopy and polarimetry
		could help distinguish stellar emission, nebular emission, and
		scattered quasar light.

		High-resolution ALMA imaging could test whether molecular gas is
		concentrated or kinematically disturbed at candidate jet--ISM
		interaction sites.  An intermediate-$J$ transition such as CO(5--4)
		would trace more highly excited gas, but AGN heating and jet-driven
		shocks can also enhance its emission
		\citep{papadopoulos+08,dasyra+16}.  Pairing it with a lower-$J$
		transition would constrain the excitation correction and better trace
		the bulk molecular-gas reservoir, reducing the uncertainty in
		$M_{\rm mol}$ \citep{carilli+13,daddi+15}.  The inferred mass would
		nevertheless remain sensitive to the choice of CO-to-H$_2$ conversion
		factor \citep{bolatto+13}.  With an independent star-formation tracer,
		the depletion time, $t_{\rm dep}=M_{\rm mol}/{\rm SFR}$, could then be
		compared on and off the radio structure.  A shorter depletion time
		co-spatial with disturbed gas would be consistent with locally enhanced
		star-formation efficiency.

		JWST/MIRI would provide a complementary test for dust-obscured star
		formation missed by F606W.  Galaxy-integrated Spitzer and Herschel SED
		modeling inferred luminous star-forming components in a substantial
		subset of $z>1$ 3CR hosts but could not locate that activity relative to
		the radio source \citep{podigachoski+15,podigachoski+16}.  For the 12
		Herschel-selected radio galaxies studied by
		\citet{podigachoski+16}, the rest-frame mid-infrared continuum was
		attributed largely to the AGN torus, and PAH features were not prominent
		in the four objects with Spitzer/IRS spectra.  
		
		Multi-band MIRI imaging could search for spatially resolved
		warm-dust continuum and excess emission in filters containing
		redshifted PAH features, identifying candidate regions of obscured
		star formation.  PAH luminosities could help constrain SFRs even
		in galaxies hosting an AGN \citep{shipley+16}.  MIRI/MRS could also
		map rotational H$_2$ and ionic emission lines accessible at each
		source's redshift.  PAH-to-H$_2$ ratios, ionic-line ratios, and
		gas kinematics could help assess the contributions of star
		formation, shocks, and AGN photoionization to the emission in
		the $z>1$ 3CR radio galaxies studied here
		\citep[e.g.,][]{pereira-santaella+22}.  Registration with the NIRSpec,
		molecular-gas, and radio maps would then show whether regions
		consistent with obscured star formation coincide with candidate
		jet impact sites.
		
		\section{Summary \& Concluding Remarks}
		\label{conclusion}
		
		In this study, we have combined HST/WFC3 imaging in F140W and F606W with high-spatial-resolution radio imaging for the statistically complete sample of type~2 3CR radio galaxies at $z>1$.  We subsequently used these data to examine the relationships among radio-source properties, host-galaxy structure in the rest-frame optical, rest-frame UV emission, and nearby candidate mergers.  Our principal conclusions are as follows.
		
		\begin{enumerate}
			\item The radio axes are preferentially aligned with the
			rest-frame UV emission, the major axes of the smooth stellar hosts, and
			the projected axes of close candidate mergers.  The close-companion
			result is based on a small sample and requires spectroscopic confirmation,
			but together these alignments indicate a common projected axis connecting
			the radio source to the structure and assembly of the host galaxy.
			
			\item Galaxies with closer host--radio alignment contain stronger
			total UV emission, both in absolute terms and relative to the
			rest-frame optical host.  The common geometry is therefore related not
			only to the orientation but also to the prominence of the aligned UV
			component.
			
			\item The UV regions become redder and fainter in surface brightness with
			increasing distance from the nucleus.  However, the surface brightness
			declines much more slowly than predicted by a simple centrally illuminated
			scattering model.  The large ionized-gas masses required for electron
			scattering further argue against it as the dominant population-wide
			mechanism, while the much smaller dust requirements leave dust-scattered
			quasar light viable in individual regions.
			
			\item The stellar-population and jet-propagation timescales identify a
			restricted subset of UV regions as the strongest candidates for recently
			initiated, jet-triggered star formation.  Their timing is consistent with
			jet triggering but does not demonstrate causation, and the interpretation
			remains conditional on the UV emission being predominantly stellar.
			Merger-triggered star formation, dust scattering, and line and nebular
			emission remain plausible contributions to the rest-frame UV emission observed in our sample of radio galaxies.
			
		\end{enumerate}
		
		These results favor a composite origin for the alignment effect.  The
		common projected geometry may trace a long-lived axis of galaxy assembly.
		Repeated mergers or anisotropic infall along a cosmic-web filament could
		establish the host major axis, potentially imprinting this direction on
		a triaxial or prolate stellar remnant, while delivering subsequent
		companions and gas along the same axis.  Because the observed companions
		have generally not yet coalesced, their positions may trace the
		continuation of this assembly history rather than the origin of the
		host's existing structure.  If the nuclear accretion flow or black-hole
		spin also became aligned with this assembly axis, the radio jet could
		emerge along the same direction, although the present data do not
		demonstrate such coupling.  The jet could then shock or compress gas
		distributed along this axis, mergers could trigger star formation
		independently of the jet, and radiation from the hidden quasar could
		produce scattered and nebular emission.  Galaxy assembly may therefore
		establish the common geometry within which several mechanisms contribute
		to the alignment effect.

		The highest-priority follow-up is JWST/NIRSpec integral-field
		spectroscopy of the strongest candidates and close companions.
		Spatially resolved Balmer-decrement, excitation, and kinematic maps
		could constrain the nebular contribution to F606W and help distinguish
		H~II-region-like emission from AGN photoionization and jet-driven
		shocks.  Mapping the emission lines that fall within F140W would allow
		us to subtract their contribution from the HST images and refit the
		host position angles, testing whether diffuse line emission
		contributed to the host--radio alignment found in this study.  Deep NIRSpec observations of stellar absorption lines could test
		for short- or long-axis rotation.  A projected stellar rotation axis
		aligned with both the host major axis and the radio axis would be
		consistent with a scenario in which filamentary accretion and galaxy
		assembly help set a common axis for stellar rotation and jet launching.  JWST/MIRI imaging and spectroscopy could search for dust-obscured
		star formation at candidate jet impact sites and assess contributions
		from shocks and AGN photoionization. 
		
		Spatially resolved optical spectroscopy could search for redshifted stellar
		photospheric absorption and wind features in the candidate UV regions
		using HST/STIS or VLT/MUSE
		adaptive optics.  HST/ACS F606W
		imaging polarimetry could test for the polarization pattern expected
		from scattered quasar light, and ALMA CO line maps could test for
		disturbed molecular gas along the radio structures.  Together, these observations
		would test whether recent star formation and disturbed molecular gas
		coincide with specific radio features, establish which candidate
		companions are physically associated, and clarify how the alignment
		effect connects powerful radio activity with host-galaxy assembly.
		
		\begin{acknowledgements}
			We thank Barbara Balmaverde, Peter D.\ Barthel, Alessandro Capetti, F.\ Duccio Macchetto, William B.\ Sparks, Carlo Stanghellini, and Grant R.\ Tremblay for their contributions as co-investigators on HST program GO-16281.
			This research is based on observations made with the NASA/ESA Hubble Space Telescope obtained from the Space Telescope Science Institute, which is operated by the Association of Universities for Research in Astronomy, Inc., under NASA contract NAS 5–26555.  These observations are associated with programs GO-16281 and SNAP-13023.  Some of the data presented in this paper were obtained from the Mikulski Archive for Space Telescopes (MAST) at the Space Telescope Science Institute.  The specific observations analyzed can be accessed via \dataset[DOI]{https://doi.org/10.17909/7656-0f61}.  This work was supported by the
			HST-GO grant No.  16281.018-A.  
		\end{acknowledgements}
		
		\bibliography{jet_alignment}{}
		\bibliographystyle{aasjournal}

		\appendix
		\label{appendix}

		\section{PSF homogenization}
		\label{sec:psf_homogenization}
		
		Our WFC3/UVIS F606W images have a natively sharper PSF than our WFC3/IR F140W images.  The WFC3 Instrument Handbook states that the core FWHMs are approximately 0.067 and 0.141 arcsec for
		F606W and F140W, respectively \citep{WFC3IHB}.  Therefore, before performing
		aperture photometry, we convolved the sharper F606W images to the broader
		F140W PSF.
		
		To construct representative PSFs, we followed the method described by
		\citet{breiding+24} for one radio galaxy from each HST observing program used in this study:
		3C~222 for the GO-16281 program and 3C~322 for the SNAP-13023 program.  We used \texttt{TinyTim} \citep{krist+11} to construct a synthetic PSF for
		every calibrated
		\texttt{flt} exposure in WFC3/IR F140W and every calibrated \texttt{flc}
		exposure in WFC3/UVIS F606W.  For each representative galaxy, we selected a
		common sky position within the UVIS/F606W emission region used for aperture
		photometry and transformed that position to the appropriate detector
		coordinates in every exposure.  We assumed a flat spectrum in flux density,
		$f_\nu\propto\nu^0$, and used the telescope focus predicted at the time of each
		exposure by the STScI HST focus model.\footnote{Available at
			\url{https://focustool.stsci.edu/cgi-bin/control.py}.}  Because the native WFC3/IR pixels do not fully sample the F140W PSF core, we
		mapped each F140W model onto the distorted detector grid using the
		\texttt{TinyTim} \texttt{tiny3} stage with \texttt{SUB=5}.  This produced a
		PSF sampled five times more finely than the native detector pixels.  We placed
		each model at the precise detector position of the selected sky location and
		then integrated it onto the native detector pixels.  After this integration,
		we applied the standard WFC3/IR detector-response kernel that
		\texttt{TinyTim} normally includes in its native-sampled PSFs.  The F606W
		PSFs were generated using the normal \texttt{TinyTim} sampling.  The
		exposure-level PSFs were then combined with
		\texttt{AstroDrizzle}, using the
		same drizzle parameters and output pixel scales as the corresponding science
		images.
		
		We used the \texttt{create\_matching\_kernel} routine in
		\texttt{photutils.psf.matching}  to derive a
		normalized kernel that transforms the representative F606W PSF into the
		F140W PSF.  The F140W science images, host-galaxy models, and PSFs were mapped
		from their 0.060 arcsec pixel$^{-1}$ grids onto the F606W grids
		(0.030 arcsec pixel$^{-1}$ for GO-16281 targets and 0.04 arcsec pixel$^{-1}$ for SNAP-13023 targets)
		using bilinear \texttt{reproject\_interp} from \texttt{reproject}
		\citep{robitaille+20}.  Pixel areas were calculated with
		\texttt{proj\_plane\_pixel\_area} in \texttt{Astropy}
		\citep{astropy} to preserve integrated flux.  The reprojected F140W PSF
		was used as the target, and the resulting kernel was applied to F606W using
		\texttt{scipy.signal.fftconvolve} \citep{scipy}.
		
		Because the observing setup was common within each program, we used the
		kernel derived from 3C~222 for the remaining GO-16281 radio galaxies and the
		kernel derived from 3C~322 for the remaining SNAP-13023 radio galaxies.  The
		validity of this program-level approach is tested in
		Section~\ref{app:psf_validation}.
		
		\subsection{Validation of the PSF homogenization}
		\label{app:psf_validation}
		
\begin{figure*}
			\centering
			\includegraphics[width=\textwidth]
			{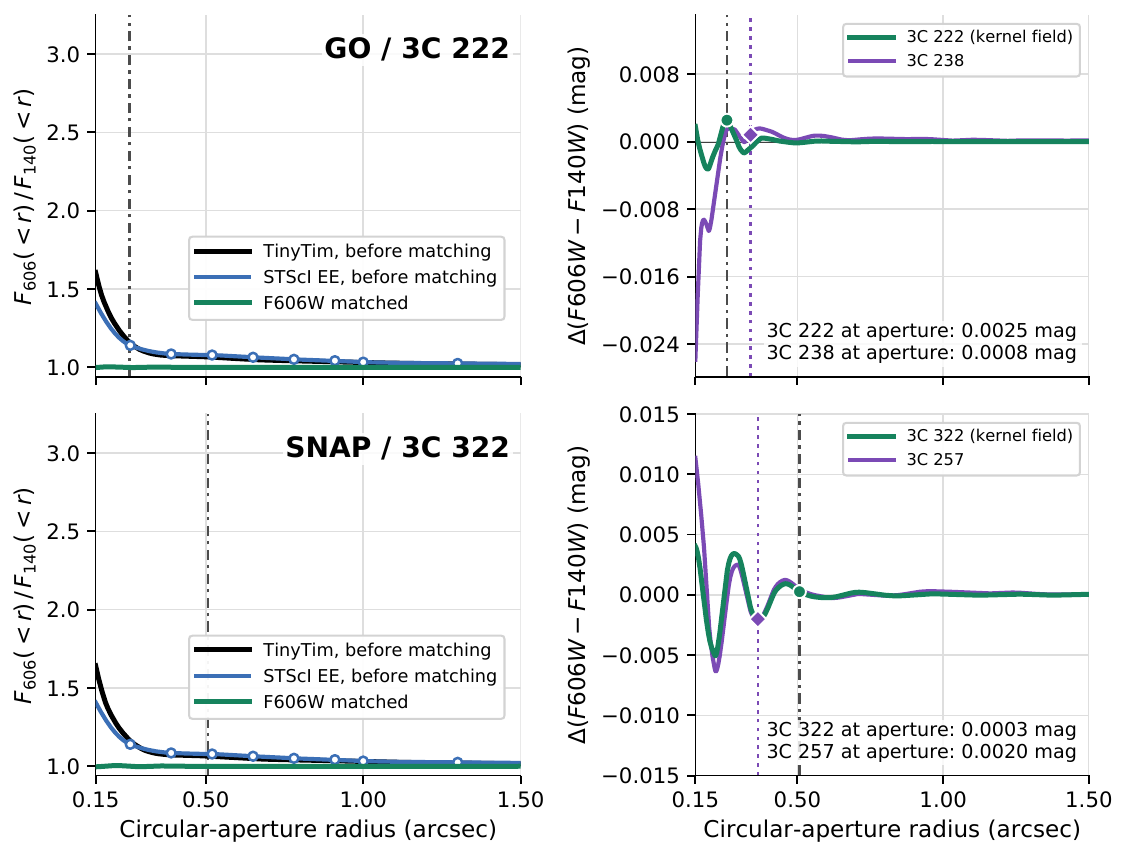}
			\caption{PSF-matching validation for the representative GO field, 3C~222
				(top), and SNAP field, 3C~322 (bottom).  The left panels show the ratio
				$R(r)$ of the normalized F606W and F140W enclosed-flux fractions.  The
				black and blue curves show the ratios before matching for the TinyTim
				PSFs and empirical STScI encircled energies, respectively.  The green
				curves show the ratios after matching F606W to F140W.  The right panels
				show the equivalent PSF-induced F606W--F140W color offset for each
				representative field (green) and after transferring the unchanged
				program kernel to the independent 3C~238 and 3C~257 PSF pairs (purple).
				The dash-dotted gray line and green circle in each row mark the
				representative-field science aperture; the dotted purple line and
				diamond mark the independent-field science aperture.  The curves are
				displayed over 0.15--1.5 arcsec, while the smallest science-aperture
				equivalent radius is 0.17 arcsec.  Perfect matching corresponds to
				$R=1$ in the left panels and $\Delta_{\rm PSF}=0$ mag in the right panels.}
			\label{fig:psf_validation}
		\end{figure*}

		We tested whether the matched F606W and F140W PSFs place the same fraction of
		their total light inside a given aperture.  After normalizing each PSF to a
		total flux of one, we measured the enclosed-flux fraction in each filter as a
		function of aperture radius.  We then calculated
		$R(r)=F_{606}(<r)/F_{140}(<r)$, which is shown in the left panels of
		Figure~\ref{fig:psf_validation}.  Before matching, $R(r)$ quantifies the
		difference between the original PSFs; after matching, it measures how well
		the convolved F606W PSF reproduces the F140W PSF.  Perfect matching
		corresponds to $R=1$.
		
		Because the F606W PSF is sharper, it places a larger fraction of its light
		inside a given aperture than the F140W PSF.  Without PSF matching, the
		measured F606W flux would therefore be biased high relative to F140W,
		producing systematically bluer colors and potentially biasing the
		stellar-population modeling toward younger ages.  To express the remaining
		post-matching error in the same units as our measured colors, we converted
		the flux ratio to the equivalent PSF-induced color offset,
		$\Delta_{\rm PSF}(F606W-F140W)=-2.5\log_{10}R$.  This quantity is shown in
		the right panels of Figure~\ref{fig:psf_validation}; perfect matching
		corresponds to $\Delta_{\rm PSF}=0$ mag.
		
		Figure~\ref{fig:psf_validation} presents the GO result for 3C~222 in the top
		row and the SNAP result for 3C~322 in the bottom row.  In the left panels,
		the black curves show the ratios calculated from the original
		\texttt{TinyTim} PSFs, while the blue curves show independent ratios derived
		from the empirical STScI encircled-energy measurements
		\citep{WFC3IHB}.  At the equivalent circular radii of the
		representative-field science apertures used for aperture photometry (as
		described in Section~\ref{sec:photometry}), the corresponding
		synthetic-PSF and STScI enclosed-energy color offsets differ by approximately
		0.015 mag for both programs.  This agreement supports our PSF construction
		and shows that the synthetic PSFs reproduce the relative F606W and F140W
		aperture losses.  The green curves show $R(r)$ after matching F606W to F140W,
		and the right panels show these same post-matching ratios converted to
		$\Delta_{\rm PSF}(F606W-F140W)$.  The green circles in the right panels mark
		the residuals at the equivalent circular radii of the representative-field
		science apertures.
		
		To test whether a single kernel could be transferred within each program, we
		independently constructed a second PSF pair for 3C~238 (GO) and 3C~257
		(SNAP), using the same procedure, and applied the corresponding 3C~222 and
		3C~322 kernels without refitting them.  These independent-field results are
		shown by the purple curves in the right panels, with purple diamonds marking
		the equivalent circular radii of their science apertures.  At the marked radii, the absolute post-matching color residuals are
		0.0025 mag for 3C~222, 0.0008 mag for 3C~238, 0.0003 mag for 3C~322, and
		0.0020 mag for 3C~257.  These values are point-source diagnostics evaluated
		from the circular-aperture curves at the equivalent circular radii of the
		science apertures; they are not corrections applied to the measured
		photometry.  Taken together, the tests show that the residuals at the marked
		radii are only a few millimagnitudes and should not materially affect the
		stellar-population synthesis results presented in this study.  The figure is
		displayed over 0.15--1.5 arcsec.  The smallest science-aperture equivalent
		radius is 0.17 arcsec, so structure below this radius is diagnostic and is
		not used as a photometric correction.

		\section{Scattering Formalism and Energetics Calculations}
		\label{app:scattering_formalism}
		
		\subsection{Scattering Intensity and Geometry}
		
		The expected radial dependence for scattering follows from geometric
		dilution of the quasar radiation field.  At the emitted-frame frequency
		$\nu_e=(1+z)\nu_o$, where $\nu_o$ is the observed-frame frequency, a
		quasar with intrinsic spectral luminosity $L_{\nu_e,{\rm QSO}}$ produces
		an incident spectral flux at a cloud a physical distance $r$ from the
		nucleus of
		
		\begin{equation}
			F_{\nu_e}
			=
			\frac{L_{\nu_e,{\rm QSO}}}{4\pi r^2}.
			\label{eq:sb_radius_incident_flux}
		\end{equation}
		
		\begin{figure}[t]
			\centering
			\includegraphics[width=\columnwidth]
			{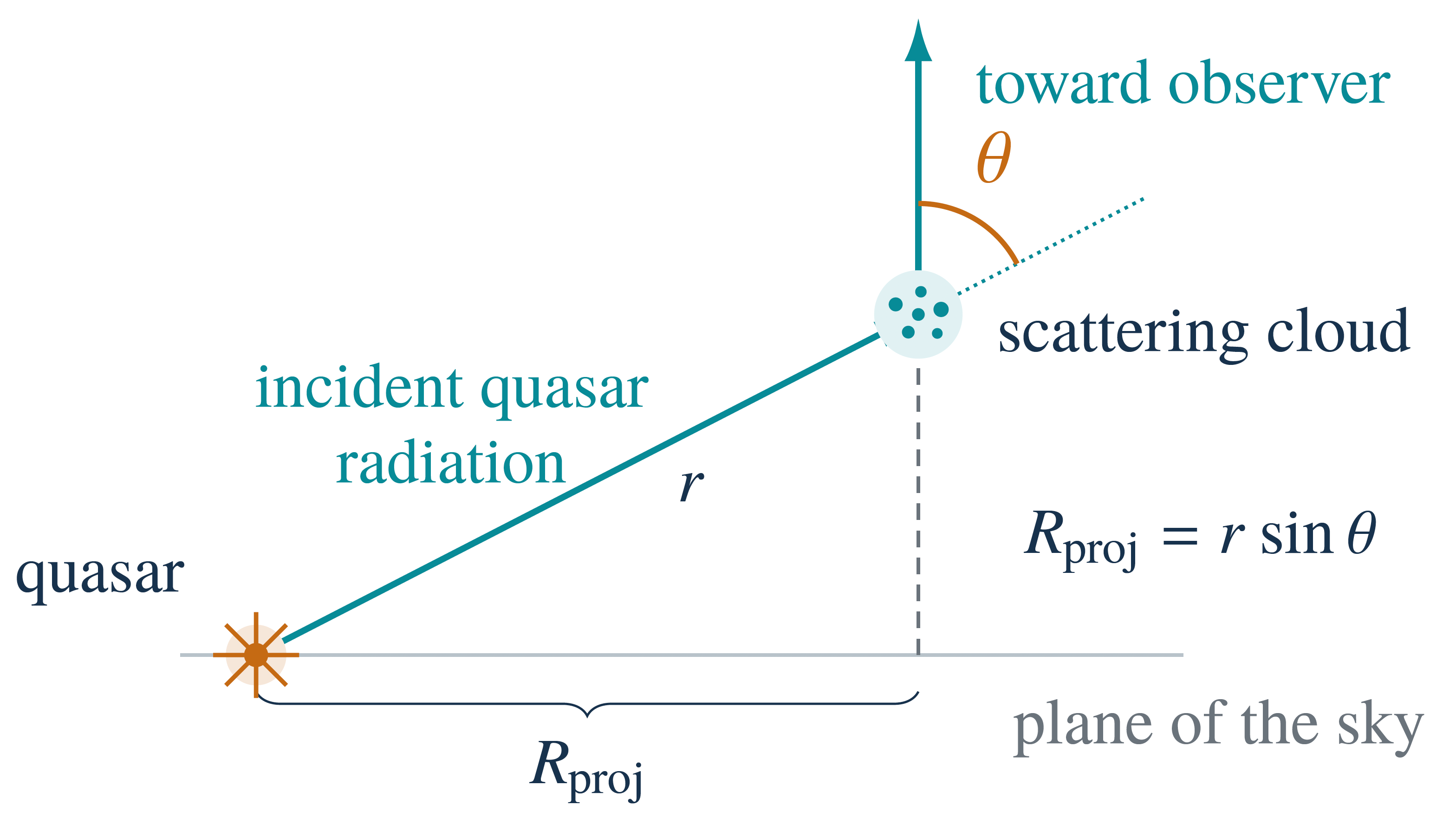}
			\caption{Geometry used in the scattering calculation.  The scattering
				cloud lies at a physical distance $r$ from the quasar and at a projected
				sky-plane distance $R_{\rm proj}$.  The scattering angle $\theta$ is
				measured between the incident photon direction and the outgoing
				direction toward the observer, such that
				$R_{\rm proj}=r\sin\theta$.  The diagram is schematic and not to scale.}
			\label{fig:scattering_geometry}
		\end{figure}
		
		For an optically thin, single-scattering cloud with scatterer number
		density $n_s$, a path length $d\ell$ contributes an emitted specific
		intensity
		$dI_{\nu_e}=F_{\nu_e}n_s
		(d\sigma_{\nu_e}/d\Omega)d\ell$ toward the observer.  Integrating through
		the cloud gives
		$I_{\nu_e}=F_{\nu_e}N_s(d\sigma_{\nu_e}/d\Omega)$, where
		$N_s=\int n_s\,d\ell$ is the line-of-sight scatterer column.  Specific
		intensity per unit frequency is cosmologically dimmed according to
		$I_{\nu_o}=I_{\nu_e}/(1+z)^3$.  Using the geometric relation
		$R_{\rm proj}=r\sin\theta$ shown in
		Figure~\ref{fig:scattering_geometry} then gives
		
		\begin{equation}
			I_{\nu_o}
			=
			\frac{L_{\nu_e,{\rm QSO}}}{4\pi(1+z)^3}
			\left[
			N_s\sin^2\theta
			\frac{d\sigma_{\nu_e}}{d\Omega}
			T_{\nu_e}
			\right]
			R_{\rm proj}^{-2}.
			\label{eq:sb_radius_scattering_intensity_appendix}
		\end{equation}
		
		Here $I_{\nu_o}$ is the observed specific intensity per unit solid
		angle.  Extinction along the incident quasar--cloud path and the outgoing
		cloud--observer path is represented by the transmission
		
		\begin{equation}
			T_{\nu_e}
			=
			e^{-\tau_{\nu_e}}
			=
			10^{-0.4A_{\nu_e}},
		\end{equation}
		
		where $\tau_{\nu_e}$ is the total extinction optical depth and
		$A_{\nu_e}=1.086\,\tau_{\nu_e}$ is the corresponding extinction in
		magnitudes.  The total extinction is
		
		\begin{equation}
			A_{\nu_e}
			=
			A_{\nu_e,{\rm QSO\rightarrow cloud}}
			+
			A_{\nu_e,{\rm cloud\rightarrow obs}}.
		\end{equation}
		
		\subsection{Region-by-Region Scattering Energetics}
		\label{app:scattering_energetics}
		
		We begin with the general relation before specifying the intrinsic
		quasar spectrum or scattering model.  Solving the scattering-intensity
		relation above for the line-of-sight scatterer column gives
		
		\begin{equation}
			N_s
			=
			\frac{4\pi(1+z)^3 I_{606}R_{\rm proj}^{2}}
			{L_\nu(\lambda_{\rm rest,piv})\sin^2\theta
				\left[d\sigma(\lambda_{\rm rest,piv},\theta)/d\Omega\right]
				T(\lambda_{\rm rest,piv})}.
		\end{equation}
		
		Here $I_{606}$ is the observed F606W flux-density intensity and
		$L_\nu(\lambda_{\rm rest,piv})$ is the intrinsic quasar specific
		luminosity at the emitted-frame wavelength sampled by F606W.  The
		remaining terms are the projected distance $R_{\rm proj}$, scattering
		angle $\theta$, differential scattering cross section
		$d\sigma/d\Omega$, and combined transmission $T$ along the
		quasar--scatterer and scatterer--observer paths.
		
		We calculate $I_{606}$ from the native, pre-PSF-matching F606W aperture
		flux density divided by the native F606W segmentation-aperture area.
		Equivalently, the intensity is related to the AB surface brightness
		$\mu_{606}$ by
		
		\begin{equation}
			I_{606}
			=
			\frac{10^{-0.4(\mu_{606}+48.60)}}{\Omega_{\rm as}},
			\label{eq:f606w_intensity_conversion}
		\end{equation}
		
		where $\mu_{606}$ is expressed in mag~arcsec$^{-2}$ and
		$\Omega_{\rm as}=2.35044\times10^{-11}\ {\rm sr}$ is the solid angle
		of one square arcsecond.
		
		We normalize the intrinsic quasar spectrum at 2500~\AA\ because our empirical bolometric correction relates the bolometric
		luminosity directly to $\nu L_\nu$ at this wavelength.  The corresponding
		frequency is
		$\nu_{2500}=c/(2500~\text{\AA})=1.20\times10^{15}\ {\rm Hz}$.  We define
		the dimensionless bolometric correction as
		
		\begin{equation}
			k_{2500}
			\equiv
			\frac{L_{\rm bol}}{\nu_{2500}L_{\nu,2500}},
		\end{equation}
		
		where $L_{\rm bol}$ is the total radiative luminosity and
		$L_{\nu,2500}$ is the intrinsic specific luminosity per unit frequency
		at 2500~\AA.  We adopt $k_{2500}=2.75$, the mean 2500~\AA\ correction
		recovered by \citet{krawczyk+13} through mid-IR-to-UV SED fitting of
		approximately 120,000 type~1 quasars at $0.06<z<5.5$.  This value is
		also broadly consistent with the UV bolometric corrections predicted
		for powerful $1<z<2$ radio-loud quasars from SED modeling of the 3CRR
		sample \citep[][the appropriate counterparts to the type~2 radio
		galaxies studied here]{azadi+22}.
		
		We adopt
		$\log(L_{\rm bol}/{\rm erg\,s^{-1}})=47.28$, the median of the 29
		$z>1$ 3CR quasars studied by \citet{breiding+24}.  Assigning this
		representative type~1 quasar luminosity to the type~2 radio galaxies
		follows the orientation-based unification picture
		\citep[i.e.,][]{lawrence_elvis_82,antonucci_miller_85,barthel89,antonucci93}.
		
		For simplicity, we represent each F606W measurement monochromatically
		at the filter pivot wavelength,
		$\lambda_{\rm piv}=5887$~\AA.\footnote{As a check, we repeated the
			calculation by weighting the quasar spectrum and scattering cross
			section over the full F606W throughput.  The inferred columns changed
			by no more than 6.6\% across the sample and scattering models,
			with no effect on our conclusions.}  For a source at redshift $z$, the
		corresponding emitted-frame wavelength is
		
		\begin{equation}
			\lambda_{\rm rest,piv}
			=
			\frac{\lambda_{\rm piv}}{1+z}.
		\end{equation}
		
		Because F606W samples different rest-frame wavelengths across the
		sample, $\lambda_{\rm rest,piv}$ ranges from approximately 1695 to
		2831~\AA.  We approximate the intrinsic quasar continuum as
		$L_\nu\propto\nu^\alpha$, with $\alpha=-0.44$
		\citep{vandenberk+01}.  The intrinsic specific luminosity at the
		emitted-frame wavelength sampled by F606W is therefore
		
		\begin{equation}
			L_\nu(\lambda_{\rm rest,piv})
			=
			\frac{L_{\rm bol}}{k_{2500}\nu_{2500}}
			\left(
			\frac{2500~\text{\AA}}{\lambda_{\rm rest,piv}}
			\right)^\alpha .
			\label{eq:quasar_lnu_pivot}
		\end{equation}
		
		If a fraction $f_{\rm UV,scat}$ of the measured F606W emission is
		attributed to scattered quasar light, the intensity produced by
		scattering is $I_{606,\rm scat}=f_{\rm UV,scat}I_{606}$.  Substituting
		this relation together with
		Equations~\ref{eq:f606w_intensity_conversion} and
		\ref{eq:quasar_lnu_pivot} into the general scatterer-column relation
		gives
		
		\begin{equation}
			\begin{aligned}
				N_s
				={}&
				\frac{4\pi(1+z)^3 k_{2500}\nu_{2500}R_{\rm proj}^{2}}
				{\Omega_{\rm as}L_{\rm bol}\sin^2\theta
					\left[d\sigma(\lambda_{\rm rest,piv},\theta)/d\Omega\right]
					T(\lambda_{\rm rest,piv})}
				\\
				&\times
				f_{\rm UV,scat}\,
				10^{-0.4(\mu_{606}+48.60)}
				\left(
				\frac{\lambda_{\rm rest,piv}}
				{2500~\text{\AA}}
				\right)^\alpha .
			\end{aligned}
			\label{eq:required_scatterer_column_general_appendix}
		\end{equation}
		
		This is the general relation presented in
		Equation~\ref{eq:required_scatterer_column_general} in the main text.
		For the fiducial calculation, we assign all of the measured F606W
		emission to scattered quasar light, so that $f_{\rm UV,scat}=1$, and
		neglect attenuation along both light paths, so that $T=1$.  Adopting
		$k_{2500}=2.75$,
		$\log(L_{\rm bol}/{\rm erg\,s^{-1}})=47.28$,
		$\lambda_{\rm piv}=5887$~\AA, and $\alpha=-0.44$ then reduces
		Equation~\ref{eq:required_scatterer_column_general_appendix} to the
		fiducial relation presented in
		Equation~\ref{eq:required_scatterer_column} in the main text and used
		in the region-by-region calculations.
		
		The measured inputs to this calculation are $\mu_{606}$,
		$R_{\rm proj}$, and $z$.  The remaining quantities specify the scattering model and geometry.  We evaluate the fiducial relation
		separately for electron and dust scattering and then convert the
		inferred columns into ionized-gas and dust masses within the UV
		apertures.  For electron scattering, $N_{s,\rm fid}=N_e$ and the
		Thomson differential cross section is evaluated at the
		scattering angle.  For dust scattering, $N_{s,\rm fid}=N_{\rm H}$ and
		the wavelength-dependent differential cross section per H nucleus is
		evaluated at $\lambda_{\rm rest,piv}$ and the scattering angle.
		
		For a given UV region and a fixed scattering model, geometry, and quasar
		continuum slope, the required scatterer column scales with the assumed
		scattered fraction, attenuation, bolometric correction, and
		hidden-quasar luminosity as
		
		\begin{equation}
			N_s
			=
			N_{s,\rm fid}
			f_{\rm UV,scat}\,
			10^{0.4A_{\rm UV}}
			\left(\frac{k_{2500}}{2.75}\right)
			\left(
			\frac{10^{47.28}\ {\rm erg\,s^{-1}}}{L_{\rm bol}}
			\right),
			\label{eq:scattering_scaling_simple}
		\end{equation}
		
		where $A_{\rm UV}$ is the combined effective attenuation along the
		quasar--scatterer and scatterer--observer paths.  The inferred
		ionized-gas and dust masses scale in the same way because, at fixed
		aperture area and scattering model, each mass is directly proportional
		to the scatterer column.  A stellar or nebular contribution therefore
		lowers the required column and mass, whereas attenuation or a less
		luminous hidden quasar raises them.
	\end{document}